\documentclass[3p,11pt,sort&compress]{elsarticle}

\usepackage[algoruled,vlined]{algorithm2e}
\usepackage{nomencl}
\usepackage{amsmath}
\usepackage{xcolor}
\usepackage{graphicx}
\usepackage{hyperref}
\usepackage{import}

\journal{arXiv}

\newlength{\myFigureHeight}
\hypersetup{colorlinks=true}

\makeatletter
\providecommand{\doi}[1]{%
  \begingroup
    \let\bibinfo\@secondoftwo
    \href{http://dx.doi.org/#1}{%
      doi:\discretionary{}{}{}%
      \nolinkurl{#1}%
    }%
  \endgroup
}
\makeatother

\newcommand{\Fref}[1]{Figure~\ref{#1}}
\newcommand{\Eref}[1]{Eq.~\eqref{#1}}

\newcommand{\Sref}[1]{Section~\ref{#1}}
\newcommand{\Tref}[1]{Table~\ref{#1}}

\newcommand{\de}[1]{\,{\mathrm d}#1} 
\newcommand{\del}[2]{\mbox{$\displaystyle\frac{#1}{#2}$}}
\newcommand{\der}[2]{\del{\de #1}{\de #2}}

\newcommand{\M}[1]{{\boldsymbol #1}}
\newcommand{\trn}{{\sf ^T}}        

\newcommand{\lay}[1]{^{(#1)}}
\newcommand{\h}[1]{h\lay{#1}}
\newcommand{\Le}{L_e}	
\newcommand{\eL}[2]{L_{#1}\lay{#2}}
\newcommand{\be}{b}	

\newcommand{\prony}{p}
\newcommand{\tp}{\theta}
\newcommand{\tre}{t_\text{true}}
\newcommand{\tad}{t}
\newcommand{\tin}{t'}
\newcommand{\aT}{a_{T}}
\newcommand{\T}{T}

\newcommand{\MQ}{\M{Q}}
\newcommand{\MDnu}{\M{D}_{\nu}}

\newcommand{\E}{E}
\newcommand{\G}{G}
\newcommand{\Kmod}{K}
\newcommand{\Pnum}{\nu}
\newcommand{\A}[1]{A\lay{#1}}%
\newcommand{\I}[1]{I\lay{#1}}%
\newcommand{\As}[1]{A_\mathrm{s}\lay{#1}}%

\newcommand{\x}[1]{x}			
\newcommand{\X}[1]{x}			
\newcommand{\z}[1]{z\lay{#1}}
\newcommand{\Z}[1]{z\lay{#1}}

\newcommand{\uFS}[1]{u\lay{#1}}
\newcommand{\wFS}[1]{w\lay{#1}}
\newcommand{\uVK}[1]{u\lay{#1}}
\newcommand{\wVK}[1]{w\lay{#1}}
\newcommand{\uc}[1]{u_0\lay{#1}}
\newcommand{\wc}[1]{w_0\lay{#1}}
\newcommand{\Mu}{\M{u}}
\newcommand{\uce}[2]{u_{0,#2}\lay{#1}}
\newcommand{\wce}[2]{w_{0,#2}\lay{#1}}

\newcommand{\rot}[1]{\varphi_y\lay{#1}}
\newcommand{\Rot}[1]{\varphi_y\lay{#1}}
\newcommand{\rote}[2]{\varphi\lay{#1}_{y,#2}}
\newcommand{\curv}[1]{\kappa_y\lay{#1}}
\newcommand{\Curv}[1]{\kappa_y\lay{#1}}
\newcommand{\curve}[2]{\kappa\lay{#1}_{y,#2}}

\newcommand{\strain}{\varepsilon}
\newcommand{\sstrain}{\gamma}
\newcommand{\strainx}[1]{\varepsilon_x\lay{#1}}
\newcommand{\sstrainxz}[1]{\gamma_{xz}\lay{#1}}
\newcommand{\strainxc}[1]{\varepsilon_{x,0}\lay{#1}}
\newcommand{\strainxce}[1]{\varepsilon_{x,0,e}\lay{#1}}
\newcommand{\sstrainxze}[1]{\gamma_{xz,e}\lay{#1}}

\newcommand{\Strainx}[1]{\varepsilon_x\lay{#1}}
\newcommand{\Sstrainxz}[1]{\gamma_{xz}\lay{#1}}
\newcommand{\Strainxc}[1]{\varepsilon_{x,0}\lay{#1}}

\newcommand{\dstrain}{e}

\newcommand{\stress}{\sigma}
\newcommand{\sstress}{\tau}
\newcommand{\dstress}{s}

\newcommand{\el}{_{e}}
\newcommand{\numel}{n^\mathrm{\el}}
\newcommand{\Md}{\M{d}}
\newcommand{\ite}[1]{{^{#1}}}
\newcommand{\Mfext}{\M{f}_{\mathrm{ext}}}

\newcommand{\Kt}[1]{K_{\mathrm{t},#1}}
\newcommand{\MKt}{\M{K}_\mathrm{t}}

\newcommand{\Mfint}{\M{f}_\mathrm{int}}

\newcommand{\half}{\tfrac{1}{2}}

\begin{document}

\begin{frontmatter}

\title{Comparison of viscoelastic finite element models for laminated glass beams}

\author{Alena Zemanov\'{a}\corref{cor1}}
\cortext[cor1]{Corresponding author}
\ead{alena.zeman@fsv.cvut.cz}

\author{Jan Zeman\corref{}}
\ead{jan.zeman@fsv.cvut.cz}

\author{Michal \v{S}ejnoha}
\ead{sejnom@fsv.cvut.cz}

\address{Department of Mechanics, Faculty of Civil Engineering,
  Czech Technical University in Prague, \\ Th\'{a}kurova 7, 166 29 Prague
  6, Czech Republic}

\begin{abstract}
Laminated glass elements, which consist of stiff elastic glass layers connected with a compliant viscoelastic polymer foil, exhibit geometrically non-linear and time/temperature-sensitive behavior. In computational modeling, the viscoelastic effects are often neglected or a detailed continuum formulation typically based on the volumetric-deviatoric elastic-viscoelastic split is used for the interlayer. Four layerwise beam theories are introduced in this paper, which differ in the non-linear beam formulation at the layer level (von K\'{a}rm\'{a}n/Reissner) and in constitutive assumptions for the interlayer (a viscoelastic solid with the time-independent bulk modulus/Poisson ratio). We perform detailed verification and validation studies at different temperatures and compare the accuracy of the selected formulation with simplified elastic solutions used in practice. We show that all the four formulations predict very similar responses. Therefore, our suggestion is to use the most straightforward formulation that combines the von K\'{a}rm\'{a}n model with the assumption of time-independent Poisson ratio. The simplified elastic model mostly provides response in satisfactory agreement with full viscoelastic solutions. However, it can lead to unsafe or inaccurate predictions for rapid changes of loading. These findings provide a suitable basis for extensions towards laminated plates and glass layer fracture, owing to the modular format of layerwise theories.
\end{abstract}


\end{frontmatter}

\section{Introduction}

Laminated glass units consist of multiple glass sheets bonded together with
polymer foils. The main reason to combine brittle glass sheets with soft polymer
layers is to enhance the post-fracture behavior, so that a unit retains the
load-bearing capacity after fracture of one or multiple glass
plies~\cite{Haldimann:2008:SUG}. Of course, the interlayer foil also influences
the overall response of the unfractured elements, which is in the focus of this
study, by mediating the shear interaction among the glass layers.

The glass layers can be accurately modeled as linear elastic, but the polymer
foil exhibits a strongly time/temperature-dependent response, e.g.~\cite[and
references
therein]{Serafinaviciusa:2013:LTLG,Huang:2014:ESLG,Shitanoki:2014:PNM}. Several
types of interlayers have found their use in practice. From the first
application of laminated glass members with polyvinyl butyral foil (PVB) in car
windscreens, laminated glass has expanded into the building constructions, and
new types of interlayer foils became available, such e.g. ionoplast polymer foil
(IP) that is stiffer and less sensitive to viscoelastic phenomena than PVB at
common temperatures.

As discussed in our previous contribution~\cite{Zemanova:2014:NMFS}, the behavior of
polymeric interlayer is often approximated as \emph{linear elastic}, assuming a
known temperature profile and duration of loading. The following methods have
been developed for \emph{elastic laminated glass beams} under this simplification:
\begin{itemize}
\item The basic approach to approximate the mechanical response of laminated
  glass units involves two limiting cases: the
  \emph{layered} case representing an assembly of
  independent glass layers and the \emph{monolithic}
  case with thickness equal to the combined thickness of glass layers
  and interlayers, under the assumption of geometric
  linearity~\cite{Behr:1985:LGU}. 

\item Single-layered approaches, such as \emph{effective thickness} methods that
approximate laminated units by monolithic, geometrically linear, systems that achieve the
  equivalent maximum deflection or
  stress~\cite{Koutsawa:2007:SFVA,Benninson:2008:HPLG,Galuppi:2012:ETL}; see
  also~\cite{Aenlle:2013:ETC,Eisentraeger:2015:AFO} for an overview.

\item \emph{Multi-layered} approaches, explicitly resolving all the layers.
  These include \emph{analytical} formulations, which are solved in a closed
  form for specific boundary
  conditions~\cite{Hooper:1973:BAL,Norville:1998:BSL,Ivanov:2006:AMO} and
  geometrically linear behavior, and \emph{numerical}
  formulations based on the finite-difference~\cite{Asik:2005:MMB} or
  finite-element methods~\cite{Schultze:2012:ALG,Zemanova:2014:NMFS,Jaskowiec:2015:XFEM} that can
  treat general boundary conditions and geometrically non-linear effects. 

\end{itemize}

Solely the \emph{multi-layered approaches} have been extended to account for
viscoelastic effects, because they offer sufficient accuracy in stresses and
strain distributions at the ply level. An \emph{analytical} model of
geometrically linear, simply supported, beams with viscoelastic interlayer
represented by a Maxwell chain was developed by \citet{Galuppi:2012:LBV,Galuppi:2013:DLG}, who approximated the
viscoelastic solution by the Fourier series in the spatial coordinate and solved
the ensuing integral equations in the closed form to account for the time
dependent effects. The same approach was later successfully used to study the stability of laminated columns~\cite{Galuppi:2014:BTL} or 
time-dependent response of cold-bent laminated beams~\cite{Galuppi:2015:LCS}. The most recent analytical study has been performed by~\citet{Wu:2016:EST}, who develop a closed-form solutions for two-dimensional stress and strain fields in two-layer simply-supported beam with interlayer response described by the standard linear model. The only detailed \emph{numerical} studies we are aware of include geometrically
non-linear three-dimensional finite element simulations of viscoelastic plates
under transverse loading by~\citet{Duser:1999:AGBL} and~\citet{Bennison:1999:FLB}, or laminated beams under
lateral-torsional buckling by~\citet{Bedon:2014:AEA}.

The goal of this paper is to propose a viscoelastic geometrically
non-linear finite element formulation for laminated glass beams based on
\emph{layerwise} refined laminate theories
initiated by~\citet{Mau:1973:RLP}.
The basis of our formulation is the elastic finite strain model
developed earlier by~\citet{Zemanova:2014:NMFS},
where we demonstrated that it offers accuracy comparable to
detailed two-dimensional simulations at much lower computational cost.
Here, this formulation is extended to account for the
time/temperature-dependent behavior of the interlayer, which is
incorporated into the numerical model by the incremental exponential
algorithm by~\citet{Zienkiewicz:1968:NVS}. In particular, we
will compare the formulations arising from the two modeling aspects:
\begin{itemize}

\item \emph{beam kinematics}, involving the finite strain shear-deformable
  \emph{Reissner} beam formulation and a simpler variant based on the Timoshenko
  beam theory complemented with the \emph{von K\'{a}rm\'{a}n} assumptions of large
  deflections,

\item constitutive assumptions for \emph{viscoelastic models}, where we will
consider models based on time/temperature-dependent shear modulus, complemented
with either \emph{constant}~(time-independent) \emph{bulk modulus} or
\emph{constant Poisson} ratio.

\end{itemize}
The differences in the mechanical results of the corresponding models will be
studied in detail and verified against detailed finite element simulations, with the
goal to determine the simplest, but still sufficiently accurate, formulation
that is suitable for the generalization of elastic models of laminated glass
plates~\cite{Zemanova:2015:FEM} to the viscoelastic regime.

The rest of the paper is organized as follows. In \Sref{sec:formulation}, we introduce the different variants of laminated glass beam models, starting from kinematics assumptions in \Sref{sec:kinematic}, proceeding to interlayer constitutive relations in \Sref{sec:constitutive}, up to the finite element discretization in \Sref{sol_FS&VK}. \Sref{sec:comp_ver_val} is devoted to the comparison of different formulations, along with an additional verification, validation, and parametric studies. The most important findings are summarized in \Sref{sec:conclusions}. Finally, in Appendix~\ref{app:sensitivity_analysis} we gather technical details regarding the finite element technology, in order to make the paper self-contained.

\section{Finite element models}\label{sec:formulation}

The models introduced in this work share the following assumptions: 

\begin{itemize}
  \item they assume planar cross sections of individual layers but not of the
  whole laminated glass unit, because the stiffness of glass layer and the effective
  stiffness of the interlayer differs by at least three orders of magnitude, e.g.~\cite{Asik:2005:MMB}, 

  \item being based on Mau's refined plate theory~\cite{Mau:1973:RLP}, each
  layer is treated independently and the inter-layer compatibility is enforced
  by the Lagrange multipliers. The main advantages of this approach is that the
  different constitutive models for each layer can be easily combined
  together and delamination phenomena can be efficiently accounted for, if
  needed, e.g.,~\cite{Kruis:2013:MII}\footnote{Note, however, that the perfect inter-layer
  adhesion is assumed in this paper, because the application of high pressures
  and temperatures during the production process result in high gluing forces of
  chemical nature~\cite{Larcher:2012:ENI}.}.
\end{itemize} 
As noted above, the formulations differ in the beam kinematics
(Finite-Strain~(FS) Reissner vs. Large-Deflection von K\'{a}rm\'{a}n~(VK) models)
and in the assumption of viscoelastic constitutive models~(constant bulk modulus
$K$ vs. the Poisson ratio $\nu$). These considerations lead to four models, which are briefly introduced in this section. 

Our notation is as follows. Scalar quantities are denoted by lightface letters,
e.g., $a$, and matrices are denoted in bold, e.g., $\M{a}$ or $\M{A}$. In
addition, $\M{A}\trn$ stands for the matrix transpose, $\M{A}^{-1}$ for the
matrix inverse, and $a\lay{i}$ denotes that the quantity $a$ is related to
the $i$-th layer. Also note that in order to avoid a profusion of notation, in
\Sref{sec:kinematic} we omit the dependence of quantities of interest on the
time variable $t$, while in \Sref{sec:constitutive} we omit the layer index
$\bullet\lay{i}$, because the constitutive description holds only for the
interlayer.

\subsection{Kinematics}\label{sec:kinematic}
%
\subsubsection{Reissner finite strain model}\label{sec:formulation_FS}
\begin{figure}[ht]
\centerline{
	\def\svgwidth{90mm}
	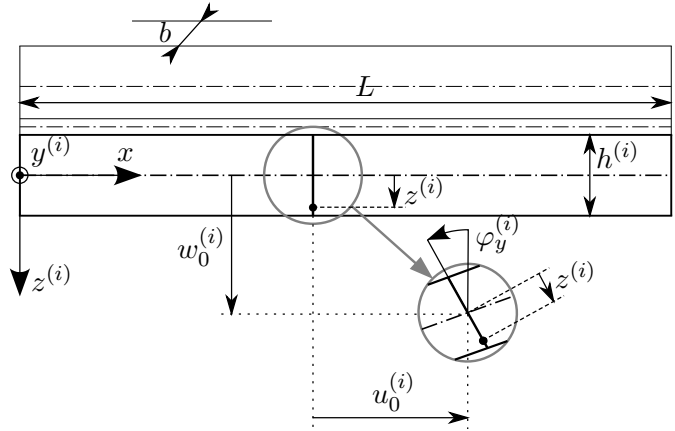
}
\caption{Kinematics of a cross section of the bottom layer of a laminated
beam~($i=3$).}
\label{fig:lam_beam}
\end{figure}

%
In the Reissner finite strain beam theory~\cite{Reissner:1972:ODFSBT}, the non-zero
displacement components of the $i$-th layer can be parameterized as, \Fref{fig:lam_beam},
\begin{subequations}\label{eq:dispFS}
\begin{align}
\uFS{i} (\X{i},\Z{i}) & = \uc{i}( \X{i} ) + \sin\left(\Rot{i}( \X{i} )\right) \Z{i}, 
\\ 
\wFS{i} (\X{i},\Z{i}) & = \wc{i}(\X{i}) + \left( \cos\left(\Rot{i}( \X{i} ) \right) - 1\right) \Z{i},
\end{align}
\end{subequations}
where $\uc{i}$ and~$\wc{i}$ are centerline displacements, $\Rot{i}$ is the
cross-section rotation, $x$ is the coordinate measured along the centerline, $\Z{i}$ is the coordinate measured along the cross-section, and $i=1,2,3$ refers to individual layers.

The inter-layer compatibility is ensured via the geometric continuity conditions
at the interfaces between the layers~(with $i=1,2$)
\begin{align*}
\uFS{i} (\X{i},\frac{\h{i}}{2}) - \uFS{i+1} (\X{i},-\frac{\h{i+1}}{2}) & =
0,
\\  
\wFS{i} (\X{i},\Z{i}) - \wFS{i+1} (\X{i},-\frac{\h{i+1}}{2}) & = 0,
\end{align*}
where the perfect horizontal and vertical adhesion is supposed.
From~\eqref{eq:dispFS}, the compatibility condition become
\begin{subequations}\label{eq:comp_conFS}
\begin{align}
\uc{i}( \X{i} ) - \uc{i+1}( \X{i} ) + \frac{\h{i}}{2} \sin\left(\Rot{i}( \X{i} )\right) + \frac{\h{i+1}}{2} \sin\left(\Rot{i+1}( \X{i} )\right) & = 0, 
\\  
\wc{i}(\X{i}) - \wc{i+1}(\X{i}) + \frac{\h{i}}{2} \cos\left(\Rot{i}( \X{i} )\right) + \frac{\h{i+1}}{2} \cos\left(\Rot{i+1}( \X{i} )\right) - \frac{\h{i}}{2} - \frac{\h{i+1}}{2} & = 0,
\end{align}
\end{subequations}

The non-zero normal, $\Strainx{i}$, and shear, $\Sstrainxz{i}$, strain
components follow from
\begin{subequations}\label{eq:strain_components}
\begin{align}
\Strainx{i}(\X{i}, \Z{i})
& = 
\Strainxc{i}(\X{i}) + \Curv{i}(\X{i})\Z{i},
\\
\Sstrainxz{i}(\X{i})
& = 
\sin \left( \Rot{i}(\X{i}) \right) 
\left( 
  1 + \der{\uc{i}(\X{i})}{\X{i}}
\right)
+ 
\cos \left( \Rot{i}(\X{i}) \right)
\der{\wc{i}(\X{i})}{\X{i}},
\label{eq:H}
\end{align}
\end{subequations}
where $\Strainxc{i}$ and $\Curv{i}$ denote the generalized axial strain
and pseudo-curvature of the layer reference axis, introduced by
Reissner~\cite{Reissner:1972:ODFSBT} in the form
\begin{subequations}\label{eq:generalized_measures}
\begin{align}
\Strainxc{i} (\x{i})
& =
\cos \left( \Rot{i}(\X{i}) \right)
\left( 
  1 + \der{\uc{i}(\X{i})}{\X{i}}
\right)
- 
\sin \left( \Rot{i}(\X{i}) \right)
\der{\wc{i}(\X{i})}{\X{i}}
-
1,
\\
\Curv{i} (\x{i})
& =  
\der{\Rot{i}(\X{i})}{\X{i}}. \label{eq:curv_FS}
\end{align}
\end{subequations}

\subsubsection{Von K\'{a}rm\'{a}n model}\label{sec:formulation_VK}
%
A simpler model is obtained when adopting the von-K\'{a}rm\'{a}n assumptions of moderate deflections $\wc{i}$ and small rotations $\Rot{i}$, e.g.~\cite[Section 4.3]{Reddy:2004:FEM}. Under these assumptions, the displacements of the $i$-th
layer~\eqref{eq:dispFS} simplify to
\begin{eqnarray*}
\uVK{i}(\x{i},\z{i}) &=& \uc{i}(\x{i}) + \rot{i}(x) \z{i}, \\
\wVK{i}(\x{i}) &=& \wc{i}(\x{i}),
\end{eqnarray*}
and the inter-layer compatibility conditions~\eqref{eq:dispFS} become
linear in $\Rot{i}$,
\begin{subequations}\label{eq:comp_conVK}
\begin{align}
\uc{i}( \X{i} ) - \uc{i+1}( \X{i} ) + \frac{\h{i}}{2} \Rot{i}( \X{i} )+ \frac{\h{i+1}}{2} \Rot{i+1}( \X{i} ) &=0,\\  
\wc{i}(\X{i}) - \wc{i+1}(\X{i}) &=0.
\end{align}
\end{subequations}

In addition, the normal and shear strains are expressed as, cf.~\eqref{eq:strain_components},
\begin{subequations}\label{eq:VKstrain}
\begin{align}
\strainx{i}(\x{i},\z{i}) 
& = 
\strainxc{i}(\x{i}) +\curv{i}(\x{i})\z{i}, 
\label{eq5:vonKarman_strain_x}
\\
\sstrainxz{i}(\x{i}) 
& = 
\rot{i}(\x{i}) + \der{\wc{i}}{\x{i}}(\x{i}),
\end{align}
\end{subequations}
where the axial strains $\Strainxc{i}$ must be measured by the Green-Lagrange
strain tensor because of large deflections $\wc{i}$,  
\begin{align}\label{eq:VK_normal_strain}
\Strainxc{i} (\x{i})
=
\der{\uc{i}}{\x{i}}(\x{i}) + 
\half \left( \der{\wc{i}}{\x{i}}(\x{i})\right)^2,
\end{align}
while the layer pseudo-curvature $\curv{i}$ still follows
from~\eqref{eq:curv_FS}.

\subsection{Constitutive relations}\label{sec:constitutive}
As indicated earlier, accurate description of time/temperature-dependent response of polymeric interlayer is decisive for the development of predictive models of laminated glass structures. In \Sref{sec:PVB} we introduce such a model in the framework of linear viscoelasticity with accelerated/retarded time to account for the temperature dependence. We will also clarify the impact of the constitutive assumptions on the structure of stress-strain relations. The integral form of constitutive equations used in \Sref{sec:PVB} is not very convenient for the numerical implementation, because all
history variables during the whole loading process must be stored in order to evaluate the integral. Therefore, in the following sections, we will develop an incremental approach based on the well-established exponential algorithm by~\citet{Zienkiewicz:1968:NVS}. The case of the constant bulk modulus is treated in detail in \Sref{sec:GK_model}, leading to a rather involved procedure because of assumptions of beam theories.  
\Sref{sec:Gnu_model} then explains how the procedure simplifies under the assumption of time-independent Poisson ratio.

\subsubsection{Time/temperature-dependent behavior of polymer foil}\label{sec:PVB}

\begin{figure}[ht]
\small (a)\includegraphics[height=\myFigureHeight]{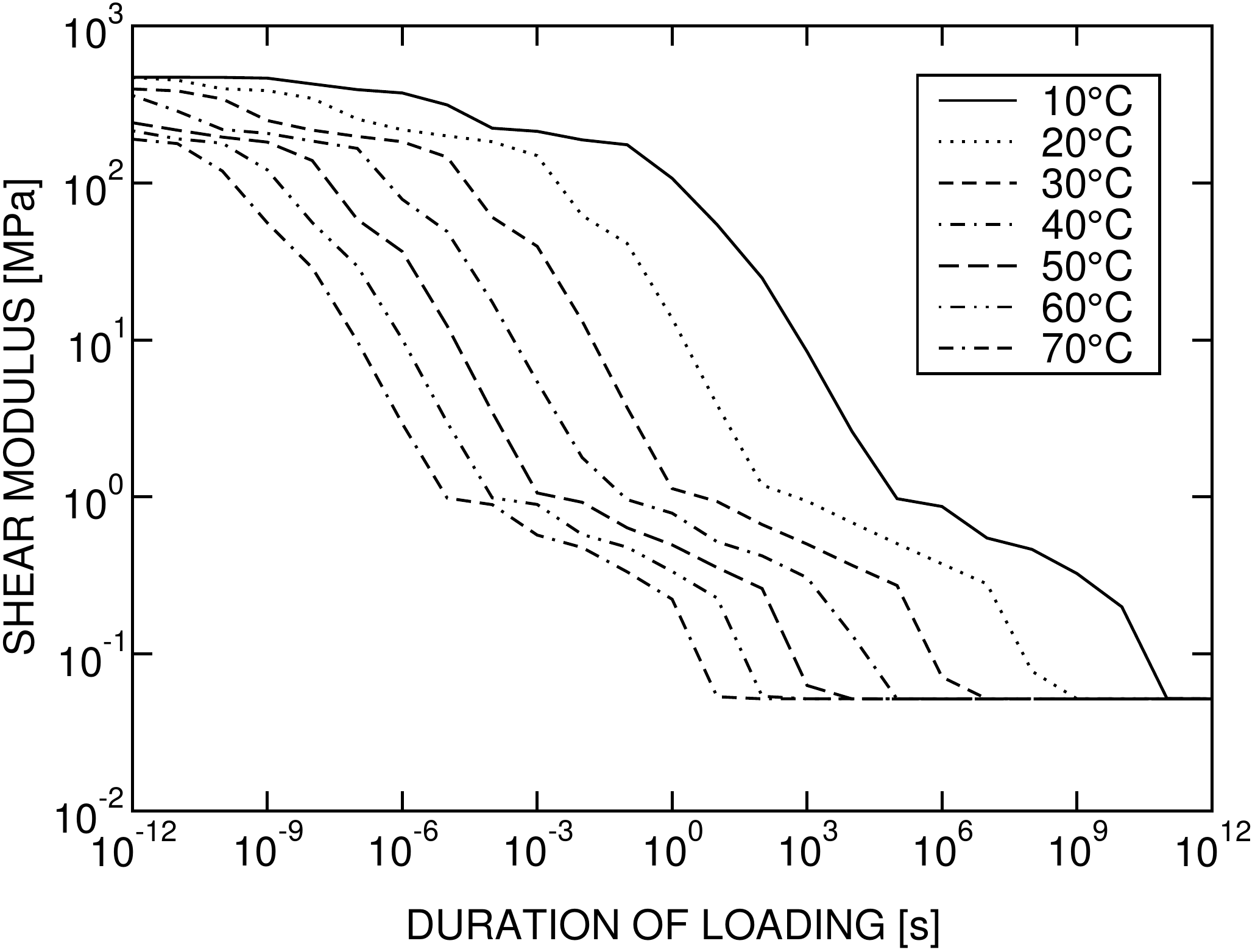}	
\hfill
\small (b)\includegraphics[height=\myFigureHeight]{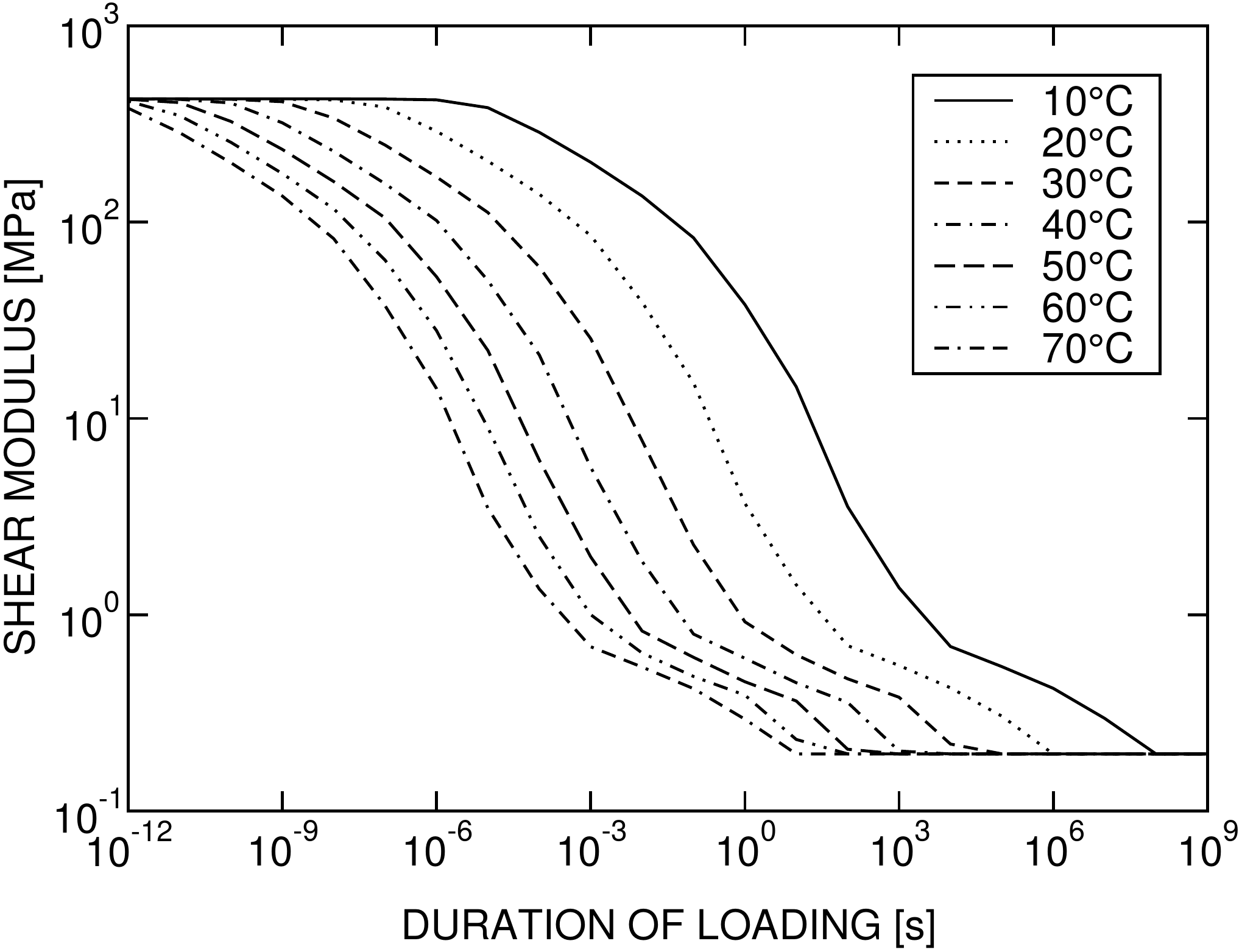}	
\caption{Shear modulus $\G$ as a function of the duration of loading and
temperature, after (a)~\citet{Bennison:1999:FLB} and
(b)~\citet{Pelayo:2013:MSLGP}.}
\label{fig:GPVB2}
\end{figure}

The main engineering property relevant to the composite behavior of the units is
the shear-stress versus shear-strain characteristics of the soft interlayer. The
shear modulus of the polymer interlayer $\G$ is experimentally determined as a
function of duration of loading and temperature, see \Fref{fig:GPVB2} for
data for two PVB foils.

The temperature dependence is taken into account by the time-temperature
superposition principle, in which the true time $\tre$ is replaced by the
adjusted value $\tad=\tre/\aT$ modified by the temperature-dependent shift
factor $\aT$. We employ the Williams-Landel-Ferry~(WLF)
equation~\cite{Williams:1955:TDR} for this purpose:
\begin{equation}
\log{\aT} = -\frac{C_1 (\T - T_0)}{C_2 + T - T_0},
\label{eq6:shiftfactor}
\end{equation}
in which $C_1$ and $C_2$ are material constants, and $\T$ and $T_0$
are the current and reference temperatures, respectively.

For further discussion, it will be convenient to
decompose the vectors of strain $\M{\strain}$ and stress~$\M{\stress}$, 
\begin{align*}
\M{\strain} = 
\begin{bmatrix}
\strain_x & \strain_y & \strain_z & \sstrain_{xy} & \sstrain_{xz} & \sstrain_{yz} \end{bmatrix}\trn,
&&
\M{\stress} = \begin{bmatrix}\stress_x & \stress_y & \stress_z & \sstress_{xy} & \sstress_{xz} & \sstress_{yz} \end{bmatrix}\trn,
\end{align*}
into the volumetric and deviatoric parts
\begin{align}\label{eq6:split}
\M{\strain}(\tad) = \frac{1}{3}\strain_\mathrm{V}(\tad) \M{i} + \M{\dstrain}(\tad), 
\\
&&
\M{\stress}(\tad) = \stress_\mathrm{m}(\tad) \M{i} + \M{\dstress}(\tad),
\end{align}
where $\tad$ denotes the time instant, $\M{i} = \begin{bmatrix}1 & 1 & 1 & 0 & 0
& 0\end{bmatrix}\trn$. The volumetric parts follow from $\strain_\mathrm{V} =
\M{i}\trn\M{\strain}$ and $\stress_\mathrm{m} = \frac{1}{3}\M{i}\trn\M{\stress}$, and
the deviatoric parts have the components
\begin{align*}
\M{\dstrain} = \begin{bmatrix}\dstrain_x & \dstrain_y & \dstrain_z & \sstrain_{xy} & \sstrain_{xz} & \sstrain_{yz} \end{bmatrix}\trn
, &&
\M{\dstress} = \begin{bmatrix}\dstress_x & \dstress_y & \dstress_z & \sstress_{xy} & \sstress_{xz} & \sstress_{yz} \end{bmatrix}\trn
.
\end{align*}

Assuming isotropic material behavior and a smooth strain history
$\M{\strain}(\tad)$ with $\M{\strain}(0) = \M{0}$, the stress at time
$\tad$ reads, e.g.,~\cite[Section 1.2]{Christensen:1982:TVI}
\begin{equation}\label{eq6:integral_law_gen}
\M\stress(\tad) 
= 
\stress_\mathrm{m}(\tad) \M{i}
+
\M{\dstress}(\tad)
=
\M{i} 
\int_{0}^{\tad} 
  \Kmod(\tad-\tin) \der{{\strain}_V}{\tin}(\tin) \de\tin
+
2 \MQ \int_{0}^{\tad} \G(\tad-\tin) \der{\M{\dstrain}}{\tin}(\tin) \de \tin,
\end{equation}
where $\MQ = \text{diag} \left[
\begin{array}{cccccc}
1 & 1 & 1 & 0.5 & 0.5 & 0.5
\end{array}
\right]$, and $\Kmod(\tad-\tin)$ and $\G(\tad-\tin)$ denotes the bulk and shear relaxation moduli that completely characterize the interlayer response.

\begin{figure}[th]
	\centerline{%
		\def\svgwidth{85mm}
		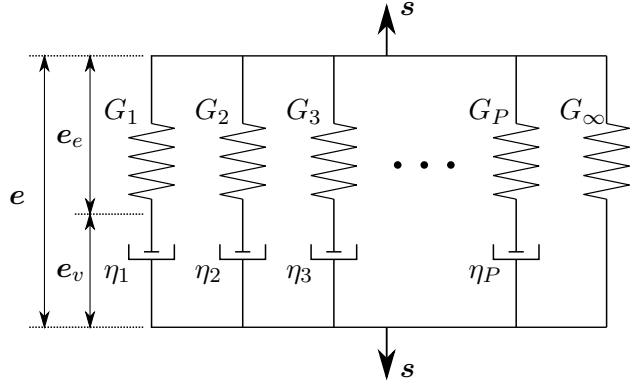}%
	\caption{Generalized Maxwell chain consisting of $P$ viscoelastic units and
		one elastic spring.}
	\label{fig6:maxwell}
\end{figure}

In what follows, the response under shear will be represented by a generalized Maxwell chain, \Fref{fig6:maxwell}, whose relaxation function is provided by the Dirichlet-Prony series, e.g.,~\cite[page 32]{Christensen:1982:TVI}
\begin{eqnarray}
\G(\tad-\tin) = \G_{\infty} +  \sum_{\prony=1}^P \G_{\prony} \exp^{-\frac{\tad-\tin}{\tp_{\prony}}}
= \G_{0} -  \sum_{\prony=1}^P \G_{\prony} (1-\exp^{-\frac{\tad-\tin}{\tp_{\prony}}}),
\label{eq6:PronyG}					
\end{eqnarray}
where $G_\infty$ is the long-term shear modulus, $P$ stands for the number of viscoelastic units, $G_{\prony}$~denotes the shear modulus of
the ${\prony}$-th unit, $\theta_{\prony} = \eta_{\prony} / G_{\prony}$ is its relaxation time related to the viscosity~$\eta_{\prony}$, and $\G_0=\G_{\infty} +  \sum_{\prony=1}^P \G_{\prony}$ is the elastic shear modulus of the chain.

In order to close the constitutive description for the interlayer, two assumptions on the bulk relaxation functions will be considered. First, the \emph{constant bulk modulus} approximation, e.g.~\cite{Duser:1999:AGBL,Bennison:1999:FLB}, under which the material behaves as a linear elastic solid for volumetric loading. The constitutive equation 
\begin{equation}\label{eq:stress_K}
\M\stress(\tad) = \Kmod \M{i} \strain_\mathrm{V}(\tad) + 2 \int\limits_{0}^{\tad} \G(\tad-\tin) \MQ \der{\M{\dstrain}}{\tin}(\tin) \de \tin,
\end{equation}
thus additionally involves the constant elastic bulk modulus. Second, \emph{the Poisson ratio} $\nu$ is taken as \emph{constant}, e.g.~\cite{Koutsawa:2007:SFVA,Wu:2016:EST}, which simplifies the constitutive relation~\eqref{eq6:integral_law_gen} into, e.g.,~\cite[Eq.~(29.63)]{Jirasek:2002:IAS}:
\begin{equation}\label{eq:stress_nu}
\M\stress(\tad) = \MDnu \int\limits_{0}^{\tad} \G(\tad-\tin) \der{\M{\strain}}{\tin}(\tin) \de \tin,
\end{equation}
with the constant matrix $\MDnu$ provided by 
\begin{align}\label{eq:Dnu_def}
\MDnu =
\frac{1}{1-2\nu}
\left[
\begin{array}{cccccc}
{2(1-\nu)}
& 
{2\nu}
& 
{2\nu}
& 0 & 0 & 0\\
{2\nu}
& 
{2(1-\nu)}
& 
{2\nu}
& 0 & 0 & 0\\
{2\nu}
& 
{2\nu}
& 
{2(1-\nu)}
& 0 & 0 & 0\\
0 & 0 & 0 & 1 & 0 & 0\\
0 & 0 & 0 & 0 & 1 & 0\\
0 & 0 & 0 & 0 & 0 & 1
\end{array}
\right].
\end{align} 

\subsubsection{Incremental formulation for constant bulk
modulus}\label{sec:GK_model}

\paragraph{Material point}

We start by dividing the time interval of interest $[ 0;
t_{\max}]$ into non-equidistant time instants $0 = t_0 < t_1 <
t_2 < \cdots < t_{N-1} < t_N = t_{\max}$ (recall that these correspond to temperature-adjusted values according to~\eqref{eq6:shiftfactor}). The strain history over the time interval $[ t_n; t_{n+1} ]$ is assumed to be known, so that  
\begin{align*}
\M{\dstrain}(t_{n+1}) = \M{\dstrain}(t_n) + \Delta \M{\dstrain}, 
\end{align*}
where $\Delta \M{\dstrain}$ denotes the increment of deviatoric strain between the
time instants $t_n$ and $t_{n+1}$. An analogous relation holds for the deviatoric stresses,
\begin{align*}
\M{\dstress}(t_{n+1}) = \M{\dstress}(t_n) + \Delta \M{\dstress}, 
\end{align*}
so that, assuming that $\M{\dstress}(t_n)$ is known, the goal is to determine the increment of the deviatoric stress $\Delta \M{\dstress}$. 

For a linear strain variation over the time interval $[t_n; t_{n+1}]$, this increment can be expressed as~\cite{Zienkiewicz:1968:NVS}
\begin{align}\label{eq6:incr_deviatoric}
\Delta\M{\dstress} = 2 \widehat{\G} \MQ \Delta\M{\dstrain} + \Delta\widehat{\M{\dstress}},
\end{align}
where $\widehat{\G}$ stands for the effective shear modulus over the time interval and
$\Delta\widehat{\M{\dstress}}$ for the stress relaxation under the zero strain increment. For the Maxwell chain model, \Fref{fig6:maxwell}, these quantities follow from
\begin{align}\label{eq6:incr_detailed}
\widehat{\G} = \G_{\infty} +  \sum_{\prony=1}^P \G_{\prony} \frac{\tp_{\prony}}{\Delta \tad} (1-\exp^{-\frac{\Delta \tad}{\tp_{\prony}}}), 
&&
\Delta \widehat{\M{\dstress}} = \sum_{\prony=1}^P \Delta
\widehat{\M{\dstress}}_{\prony} = 
- \sum_{\prony=1}^P
\M{\dstress}_{\prony}(t_n) (1 - \exp^{-\frac{\Delta \tad}{\tp_{\prony}}}), 
\end{align}
where $\Delta \tad = t_{n+1} - t_n$ is the temperature-adjusted time step~(we assume for simplicity that the temperature remains
constant over the interval), and $\M{\dstress}_\prony$ and $\Delta
\widehat{\M{\dstress}}_{\prony}$ stand for the values of the deviatoric stresses
and its relaxation in the $\prony$-th unit.

Because the volumetric response is assumed to be elastic, the increment of the mean strain, $\Delta \stress_\mathrm{m}$, can be obtained from the given increment of the volumetric strain $\Delta \strain_\mathrm{V}$ directly as
\begin{align}\label{eq6:incr_volumetric}
\Delta \stress_\mathrm{m} 
=
\Kmod 
\Delta \strain_\mathrm{V}.
\end{align}

\paragraph{Beams} As follows from the standard assumptions of the conventional beam theory, e.g.~\cite[Section 29]{Sokolnikoff:1956:MTE}, the only non-zero stress components are the increments of normal and shear stresses $\Delta \sigma_x$ and $\Delta \tau_{xz}$; the corresponding volumetric and deviatoric stress components entering~\eqref{eq6:incr_volumetric} and~\eqref{eq6:incr_deviatoric} thus become $\Delta \sigma_{\mathrm m} = 1/3 \Delta \sigma_x$, $\Delta s_x = 2/3 \Delta \sigma_x$, and $\Delta s_{xz} = \Delta \tau_{xz}$. 

In order to obtain the corresponding strain increments $\Delta\strain_x$ and $\Delta\gamma_{xz}$ needed in the beam formulations, we employ the volumetric-deviatoric split~\eqref{eq6:split} and invert the relations~\eqref{eq6:incr_deviatoric} and~\eqref{eq6:incr_volumetric} to obtain
\begin{subequations}\label{eq:strain_increments}
\begin{align}
\Delta\strain_x 
& = 
\frac{1}{3} \Delta \strain_\mathrm{V}
+
\Delta\dstrain_x
=
\frac{1}{3\Kmod}\Delta \stress_\mathrm{m}
+
\frac{1}{2 \widehat{\G}}
\left( 
  \Delta\dstress_x - \Delta\widehat{\dstress}_x 
\right)
=
\frac{1}{\widehat{\E}} \Delta \stress_x
-
\frac{1}{2 \widehat{\G}} \Delta\widehat{\dstress}_x,
\\
\Delta\gamma_{xz}
& = 
\frac{1}{\widehat{\G}}
\left( 
  \Delta\tau_{xz} - \Delta\widehat{\tau}_{xz} 
\right),
\end{align}
\end{subequations}
from which we directly obtain the incremental stress-strain relations in the form
\begin{align}
\Delta \stress_{x}
= 
\widehat{\E} \Delta \strain_{x} 
+
(1+\widehat{\nu})
\Delta \widehat{\dstress}_{x},
&&
\Delta \tau_{xz}
= 
\widehat{\G} \Delta \gamma_{xz} 
+
\Delta \widehat{\tau}_{xz}.
\label{eq:stresses_beam}
\end{align}
The symbols $\widehat{\E}$ and $\widehat{\nu}$ in Eqs.~\eqref{eq:strain_increments} and~\eqref{eq:stresses_beam} stand for the effective Young modulus and Poisson ratio, obtained as $\widehat{\E} = (9K \widehat{\G}) / (\widehat{\G} + 3 K)$ and $\widehat{\nu} = (3K - 2\widehat{\G}) / (2(3K + \widehat{\G}))$, e.g.~\cite[Table~D.1]{Jirasek:2002:IAS}. It will also prove useful to relate the deviatoric normal strain to the full normal strain via 
\begin{equation}\label{eq:incr_normal_strain}
\Delta \dstrain_{x}
= 
\frac{2}{3} (1+\widehat{\nu}) \Delta \strain_{x} 
-
\frac{(1+\widehat{\nu})}{9\Kmod}
\Delta \widehat{\dstress}_{x}.
\end{equation}

\paragraph{Cross-section}

In order to obtain the constitutive equations for a cross-section located at $x$, recall from~\eqref{eq:generalized_measures} or~\eqref{eq:VKstrain} that the normal strain $\strain_x$ is an affine function in the through-the-thickness coordinate $z$, whereas the shear strain $\sstrain_{xz}$ is independent of $z$. Upon integrating the stress increments~\eqref{eq:stresses_beam} over the cross-section, we obtain
\begin{subequations}\label{eq6:incr_int_forces_K}
\begin{align}
\Delta N_x(x)
& =
\widehat{E}A \Delta \varepsilon_{x,0}(x)
+
(1+\widehat{\nu}) \int_A \Delta \widehat{\dstress}_{x} (x,z) \de A
=
\widehat{E}A \Delta \varepsilon_{x,0}(x)
+
\Delta \widehat{N}_x(x),
\\
\Delta M_y(x)
& =
\widehat{E}I_y \Delta \kappa_y(x)
+
(1+\widehat{\nu}) \int_A \Delta \widehat{\dstress}_{x} (x,z) z \de A
=
\widehat{E}I_y \Delta \kappa_y(x)
+
\Delta \widehat{M}_y(x),
\\
\Delta V_z(x)
& =
\widehat{\G}A_\mathrm{s} \Delta \gamma_{xz}(x)
+
\int_{A_\mathrm{s}} \Delta \widehat{\tau}_{xz} (x) \de A
=
\widehat{G} A_\mathrm{s} \Delta \gamma_{xz}(x)
+
\Delta \widehat{V}_z(x),
\end{align}
\end{subequations}
where $\Delta N_x$, $\Delta V_z$, and $\Delta M_y$ stand for the increments of the normal and shear forces, and the bending moment~(the corresponding relaxation forces are distinguished by a hat); $\Delta \varepsilon_{0,x}$, $\Delta \gamma_{xz}$, and $\Delta \kappa_y$ are the corresponding increments of the centerline strain, the transverse shear, and the pseudo-curvature determined from~\eqref{eq:generalized_measures} or~\eqref{eq:VKstrain}; and $A$, $I_y$, and $A_\mathrm{s}$ denote the cross-section area, the second moment of area, and the shear-effective area.\footnote{%
For three-layered units (glass/foil/glass) we set the value of the shear-effective area to $\As{i}=5/6 \A{i}$ for glass layers with $i=1,3$ and $\As{2}=\A{2}$ for the foil. For these values, which correspond to a quadratic shear stress variation in glass and to constant shear in the interlayer, e.g.~\cite{Birman:2002:CSC}, we achieved the best agreement with the response of the detailed two-dimensional model from the ADINA solver~(not shown).} Note that in~\eqref{eq6:incr_int_forces_K}, we utilized the fact that the cross-section and the effective material constants $\widehat{\E}$, $\widehat{\G}$, and $\widehat{\nu}$ are identical for the whole layer.

As follows from Eq.~\eqref{eq6:incr_int_forces_K}, the increments of internal forces involve the relaxation contributions $\widehat{N}_x$, $\widehat{M}_y$, and $\widehat{V}_z$, which in turn depend on the values of the deviatoric stresses in elements of the Maxwell chain~(\ref{eq6:incr_deviatoric}, \ref{eq6:incr_detailed}) across the cross-section. Assuming that $\M{\dstress}_\prony(t_0) = \M{0}$, a closer inspection of these relations together with Eq.~\eqref{eq:incr_normal_strain} reveals that the normal and shear deviatoric stress components exhibit the affine-constant distribution across the cross-section at $x$ at time $t_n$:
\begin{align}\label{eq:Maxwell_stress_K}
\dstress_{x,\prony}(t_n, x, z)
=
\dstress_{x,\prony}^N(t_n, x)
+
\dstress_{x,\prony}^M(t_n, x) z, && 
\tau_{xz,\prony}(t_n, x, z)
=
\tau_{xz,\prony}^V(t_n, x)
\text{ for }
\prony = 1, 2, \ldots, P;
\end{align}     
therefore, the three quantities $\dstress_{x,p}^N$, $\dstress_{x,p}^M$, and $\tau^V_{xz,p}$ completely determine the stress distribution in the $p$-th elements of the Maxwell chain over the cross-section. They also allow for a compact expression of the incremental relaxation forces in the form, 
\begin{align}\label{eq6:incr_fintV}
\Delta \widehat{N}_x( x )
=
(1+\widehat{\nu}) A \Delta \widehat{\dstress}_x^N( x ), 
&& 
\Delta \widehat{M}_y( x )
= 
(1+\widehat{\nu}) I_y \Delta \widehat{\dstress}_x^M( x ),
&& 
\Delta \widehat{V}_z( x )
=
A_\mathrm{s} \Delta\widehat{\tau}^V_{xz}( x ),
\end{align}
where, e.g., 
\begin{align*}
\Delta \widehat{\dstress}_x^N( x )
=
\sum_{\prony=1}^{P}
\Delta \widehat{\dstress}_{x,\prony}^N( x ), 
&&
\Delta \widehat{\dstress}_{x,\prony}^N( x )
=
-
\dstress_{x,\prony}^N( t_n, x ) (1 - \exp^{-\frac{\Delta \tad}{\tp_{\prony}}})
,
\end{align*}
recall~\eqref{eq6:incr_detailed}. The remaining quantities $\Delta \widehat{\dstress}^M_x$ and $\Delta \widehat{\tau}^V_{xz}$ follow by analogy.

At the end of the time interval, the stresses in the $p$-th unit must be updated on the basis of increments of the generalized strains $\Delta \varepsilon_{0,x}$, $\Delta \kappa_y$, and $\Delta \gamma_{xz}$ for a cross-section at $x$. Combining~\eqref{eq:incr_normal_strain} and~\eqref{eq6:incr_deviatoric} with~\eqref{eq:strain_components} or~\eqref{eq:VKstrain}, one obtains
\begin{subequations}\label{eq6:dev_increment_final_K}
\begin{align}
\Delta \dstress_{x,\prony}^N( x )
& = 
\frac{4}{3}(1+\widehat{\nu}) \widehat{G}_{\prony} \Delta \varepsilon_{0,x}( x ) +
\Delta \widehat{\dstress}_{x,\prony}^N( x ) 
-
\frac{2}{9}\frac{\widehat{G}_{\prony} (1+\widehat{\nu})}{K} \Delta
\widehat{\dstress}_{x}^N( x ),
\\
\Delta \dstress_{x,\prony}^M( x )
& = 
\frac{4}{3}(1+\widehat{\nu}) \widehat{G}_{\prony} \Delta \kappa_y( x ) 
+
\Delta \widehat{\dstress}_{x,\prony}^M( x ) 
-
\frac{2}{9} \frac{\widehat{G}_{\prony} (1+\widehat{\nu})}{K} 
\Delta \widehat{\dstress}_{x}^M( x ),
\\
\Delta \tau^V_{xz,\prony}( x )
& = 
\widehat{G}_{\prony} \Delta \gamma_{xz}( x ) 
+
\Delta \widehat{\tau}^V_{xz,\prony}( x ).
\label{eq6:dev_increment_final_tau_xz}
\end{align}
\end{subequations}

\subsubsection{Incremental formulation for constant Poisson ratio}\label{sec:Gnu_model}
%
\paragraph{Material point}
Assuming the same time discretization as in \Sref{sec:GK_model}, the strain and stress increments over the interval $[t_n; t_{n+1}]$ are expressed as 
\begin{align*}
\M{\strain}(t_{n+1}) = \M{\strain}(t_n) + \Delta \M{\strain}, &&
\M{\stress}(t_{n+1}) = \M{\stress}(t_n) + \Delta \M{\stress},
\end{align*}
which involve the full strain and stress tensors. Comparing~\eqref{eq6:integral_law_gen} with~\eqref{eq:stress_nu}, we see that the incremental formula~\eqref{eq6:incr_deviatoric} from the previous section becomes 
\begin{align}\label{eq:inc_stress_nu}
\Delta \M\stress = \MDnu \widehat{\G} \Delta \M{\strain} + \Delta \widehat{\M\stress},
&&
\Delta \widehat{\M{\stress}} = \sum_{\prony=1}^P \Delta
\widehat{\M{\stress}}_{\prony} = - \sum_{\prony=1}^P
\M{\stress}_{\prony}(t_n) (1 - \exp^{-\frac{\Delta \tad}{\tp_{\prony}}}),
\end{align}
where matrix $\MDnu$ depends on the constant Poisson ratio~$\nu$ via~\eqref{eq:Dnu_def} and $\widehat{\G}$ is defined with~\eqref{eq6:incr_detailed}.

\paragraph{Beams} Increments of the non-zero components of stress vector, $\Delta \stress_{x}$ and $\Delta \tau_{xz}$, follow directly from~\eqref{eq:inc_stress_nu} in the form
\begin{align*}\label{eq:stresses_beam_nu}
\Delta \stress_{x}
= 
\widehat{\E} \Delta \strain_{x} 
+
\Delta \widehat{\stress}_{x},
&&
\Delta \tau_{xz}
= 
\widehat{\G} \Delta \gamma_{xz} 
+
\Delta \widehat{\tau}_{xz},
\end{align*}
with the effective Young modulus now given by $\widehat{\E} = 2(1 + \nu) \widehat{\G}$, cf.~\eqref{eq:stresses_beam}

\paragraph{Cross-section}

Proceeding in the same way as in the previous section, the incremental constitutive relations at the cross-section level, Eq.~\eqref{eq6:incr_int_forces_K}, attain a similar form:
\begin{subequations}\label{eq6:incr_int_forces_nu}
\begin{align}
\Delta N_x(x)
& =
\widehat{E}A \Delta \varepsilon_{x,0}(x)
+
\int_A \Delta \widehat{\stress}_{x} (x,z) \de A
=
\widehat{E}A \Delta \varepsilon_{x,0}(x)
+
\Delta \widehat{N}_x(x),
\\
\Delta M_y(x)
& =
\widehat{E}I_y \Delta \kappa_y(x)
+
\int_A \Delta \widehat{\stress}_{x} (x,z) z \de A
=
\widehat{E}I_y \Delta \kappa_y(x)
+
\Delta \widehat{M}_y(x),
\\
\Delta V_z(x)
& =
\widehat{G}A_\mathrm{s} \Delta \gamma_{xz}(x)
+
\int_{A_\mathrm{s}} \Delta \widehat{\tau}_{xz} (x) \de A
=
\widehat{G} A_\mathrm{s} \Delta \gamma_{xz}(x)
+
\Delta \widehat{V}_z(x),
\end{align}
\end{subequations}

Again, the increments of relaxation forces $\Delta \widehat{N}_x$, $\Delta \widehat{V}_z$, and $\Delta \widehat{M}_y$ depend on the distribution of stresses in the Maxwell units over the cross-section. By analogy to~\eqref{eq:Maxwell_stress_K}, the normal and shear stresses in the $p$-th unit admit the decomposition:
\begin{align*}
\stress_{x,\prony}(t_n, x, z)
=
\stress_{x,\prony}^N(t_n, x)
+
\stress_{x,\prony}^M(t_n, x) z, 
&& 
\tau_{xz,\prony}(t_n, x, z)
=
\tau_{xz,\prony}^V(t_n, x)
\text{ for }
\prony = 1, 2, \ldots, P,
\end{align*}     
from which we obtain
\begin{align*}
	\Delta \widehat{N}_x ( x )
	=
	A \Delta \widehat{\stress}_x^N  ( x ),
	&&
	\Delta \widehat{M}_y  ( x )
	=
	I_y \Delta \widehat{\stress}_x^M  ( x ),
	&&
	\Delta \widehat{V}_z  ( x )
	=
	A_\mathrm{s} \Delta\widehat{\tau}^V_{xz} ( x ),
\end{align*}
with, e.g.,
\begin{align*}
\Delta \widehat{\stress}_x^N( x )
=
\sum_{\prony=1}^{P}
\Delta \widehat{\stress}_{x,\prony}^N( x ), 
&&
\Delta \widehat{\stress}_{x,\prony}^N( x )
=
-\stress_{x,\prony}^N( t_n, x ) (1 - \exp^{-\frac{\Delta \tad}{\tp_{\prony}}}).
\end{align*}

Finally, we need to relate the increments of the history variables $\Delta\stress_{x,p}^N$, $\Delta\stress_{x,p}^M$, and $\Delta\tau^V_{xz,\prony}$ to the increments of the generalized strains $\Delta \varepsilon_{x,0}$, $\Delta \kappa_y$, and $\Delta \gamma_{xz}$. The relations for the normal components follow directly from~\eqref{eq:inc_stress_nu} as
\begin{subequations}\label{eq6:dev_increment_final_nu}
\begin{align}
\Delta\stress_{x,p}^N( x ) 
& =
2 ( 1 + \nu ) \widehat{G}_p \Delta \varepsilon_{x,0}( x ) 
+
\Delta \widehat{\stress}_{x,\prony}^N( x ),
\\
\Delta\stress_{x,p}^M( x ) 
& =
2 ( 1 + \nu ) \widehat{G}_p \Delta \kappa_y( x ) 
+
\Delta \widehat{\stress}_{x,\prony}^M( x ).
\end{align}
\end{subequations}
The shear stress increment  $\Delta \tau^V_{xz,p}$ is obtained from~\eqref{eq6:dev_increment_final_tau_xz}.

\subsection{Finite element model}\label{sol_FS&VK}

In this section, we extend the geometrically nonlinear Reissner elastic solver from~\cite{Zemanova:2014:NMFS} to the incremental viscoelastic formulation and discuss the simplifications due to the von-K\'{a}rm\'{a}n assumptions. 

\subsubsection{Variational setting}\label{sec:variational}

Similarly to~\cite{Zemanova:2014:NMFS}, we derive the model using variational arguments. Because the glass layers are assumed to be elastic, their response is governed by a quadratic stored energy density introduced in~\cite[Eq.~(13)]{Zemanova:2014:NMFS}:
\begin{align*}
W\lay{i} 
\left( \M{\strain}\lay{i} \right)
=
\half
\left( 
\E\lay{i}\A{i} \left( \strainxc{i} \right)^2
+
\G\lay{i}\As{i}
\left( \curv{i} \right)^2
+
\E\lay{i}I_y\lay{i}
\left( \sstrainxz{i} \right)^2
\right),
\end{align*}
with~$i={1,3}$ and with the column matrix $\M{\strain}\lay{i} = \begin{bmatrix} \strainxc{i} & \curv{i} & \sstrainxz{i} \end{bmatrix}\trn$ collecting the generalized strain components. For the interlayer, $i=2$, the energy density is history dependent to account for the incremental form of the cross-section constitutive relations~\eqref{eq6:incr_int_forces_K} or~\eqref{eq6:incr_int_forces_nu}. For the time interval $[t_n; t_{n+1}]$, it reads 
\begin{align*}
W\lay{2}_{n+1}
\left( \M{\strain}\lay{2} \right)
=
W\lay{2}_{n}
\left( \M{\strain}\lay{2}(t_n) \right)
+
\Delta W\lay{2}
\left( \M{\strain}\lay{2} \right),
\end{align*}
with $W\lay{2}_{0} = 0$ and 
\begin{align}\label{eq:incr_energy_density}
\Delta W\lay{2}
\left( \M{\strain}\lay{2} \right)
& =
\half
\left( 
\widehat{\E}\lay{2}\A{2} \left( \Delta \strainxc{2} \right)^2
+
\widehat{\G}\lay{2}\As{2}
\left( \Delta \curv{2} \right)^2
+
\widehat{\E}\lay{2}I_y\lay{2}
\left( \Delta \sstrainxz{2} \right)^2
\right)
\nonumber \\
& + 
\left( N\lay{2}_x(t_n) + \Delta \widehat{N}\lay{2}_x \right)
\Delta \strainxc{2}
+
\left( V\lay{2}_z(t_n) + \Delta \widehat{V}\lay{2}_z \right)
\Delta \sstrainxz{2}
\nonumber\\
& +
\left( M\lay{2}_y(t_n) + \Delta \widehat{M}\lay{2}_y \right)
\Delta \curv{2}.
\end{align}
Here, e.g., $\Delta \strainxc{2} = \strainxc{2} - \strainxc{2}(t_n)$, and the remaining terms follow by analogy.

The energy functional at the structural level collects the contributions of layers
\begin{align*}
\Pi_{\mathrm{int}, n+1}( \Mu)
=
\sum_{i = 1, 3}
\int_{\Omega\lay{i}} 
W\lay{i} \Bigl( \M{\strain} \bigl( \Mu \lay{i} ( x ) \bigr) \Bigr)
\de x
+
\int_{\Omega\lay{2}} 
W_{n+1}\lay{2} \Bigl( \M{\strain} \bigl( \Mu \lay{2} ( x ) \bigr) \Bigr)
\de x,
\end{align*}
where $\Mu = \begin{bmatrix} \uc{i}, \wc{i}, \rot{i} \end{bmatrix}_{i=1}^3$ stands for the centerline displacements and cross-section rotations and $\M{\strain}( \bullet )$ abbreviates the strain-displacement relations for the Reissner or the von-K\'{a}rm\'{a}n kinematics introduced in Sections~\ref{sec:formulation_FS} and~\ref{sec:formulation_VK}, respectively. Complementing the internal energy functional with the energy of external forces provides
\begin{align}\label{eq:energy_total}
\Pi_{n+1}( \Mu)
=
\Pi_{\mathrm{int}, n+1}( \Mu)
+
\sum_{i=1}^3 \Pi\lay{i}_\mathrm{ext}(t_{n+1}, \Mu\lay{i}).
\end{align}
Note that the admissible generalized displacements $\Mu$ must satisfy the inter-layer continuity conditions~\eqref{eq:comp_conFS} or~\eqref{eq:comp_conVK}, written compactly as
\begin{align}\label{eq:continutity}
\M{c}\bigl( \Mu( x ) \bigr) = \M{0} \text{ for } x \in [0;L].
\end{align}
The generalized displacements at time $t_{n+1}$, $\Mu(t_{n+1})$, then follow by the minimization of the total energy functional~\eqref{eq:energy_total} under the equality constraints~\eqref{eq:continutity}.

\subsubsection{Discretization}

Discretization of the model is accomplished using the standard finite element procedures. To that purpose, each layer is discretized into the identical number of $\numel$ elements $\Omega\lay{i}\el$ and the generalized displacements are approximated as
\begin{align}\label{eq:displ_approx}
\Mu\lay{i} 
\approx 
\M{N}\el(x)
\Md\el\lay{i}
\text{ for } 
x \in \Omega\lay{i}\el,
\end{align}
where $\M{N}_e$ stands for the $3 \times 6$ matrix of linear basis functions and $\Md\el\lay{i}$ is the $6 \times 1$ column matrix of element degrees of freedom; see Appendix~\ref{app:sensitivity_analysis} for details. After inserting the approximation~\eqref{eq:displ_approx} into~\eqref{eq:energy_total}, the internal energy becomes the function of the nodal degrees of freedom
\begin{align*}
\Pi_{\mathrm{int},n+1}( \Md )
= 
\sum_{i=1,3}
\sum_{e=1}^{\numel}
\Pi_{\mathrm{int},e}\lay{i}\Bigl( \Md\el\lay{i} \Bigr)
+
\sum_{e=1}^{\numel}
\Pi_{\mathrm{int},n+1,e}\lay{2}\Bigl( \Md\el\lay{2} \Bigr),
\end{align*}
where $\Md = \begin{bmatrix} \Md_j\lay{i} \end{bmatrix}_{i=1,j=1}^{3,\numel+1}$ collects the layer nodal displacements and, e.g., 

\begin{align*}
\Pi_{\mathrm{int},n+1,e}\lay{2}\Bigl( \Md\el\lay{2} \Bigr)
=
\int_{\Omega\lay{2}\el} 
W_{n+1}\lay{2} \Bigl( \M{\strain} \bigl(  \M{N}\el(x) \Md\el\lay{i} \bigr) \Bigr)
\de x.
\end{align*}

In analogy to the previous section, the column matrix of nodal displacements $\Md(t_{n+1})$ follows from minimization of the total energy function
\begin{align*}
\Pi_{n+1}( \Md )
=
\Pi_{\mathrm{int},n+1}( \Md )
-
\sum_{i=1}^3 
\sum_{e=1}^{\numel}
\Md\el\lay{i}{}\trn \M{f}\lay{i}_{\mathrm{ext},e} \left( t_{n+1} \right),
\end{align*}
in which $\M{f}\lay{i}_{\mathrm{ext},e}$ denotes the generalized nodal forces due to external loading of the $i$-th layer. The kinematically admissible displacements must satisfy the constraint
\begin{align}\label{eq:constraint}
\M{c}( \Md ) = \M{0},
\end{align}
with the meaning of the inter-layer continuity conditions~\eqref{eq:continutity} enforced at the nodes.

\subsubsection{Solution procedure}

Following the standard procedures of equality-constrained optimization, e.g.,~\cite[Chapter~14]{Bonnans:2003:NOTPA}, we consider the Lagrange function 
\begin{align*}
\mathcal{L}_{n+1}( \Md, \M{\lambda} )
=
\Pi_{n+1}( \Md )
+
\M{\lambda}\trn \M{c}( \Md ),
\end{align*}
where the column matrix of Lagrange multipliers $\M{\lambda}$ has the meaning of interface nodal forces that ensure the inter-layer compatibility, \eqref{eq:constraint}. The nodal displacements $\Md(t_{n+1})$ and the corresponding Lagrange multipliers $\M{\lambda}(t_{n+1})$ then arise as the saddle point of the Lagrange function. 

The optimality conditions lead to a system of equations non-linear in $\Md(t_{n+1})$ which is solved iteratively using the Newton method. To this purpose, we express the $(k+1)$-th approximation to $\Md(t_{n+1})$ as
\begin{align*}
\ite{k+1}\Md = \ite{k}\Md + \ite{k+1}\delta \Md
\text{ for } 
k = 0, 1, \ldots
\end{align*}
where $\ite{k} \Md$ denotes a known previous iterate; for $k=0$, we set $\ite{0} \Md = \Md(t_{n})$. The displacement increment $\ite{k+1} \delta \Md$ and the vector of Lagrange multipliers $\ite{k+1} \M{\lambda}$ are determined from the linearized system
\begin{align}\label{eq6:systemFS}
\begin{bmatrix}
 \ite{k}\M{K} & \ite{k}\M{C}\trn \\
 \ite{k}\M{C} & \M{0}
\end{bmatrix}
\begin{bmatrix}
 \ite{k+1} \delta \Md \\
 \ite{k+1} \M{\lambda}
\end{bmatrix}
=
-
\begin{bmatrix}
 \ite{k}\M{f}_{\mathrm{int}}  - \Mfext (t_{n+1})
 \\
 \ite{k}\M{c} 
\end{bmatrix},
\end{align}
in which the individual terms correspond to, cf.~\cite[Eq.~(31)]{Zemanova:2014:NMFS}
\begin{subequations}\label{eq6:system_terms}
\begin{align}
\ite{k}\M{f}_{\mathrm{int}} 
& =  
\nabla \Pi_{\mathrm{int}, n+1} \bigl( \ite{k}\Md \bigr) 
=
\M{f}_{\mathrm{int}}\left(\ite{k}\Md \right),
& 
\ite{k}\M{c} 
& = 
\M{c}\bigl( \ite{k} \Md \bigl),
\\
\ite{k}\M{K} 
& =  
\nabla^2 \Pi_{\mathrm{int}, n+1} \bigl( \ite{k}\Md \bigr)
=
\M{K}_{\mathrm{t}} \bigl( \ite{k}\Md \bigr) 
+ 
\M{K}_\lambda \bigl( \ite{k}\Md, \ite{k}\M{\lambda} \bigr),
&
\ite{k}\M{C} 
& =  
\nabla\M{c} \bigl( \ite{k}\Md \bigr).
\end{align}
\end{subequations}
Because the unknown displacements are approximated independently at each layer, recall Eq.~\eqref{eq:displ_approx}, the internal forces $\Mfint$ and tangent stiffness $\MKt$ matrices possess the block-diagonal structure: 
\begin{subequations}\label{eq:matrices_block_diagonal}
\begin{align}\label{eq:matrices_block_diagonalf}
\M{f}_{\mathrm{int}} ( \Md )
& =
\begin{bmatrix}
\Mfint\lay{1}( \Md\lay{1} ) \\
\widehat{\M{f}}_\mathrm{int}\lay{2} ( \Md\lay{2} ) 
+
\Delta \widehat{\M{f}}_{\mathrm{int}}\lay{2} ( \Md\lay{2} )\\
\Mfint\lay{3}( \Md\lay{3} )
\end{bmatrix}
\\
\M{K}_{\mathrm{t}}\left(\Md\right)
& =
\begin{bmatrix}
\MKt\lay{1}( \Md\lay{1} )  \\ 
& 
\widehat{\M{K}}_\mathrm{t}\lay{2} ( \Md\lay{2} ) 
+ 
\Delta \widehat{\M{K}}_{\mathrm{t}}\lay{2} ( \Md\lay{2} )\\
&&  
\MKt\lay{3}( \Md\lay{3} )
\end{bmatrix}
.\label{eq:matrices_block_diagonalK}
\end{align}
\end{subequations}
The sub-matrices for the interlayer must be determined using the effective values $\widehat{G}$ and $\widehat{E}$~(as emphasized by the hat), and additionally contain the history-dependent contributions $\Delta \widehat{\M{K}}_\mathrm{t}$ and $\Delta \widehat{\M{f}}_\mathrm{int}$ that result from the incremental energy split~\eqref{eq:incr_energy_density}. As usual, all sub-matrices in~\eqref{eq6:systemFS} are obtained by the assembly of element contributions, e.g.~\cite[Section 2.3.6]{Reddy:2004:FEM}. For the readers' convenience, these are discussed in 
Appendix~\ref{app:sensitivity_analysis} and for the Reissner model particularly in Appendix~\ref{app:sensitivity_analysisFS}. 

To the same purpose, in Algorithm~\ref{alg:impl_incr_beam_nonlinear} we present a pseudo-code of one-step geometrically non-linear viscoelastic problem that summarizes the developments of the current section. Notice that the algorithm is terminated by two residuals, cf.~\cite[Eq.~(34)]{Zemanova:2014:NMFS},
\begin{align*}\label{eq6:residuals}
\ite{k}\eta_1 
= 
\frac{%
	\| \ite{k}\M{f}_\mathrm{int} - \Mfext + \ite{k}\M{C}\trn
	\ite{k}\M{\lambda} \|_2 }{%
	\max \left(\| \Mfext \|_2, 1 \right)
},
&&
\ite{k}\eta_2 
= 
\frac{%
	\| \ite{k} \M{c} \|_2
}{%
\min_{i=1}^3 h\lay{i}
},
\end{align*}
related to the nodal equilibrium and continuity conditions, respectively.

\begin{algorithm}[h]
 \KwData{\\
	tolerances $\epsilon_1$ and~$\epsilon_2$, load $\Mfext(t_{n+1})$, \\ 
	displacements $\Md(t_n)$, interlayer internal forces and generalized strains at $t_n$
	$\big[ 
		N_{x,e} \lay{2}(t_n), 
		V_{z,e}\lay{2}(t_n), 
		M_{y,e}\lay{2}(t_n) 
	\big]_{e=1}^{\numel}$, \\
	$\triangleright$~constant~$K$: Maxwell chain stresses
	$\big[ 
		\dstress^N_{x,p,e}(t_n), 
		\dstress^M_{x,p,e}(t_n), 
		\tau^V_{xz,p,e}(t_n)  
	\big]_{\prony=1, e=1}^{P, \numel}$,
	\\
	$\triangleright$~constant~$\nu$: Maxwell chain stresses
	$\big[ 
	\stress^N_{x,p,e}(t_n), 
	\stress^M_{x,p,e}(t_n), 
	\tau^V_{xz,p,e}(t_n)  
	\big]_{\prony=1, e=1}^{P, \numel}$} 
 {\bf Initialization:}\\ 
 $k \leftarrow 0, \ite{0}\M{\lambda} \leftarrow \M{0}, \ite{0} \Md \leftarrow \Md(t_n)$,
 assemble $\ite{0}\M{f}_\mathrm{int}$, $\ite{0}\M{c}$, and $\ite{0}\M{C}$ in~\eqref{eq:matrices_block_diagonalf} and~\eqref{eq6:systemFS}
 \\
 \While{$(\ite{k}\eta_1 > \epsilon_1)$ or $(\ite{k}\eta_2 > \epsilon_2)$}{%
 assemble $\ite{k}\M{K}$ from~\Eref{eq:matrices_block_diagonalK} \\
 solve for $(\ite{k+1}\delta \Md, \ite{k+1}\M{\lambda})$ 
 from \Eref{eq6:systemFS} 
 \\ 
 $\ite{k+1} \Md \leftarrow \ite{k}\Md + \ite{k+1}\delta \Md$
 \\
 assemble 
 $\ite{k+1}\M{f}_\mathrm{int}$, $\ite{k+1}\M{c}$, and
 $\ite{k+1}\M{C}$ in~\eqref{eq:matrices_block_diagonalf} and~\eqref{eq6:systemFS}
 \\ 
 $k\leftarrow k+1$ \\
 } 
 $\Md(t_{n+1}) \leftarrow \ite{k+1}\Md$,
 $\M{\lambda}(t_{n+1}) \leftarrow \ite{k+1}\M{\lambda}$
 \\
 $\triangleright$~constant~$K$: 
 update internal forces from~\eqref{eq6:incr_int_forces_K} and 
 Maxwell chain stresses from~\eqref{eq6:dev_increment_final_K}
 \\
  $\triangleright$~constant~$\nu$: 
  update internal forces from~\eqref{eq6:incr_int_forces_nu} and 
  Maxwell chain stresses from~\eqref{eq6:dev_increment_final_nu}
 \caption{One-step exponential algorithm for
 geometrically nonlinear laminated beams}
 \label{alg:impl_incr_beam_nonlinear}
\end{algorithm}

A similar procedure applies also to the von K\'{a}rm\'{a}n model with a few simplifications. Indeed, because the continuity constraint~\eqref{eq:constraint} is linear in $\Md$, recall~\eqref{eq:comp_conVK}, we have $\ite{k} \M{c} = \M{0}$, matrix $\M{C}$ remains constant during iterations, and matrix $\M{K}_{\lambda} = \M{0}$. The remaining terms needed in~\eqref{eq6:systemFS} are provided in Appendix~\ref{app:sensitivity_analysisVK}. 

\section{Comparison, verification, and validation}\label{sec:comp_ver_val}

The developments described in the previous section resulted in four geometrically non-linear finite element models summarized in \Tref{tab:scheme}; recall that they emerge as the combinations of Reissner~(FS) or von K\'{a}rm\'{a}n~(VK) kinematics and the assumptions of constant bulk modulus~(K) or Poisson ratio~($\nu$) of the interlayer. We also include the results of geometrically linear formulation~(LIN)~\cite{Zemanova:2008:SNM}, in order to highlight the importance of geometric non-linearity where relevant.

\begin{table}[h]
\caption{Overview of finite element models of laminated glass beams with viscoelastic interlayer.} 
\label{tab:scheme}
\centering
\begin{tabular}{lcccccc}
\hline
Abbreviation &
FS$_K$ & FS$_\nu$ & VK$_K$ & VK$_\nu$ & LIN$_K$ & LIN$_\nu$ \\
\hline
Kinematics &
\multicolumn{2}{c}{finite strains} &\multicolumn{2}{c}{large deflections} &\multicolumn{2}{c}{geometric linearity} \\
Constant & 
$\Kmod\lay{2}$ & $\Pnum\lay{2}$ & $\Kmod\lay{2}$ & $\Pnum\lay{2}$ & $\Kmod\lay{2}$ & $\Pnum\lay{2}$\\
\hline
\end{tabular}
\end{table}

In addition, the following data are considered in the studies reported in this section:

\begin{itemize}
	\item detailed two-dimensional viscoelastic finite element simulations~(2D), performed in ADINA~9.0.2 (ADINA R\&D, Inc.) finite element system, 
	\item experimental data, obtained by~\citet{Aenlle:2013:ETC} for three simply-supported beams and one two-span continuous beam with different layer thicknesses; 
	\item results of our earlier finite strain model with elastic interlayer~\cite{Zemanova:2014:NMFS}, under the assumption of the constant bulk modulus $K\lay{2}$; and
	\item responses of the two limiting cases, corresponding to the monolithic model, which assumes the full layer interaction, and the layered model with no inter-layer interaction, both in the small strain regime.  
\end{itemize}

The remainder of this section is organized as follows. In \Sref{sec:over_compE}, we provide dimensions and boundary conditions for beams used in the analysis and specify material data, loading procedures, and details on finite element models. \Sref{sec:comp} is dedicated to the comparison of the four finite element models from \Tref{tab:scheme}. Based on these results, the model combining the von K\'{a}rm\'{a}n kinematics with the time-independent Poisson ratio is verified against a two-dimensional finite element model and partially validated against experiments in \Sref{sec:Ver_val}. We proceed with a parametric study concerning the effects of geometric nonlinearity and temperature, \Sref{sec:GNE_TE} and conclude our investigations by highlighting the difference in full viscoelastic and simplified elastic analyses, \Sref{sec:Visco_simp}.

\subsection{Overview of examples}\label{sec:over_compE}

\subsubsection{Geometry and finite element model}\label{sec:over_geom}

The most common laminated beams with three layers (glass/PVB/glass) are considered, either simply supported or clamped. Note that 
in practice, the laminated glass elements are not perfectly fixed or simply supported, but these two cases represent limits of the real support conditions concerning the effects of geometric non-linearity, e.g.~\cite{Asik:2003:LGP,Koutsawa:2007:SFVA}. The layer thicknesses and spans were selected to ensure that the maximum deflection does not exceed $1/200$ of the span, which we estimated as a suitable serviceability limit for load-bearing glass structures. 

Similarly to~\cite{Zemanova:2014:NMFS}, each layer was discretized with $500$~elements and linear stress smoothing was adopted, in order to reach the four-digit accuracy of displacements and stress values reported below. Both tolerances in Algorithm~\ref{alg:impl_incr_beam_nonlinear} were set $\epsilon_1 = \epsilon_2 = 10^{-5}$.

\begin{table}[h]
\caption{Overview of geometry of laminated glass beams}
\label{tab:geom}
\centering
\begin{tabular}{ccccc}
\hline
Geometry & Supports & Length [m] & Width [m] & Thicknesses [mm] \\
&&&& glass/PVB/glass  \\
\hline
I  & fixed-end & 3 & 0.15 & 3/0.76/3\\ 
II & simply-supported & 3 & 0.15 & 6/0.76/6 \\ 
\hline
\end{tabular}
\end{table}

\subsubsection{Materials}\label{sec:over_mat}

The elastic constants of glass, the Young modulus $E\lay{1} = E\lay{2} = 72$~GPa and the Poisson ratio $\nu\lay{1} = \nu\lay{3} =  0.23$, correspond to the values determined by~\citet{Aenlle:2013:ETC} from a static bending test. The experimental viscoelastic characterization of PVB was performed by~\citet{Pelayo:2013:MSLGP} on specimens of thickness $0.38$~mm by tensile relaxation tests with the duration of 10~min at different temperatures ranging from $-15$ to $50^\circ$~C. The resulting parameters of the generalized Maxwell chain are summarized in \Tref{tab:MaxwellSeries}, the parameters of the Williams-Landel-Ferry~(WLF) equation~\eqref{eq6:shiftfactor} were set to $C_1 = 12.6$, $C_2=74.46$, and $T_0=20\,^\circ$C. The corresponding shear relaxation functions $G\lay{2}$ are plotted in~\Fref{fig:GPVB2}(b). According to the adopted constitutive assumption, we further set the bulk modulus $K\lay{2} = 2$~GPa, e.g.~\cite{Duser:1999:AGBL,Bennison:1999:FLB}, or the Poisson ratio $\nu\lay{2} = 0.49$, e.g.~\cite[Section 1.1.3]{Haldimann:2008:SUG}.

\begin{table}[ht]
\caption{Generalized Maxwell chain description of the shear relaxation modulus for PVB, $G_{\infty}$ = 1.9454$\times 10^{-4}$~GPa~(after~\cite{Pelayo:2013:MSLGP})}
\label{tab:MaxwellSeries}
\centerline{
\begin{tabular}{ccccccc}
\hline
$\prony$ & $\tp_{\prony}$ [s] & $G_{\prony}$ [GPa] 
&&
$\prony$ & $\tp_{\prony}$ [s] & $G_{\prony}$ [GPa] \\
\hline
1 & 2.3660$\times 10^{-7}$ & 9.9482$\times 10^{-2}$
& \hspace{10mm} &
8 & \hspace{-3.5mm} 1.7382$\times 10^{0}$ & 3.0802$\times 10^{-3}$
\\
2 & 2.2643$\times 10^{-6}$ & 9.0802$\times 10^{-2}$
&&
9 & \hspace{-3.5mm} 1.6633$\times 10^{1}$ & 1.2001$\times 10^{-3}$
\\
3 & 2.1667$\times 10^{-5}$ & 7.4140$\times 10^{-2}$
&&
10 & \hspace{-3.5mm} 1.5916$\times 10^{2}$ & 1.1523$\times 10^{-4}$
\\
4 & 2.0733$\times 10^{-4}$ & 5.0772$\times 10^{-2}$
&&
11 & \hspace{-3.5mm} 1.5230$\times 10^{3}$ & 1.8237$\times 10^{-4}$
\\
5 & 1.9839$\times 10^{-3}$ & 5.7856$\times 10^{-2}$
&&
12 & \hspace{-3.5mm} 1.4573$\times 10^{4}$ & 4.1645$\times 10^{-5}$
\\
6 & 1.8984$\times 10^{-2}$ & 2.9055$\times 10^{-2}$
&&
13 & \hspace{-3.5mm} 1.3945$\times 10^{5}$ & 2.2405$\times 10^{-4}$
\\
7 & 1.8165$\times 10^{-1}$ & 1.7601$\times 10^{-2}$
\\
\hline
\end{tabular}}
\end{table} 

\subsubsection{Loading}\label{sec:over_load}

\begin{figure}[h]
\small (a)\includegraphics[height=\myFigureHeight]{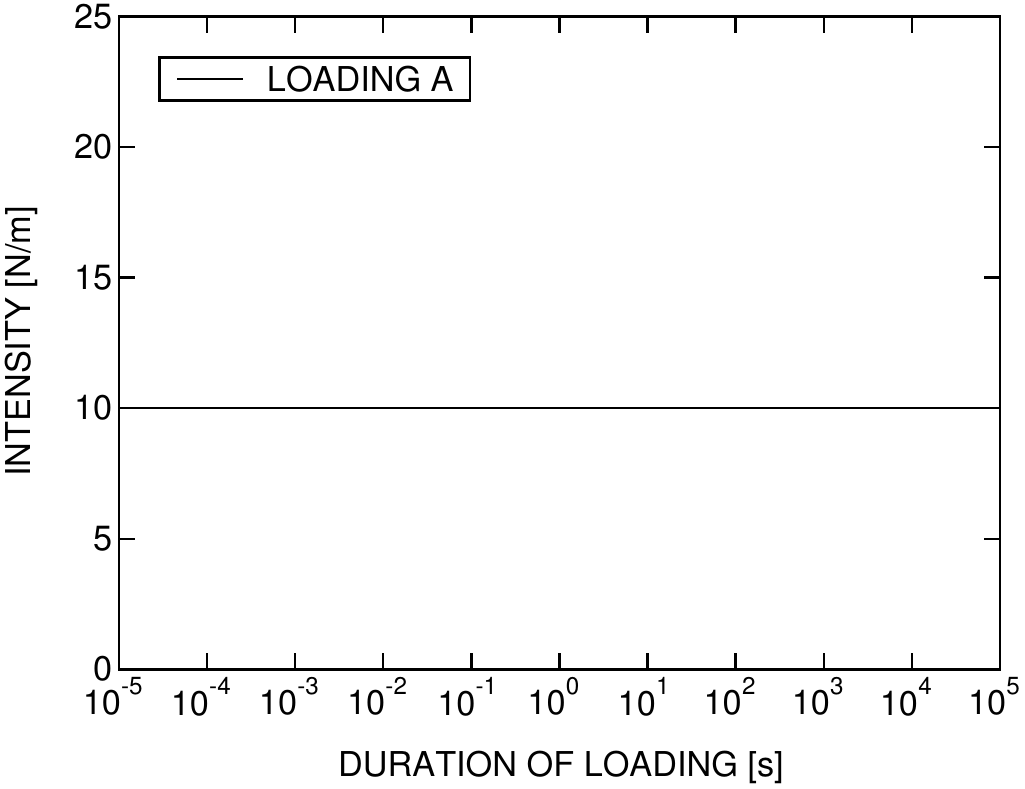}	
\hfill
\small (b)\includegraphics[height=\myFigureHeight]{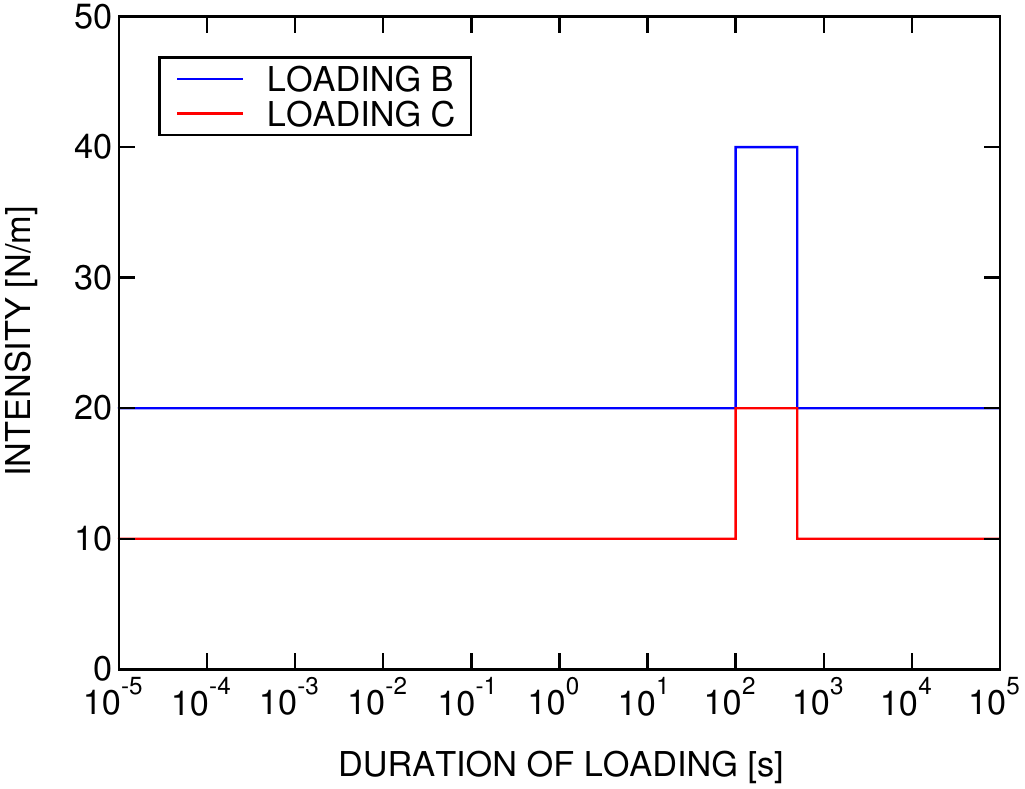}	
\caption{Intensity of (a)~one- and~(b) two-load history.} 
\label{fig:load_hist}
\end{figure}

We considered loading scenarios with spatially uniform distributed loads of intensities shown in~\Fref{fig:load_hist}. Scenario~A, \Fref{fig:load_hist}~(a), approximates the instantaneous loading reported in~\cite{Pelayo:2013:MSLGP} by a loading branch, in which the intensity linearly increases from $0$ to $10$~Nm$^{-1}$ in $10^{-5}$~s, followed by the constant load of intensity $10$~Nm$^{-1}$ up to $10^5$~s. The time steps in the incremental algorithms were distributed uniformly in the logarithmic scale, in particular $6$ time steps were used to discretize the interval $\langle 10^{-6}; 10^{-5} \rangle$~s, whereas the interval $\langle 10^{-5}; 10^5 \rangle$~s was discretized into $24$ time steps.

Histories~B and~C represent loading-unloading scenarios shown in \Fref{fig:load_hist}~(b), where the steps in intensity were approximated by linearly varying loading within $10^{-5}$~s. The overall loading program was discretized into $90$ time steps uniformly distributed in the logarithmic scale and refined in the vicinity of jumps. 

\begin{figure}[p]
\small (a)\includegraphics[height=\myFigureHeight]{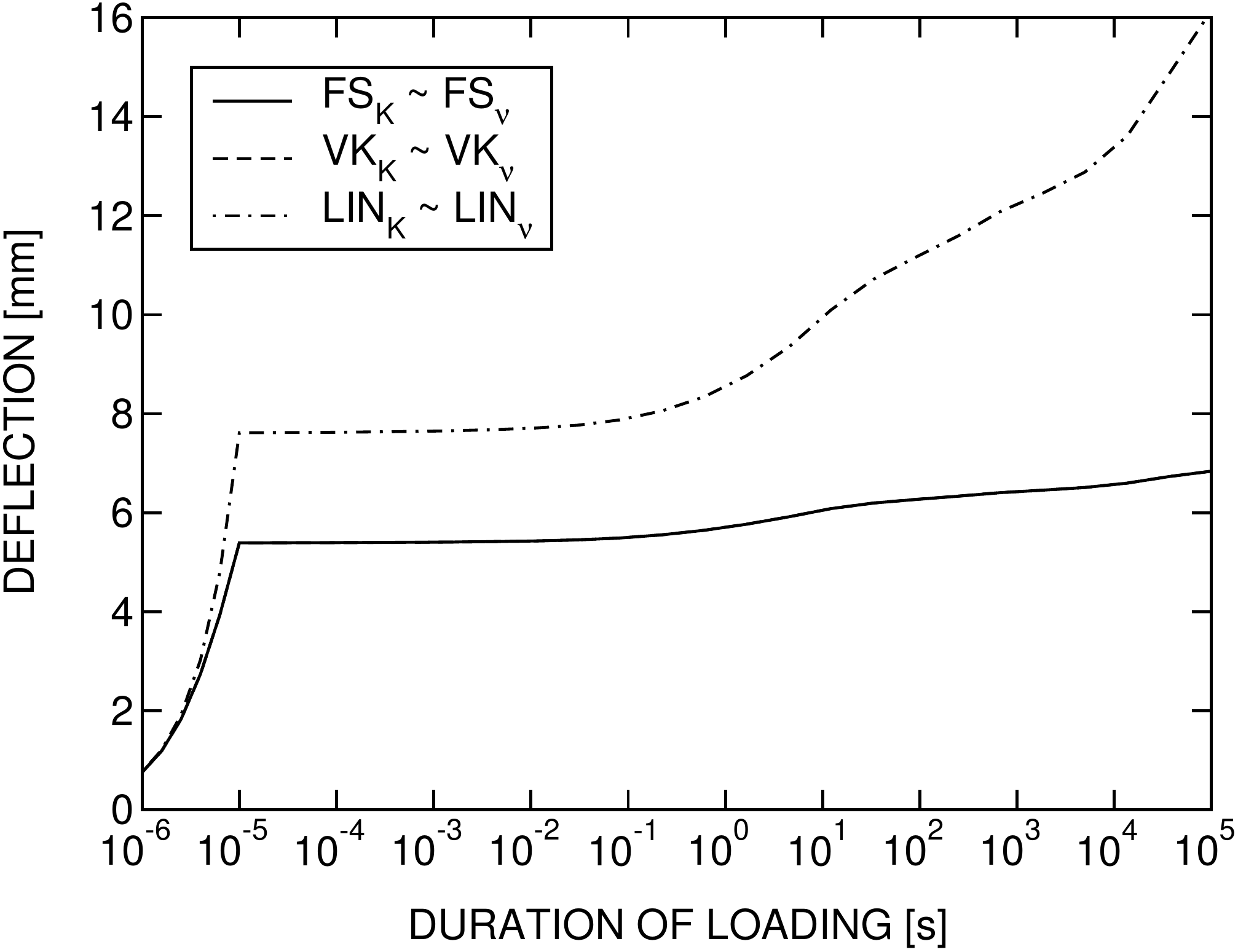}	
\hfill
\small (b)\includegraphics[height=\myFigureHeight]{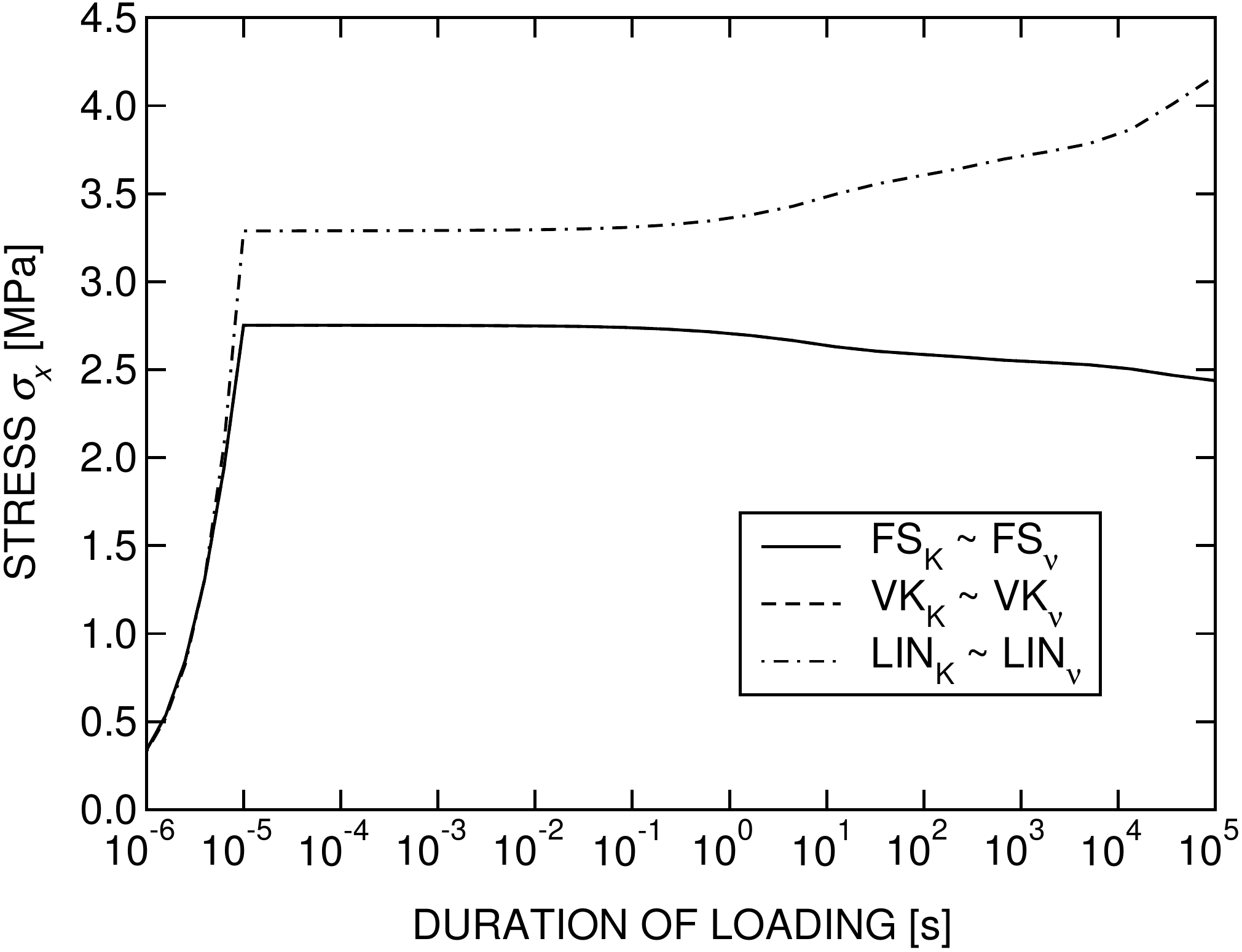}	
\caption{Comparison of mid-point (a)~deflections and (b)~maximum normal stresses for a fixed-end beam with geometry~I under loading~A, as predicted by finite element models summarized in \Tref{tab:scheme}.}
\label{fig:compw1}	
	
\bigskip

\small (a)\includegraphics[height=\myFigureHeight]{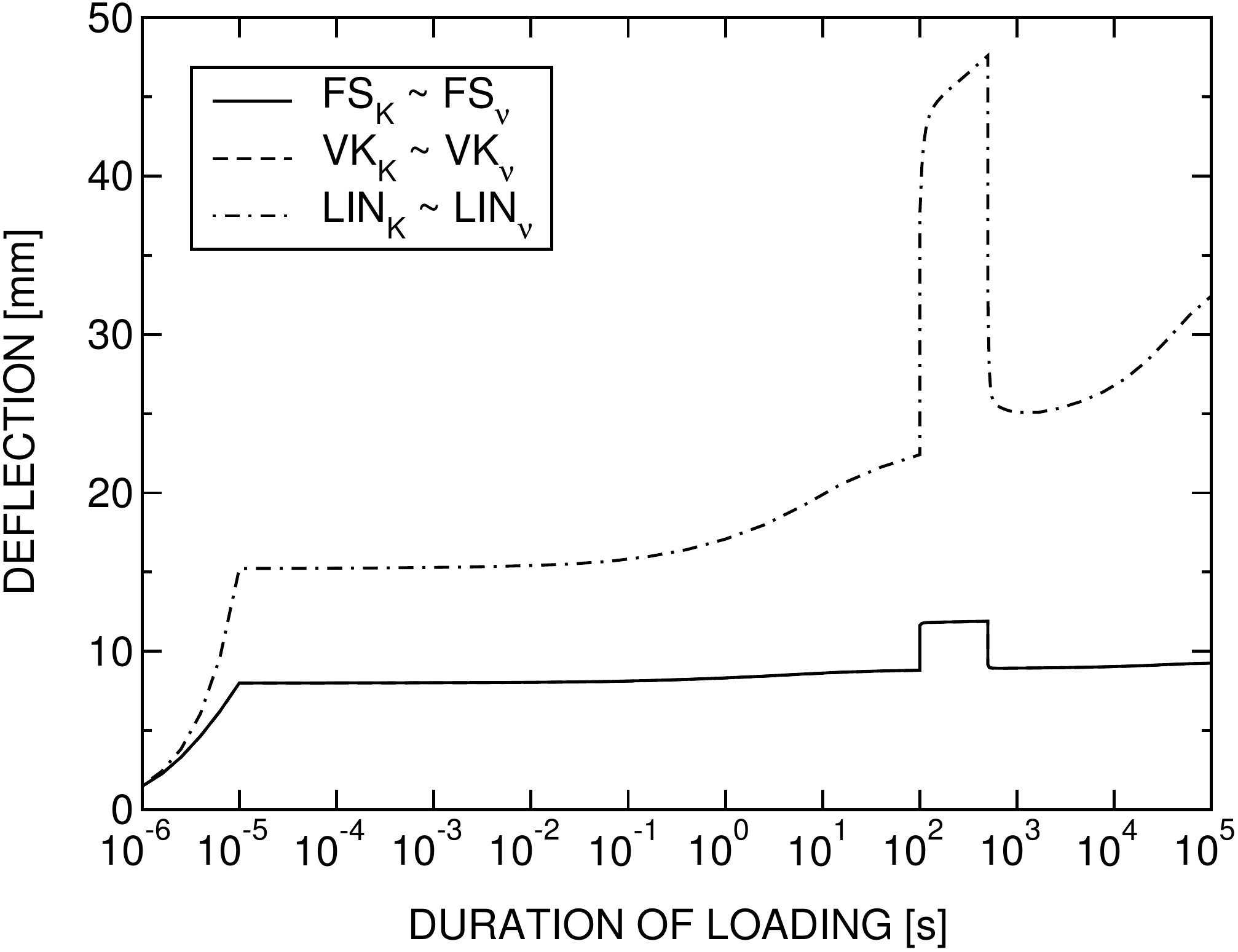}	
\hfill
\small (b)\includegraphics[height=\myFigureHeight]{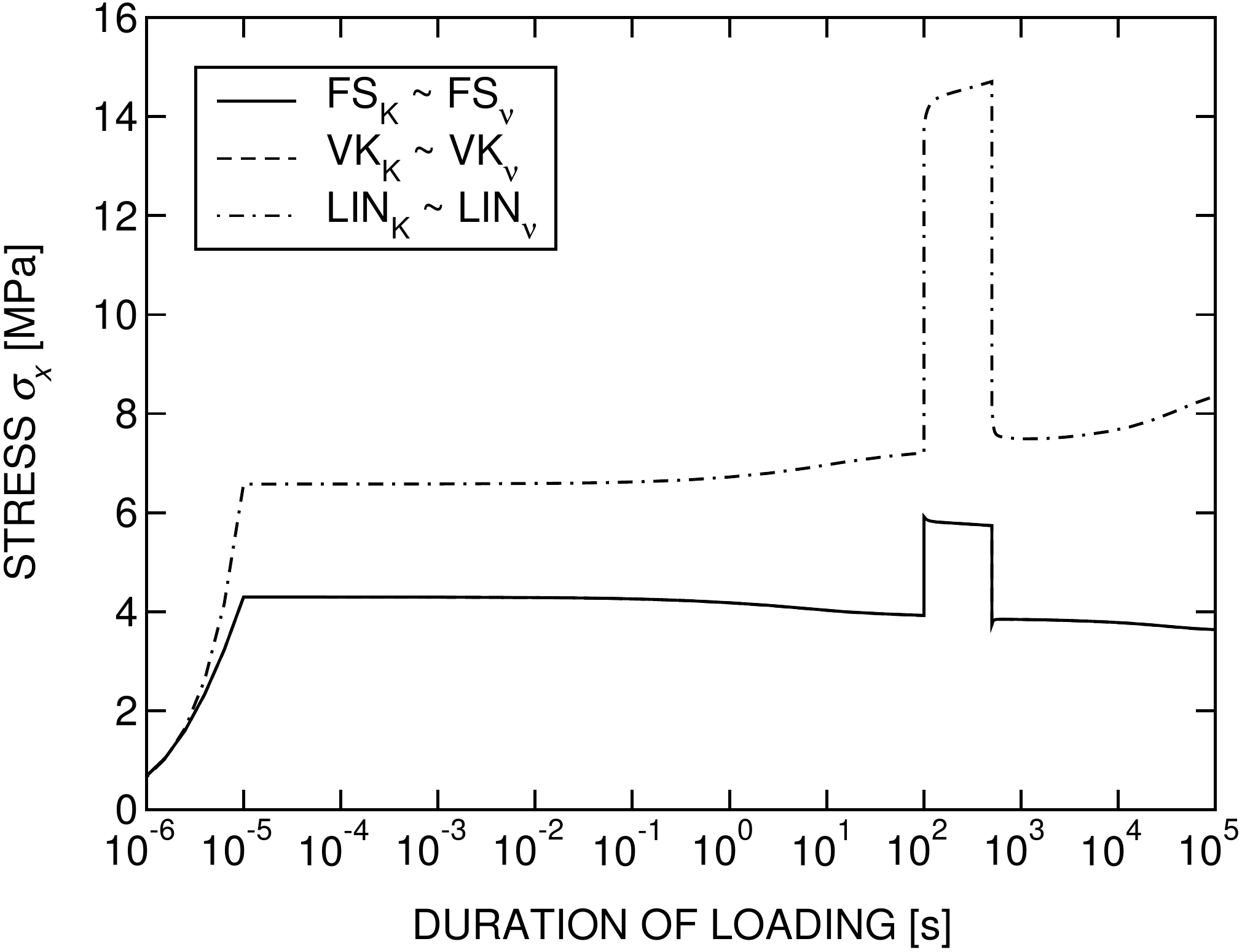}	
\caption{Comparison of mid-point (a)~deflections and (b)~maximum normal stresses for a fixed-end beam with geometry~I under loading~B, as predicted by finite element models summarized in \Tref{tab:scheme}.}
\label{fig:compw2}
	
\bigskip

\small (a)\includegraphics[height=\myFigureHeight]{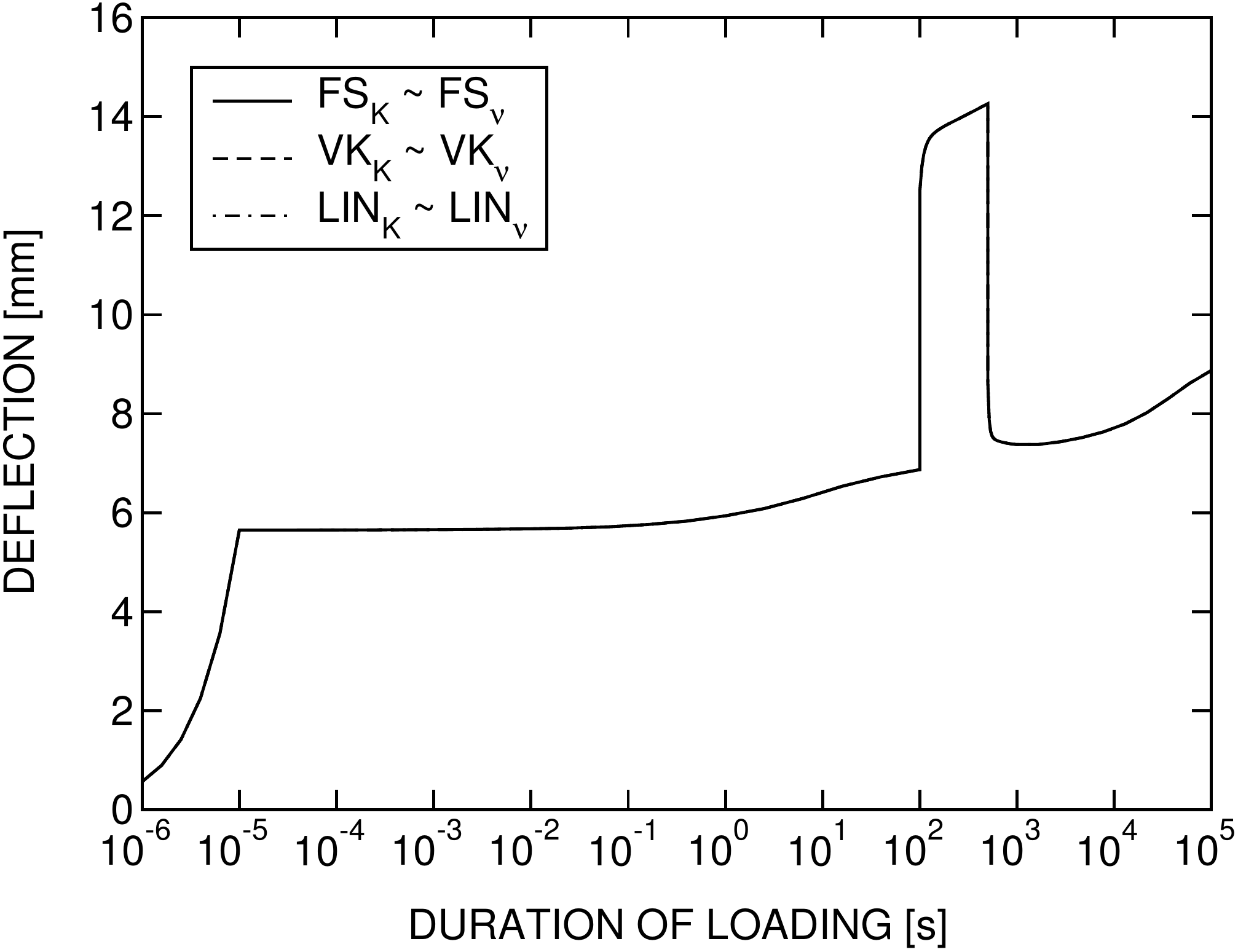}	
\hfill
\small (b)\includegraphics[height=\myFigureHeight]{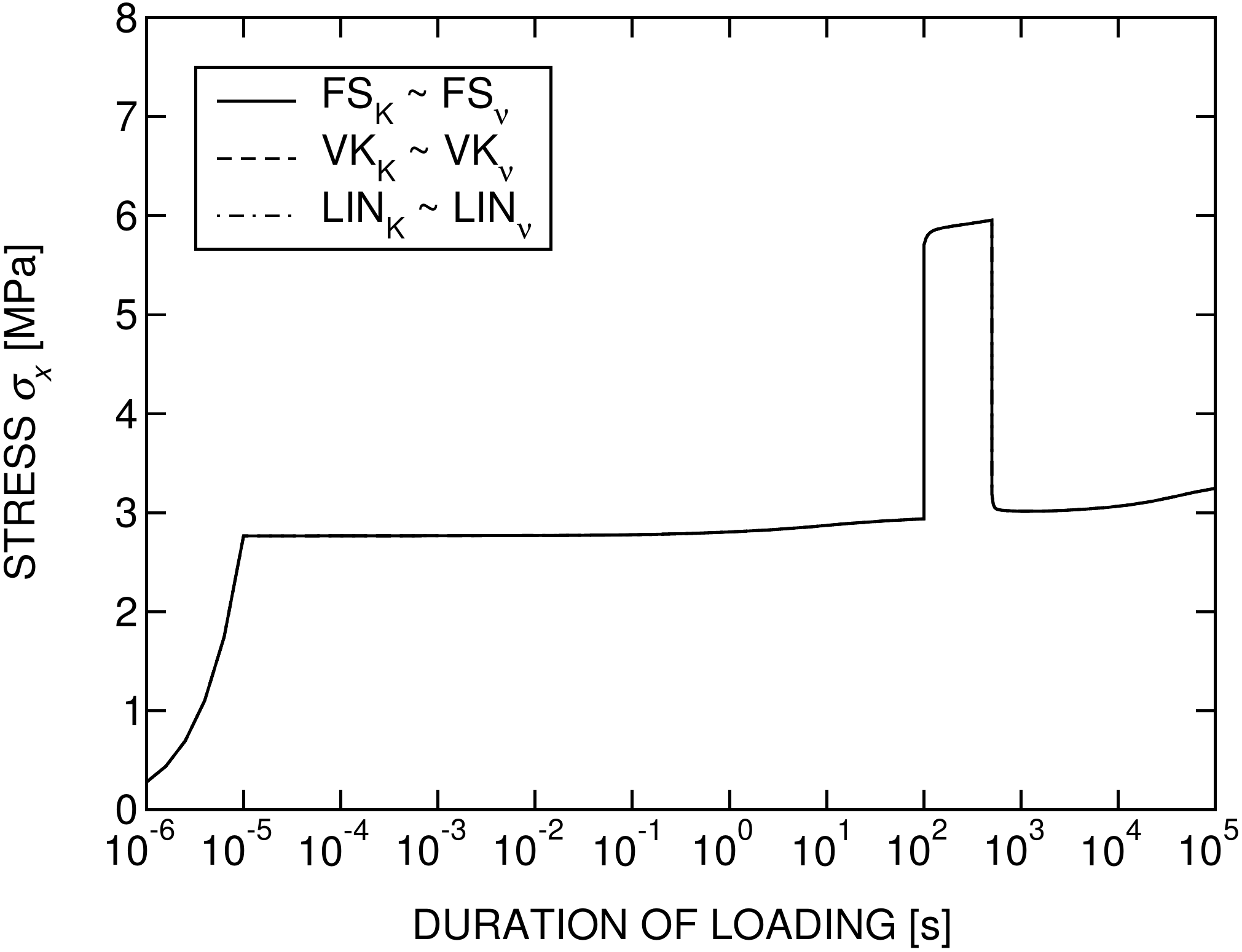}	
\caption{Comparison of mid-point (a)~deflections and (b)~maximum normal stresses for a simply-supported beam with geometry~II under loading~C, as predicted by finite element models summarized in \Tref{tab:scheme}.}
\label{fig:compw3}
\end{figure}

\subsection{Comparison}\label{sec:comp}
%
Comparison of the proposed models and of the effects of geometric nonlinearity are discussed in this section. Three examples are considered, namely fixed-end beam (geometry I) under one-load history~(loading A), fixed-end beam (geometry I) under two-load history~(loading B), and simply-supported beam (geometry II) under two-load history~(loading C). The temperature of $25\,^\circ$C was kept constant in all examples, as the temperature effects are in the focus of~\Sref{sec:GNE_TE}.

In Figures~\ref{fig:compw1}--\ref{fig:compw3}, we plot the deflections and the maximum of stresses predicted by four geometrically nonlinear (FS$_K$, FS$_\nu$, VK$_K$, VK$_\nu$) and two geometrically linear~(LIN$_K$, LIN$_\nu$) approaches. Their comparison reveals that:
\begin{itemize}
	\item The viscoelastic approaches based on the assumption of constant bulk modulus $K$ (FS$_K$, VK$_K$, and LIN$_K$), or of the Poisson ratio $\nu$ (FS$_{\nu}$, VK$_{\nu}$, resp. LIN$_{\nu}$) provide the same results for all examples because the interlayer mostly transfers shear loading (compare cf. \Eref{eq6:dev_increment_final_K} and \Eref{eq6:dev_increment_final_nu}). The errors in deflections and stresses are much smaller than 0.1\%.
	
	\item The geometrically nonlinear approaches based on the assumptions of large deflections (VK$_K$, VK$_{\nu}$), or finite strains (FS$_K$, FS$_{\nu}$) give again almost identical results for tested examples; the errors in deflections and stresses do not exceed $0.1\%$. 
	
	\item For statically determinate example, the geometrically linear and nonlinear solver provide the same results, whereas for statically indeterminate examples, the error of linear approach can reach 100\% up to 300\%; cf.~\cite{Zemanova:2014:NMFS}.
\end{itemize}

Recall that in the previous examples, the load level corresponds to maximum deflections about 1/200 of the span of the beam, which might conceal the effects of geometric non-linearity. The results collected in \Tref{tab:comparFSVK} demonstrate that predictions of VK and FS models remain very close for increased load levels; the difference in deflection is smaller than $0.1\%$ and in stresses than $2\%$. Notice that for the largest load intensity of $5,000$~Nm$^{-1}$ in \Tref{tab:comparFSVK}, the maximum deflection reaches 1/50 of the span, which exceeds the limit value of 1/65 suggested in~\cite[Paragraph 9.1.4]{prEN:16612:2013} for non-load bearing glass elements. 

\begin{table}[ht]
	\caption{Comparison of the maximum values of deflections and normal stresses at time $10^{5}$~s}
	\centerline{
		\begin{tabular}{ccccccc}
			\hline
			& \multicolumn{3}{c}{Deflections [mm]}
			&
			\multicolumn{3}{c}{Stresses $\stress_{x}$ [MPa]}
			\\ 
			load [N/m] &	 VK$_K$ $\approx$ VK$_\nu$ &	 FS$_K$ $\approx$	 FS$_\nu$ &	 error
			&	 VK$_K$ $\approx$ VK$_\nu$ &	 FS$_K$ $\approx$	 FS$_\nu$ &	 error\\
			\hline
			50	&	 13.239	& 13.239 &	  -0.00\% & 22.901	& 22.919	&  -0.08\% \\
			500	&	 30.064	& 30.067 &	  -0.01\% &	 105.49	& 105.92	&  -0.41\% \\
			5,000 &	 65.783	& 65.820 &	  -0.06\% &	 471.62	& 480.22	&  -1.79\% \\
			\hline
		\end{tabular}
	}
	\label{tab:comparFSVK}
\end{table}

In conclusion, it follows from the results presented in this section that all four geometrically non-linear models provide results indistinguishable from the practical design viewpoint. Therefore, we will report only the results of VK$_\nu$ model in what follows, because its formulation is the most transparent from the four options.

\subsection{Verification and validation}\label{sec:Ver_val}

\begin{figure}[p]
\small (1a)\includegraphics[height=\myFigureHeight]{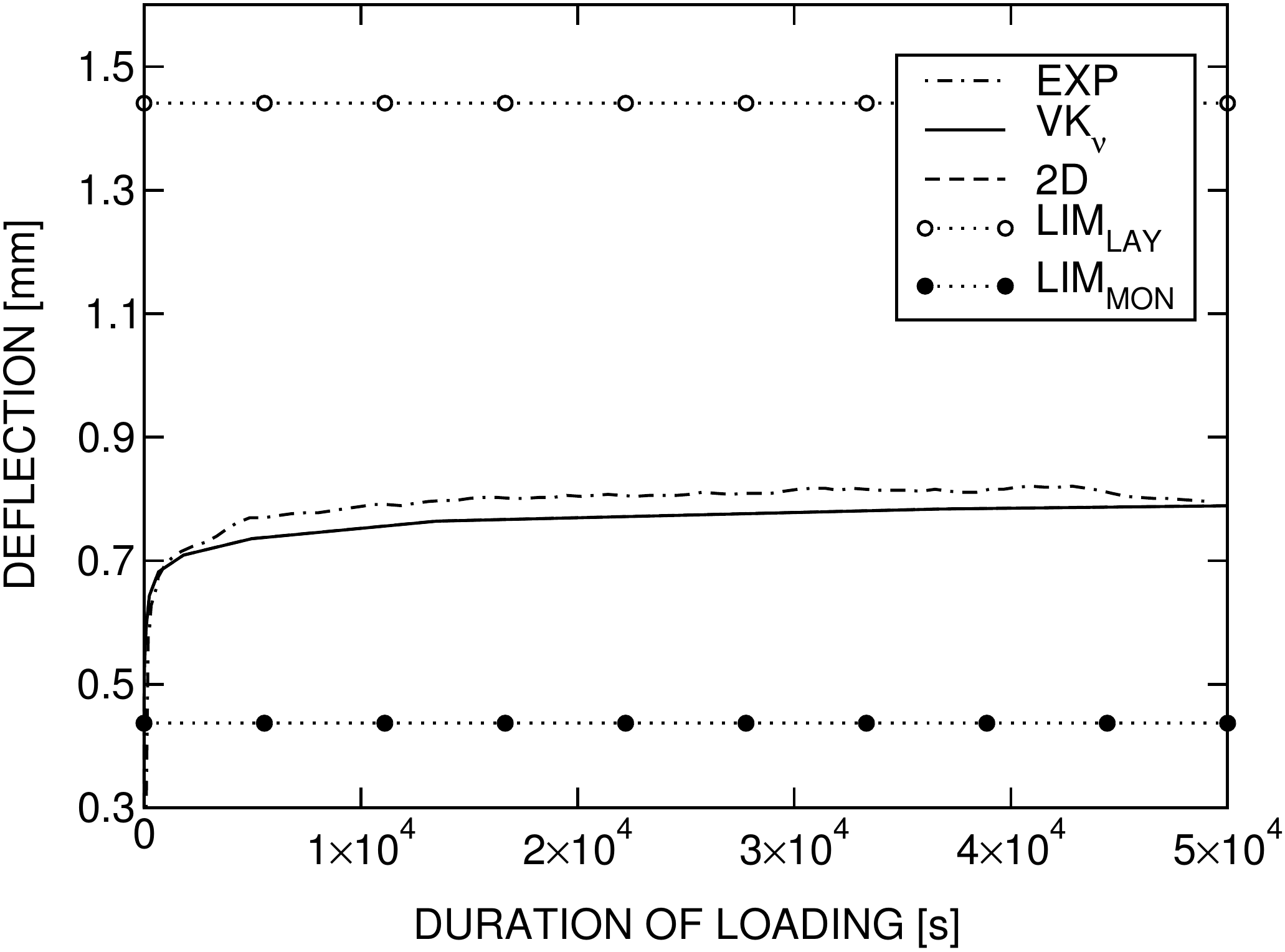}
\hfill
\small (1b)\includegraphics[height=\myFigureHeight]{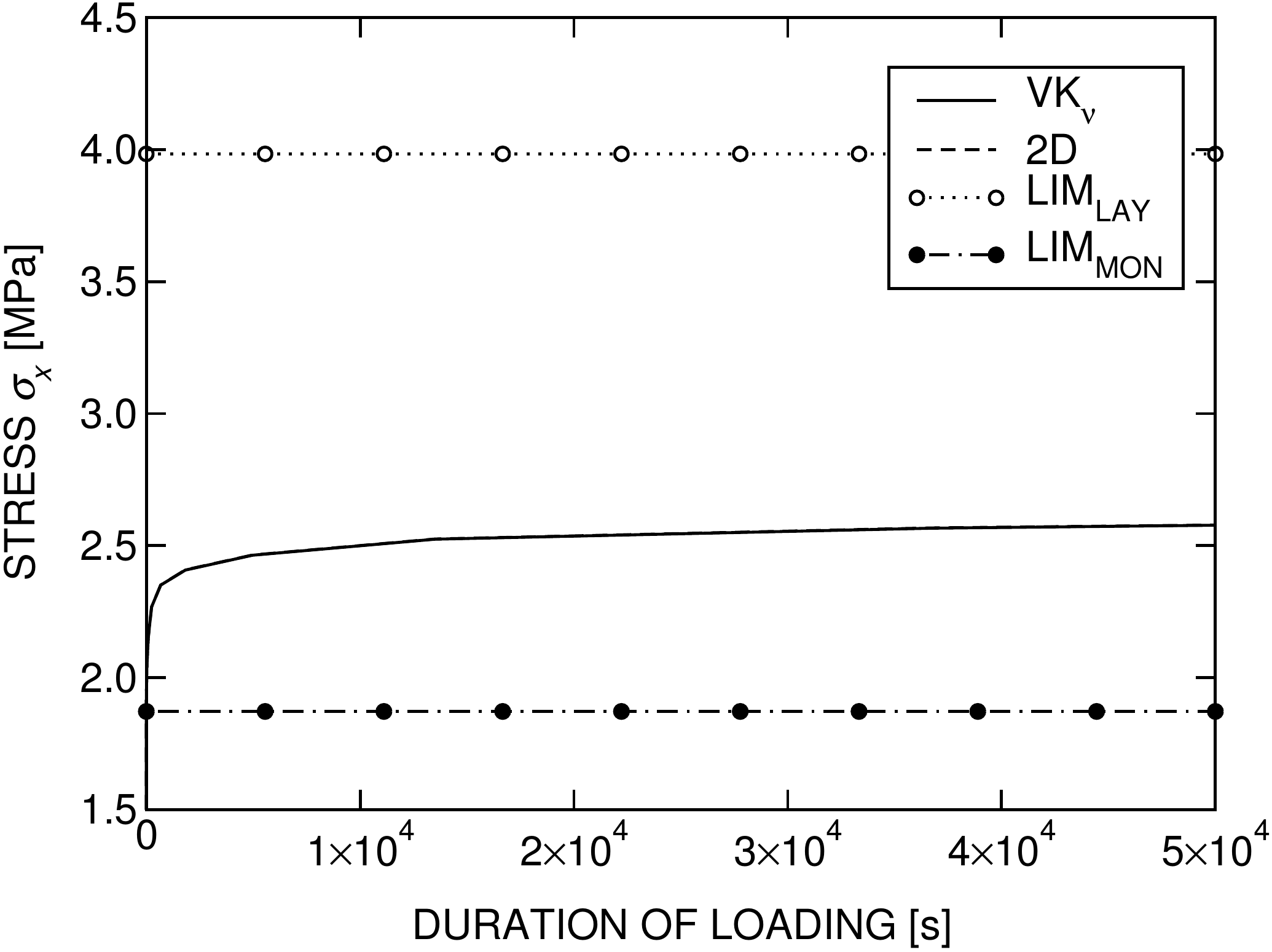}

\bigskip

\small (2a)\includegraphics[height=\myFigureHeight]{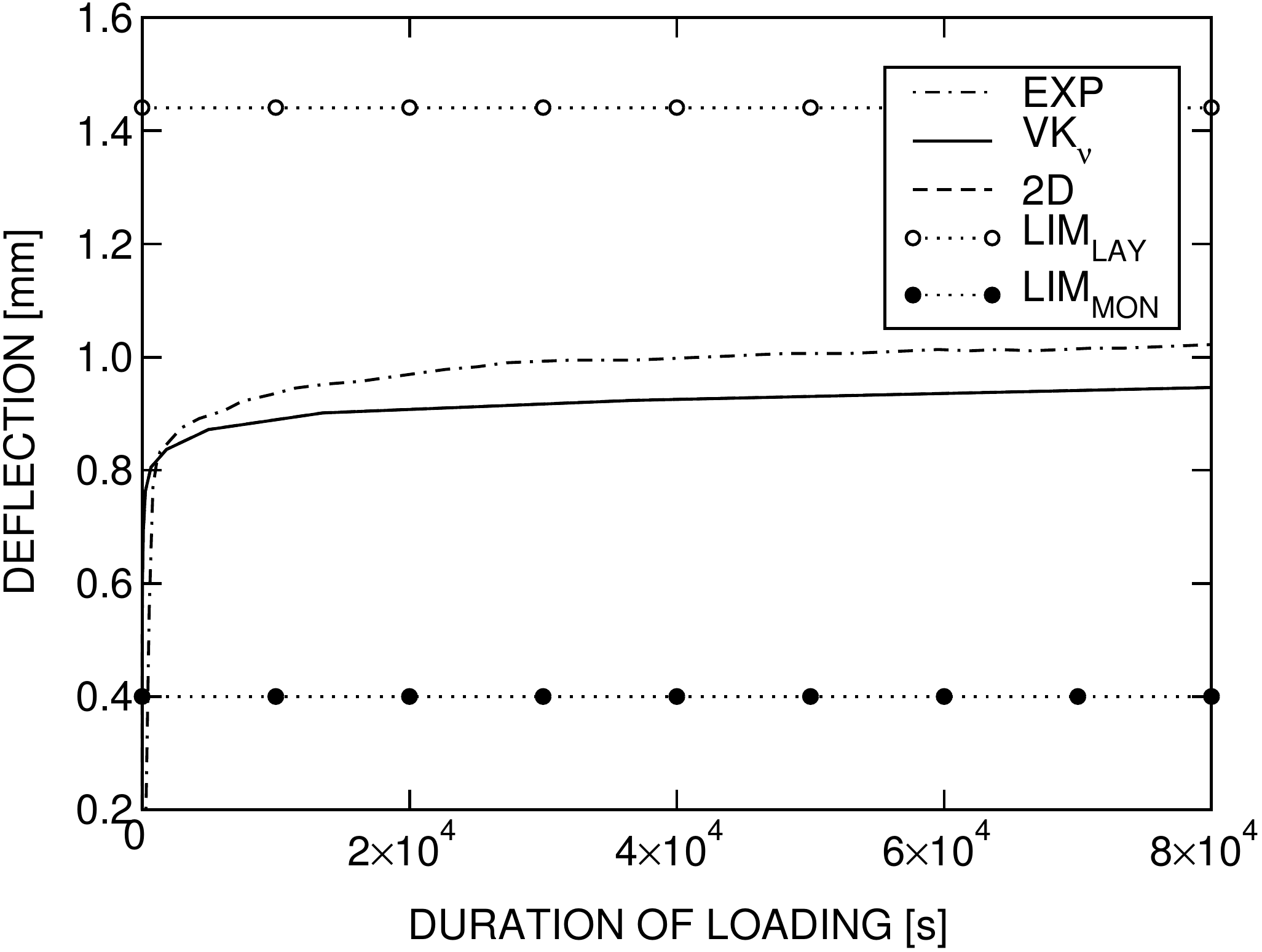}
\hfill
\small (2b)\includegraphics[height=\myFigureHeight]{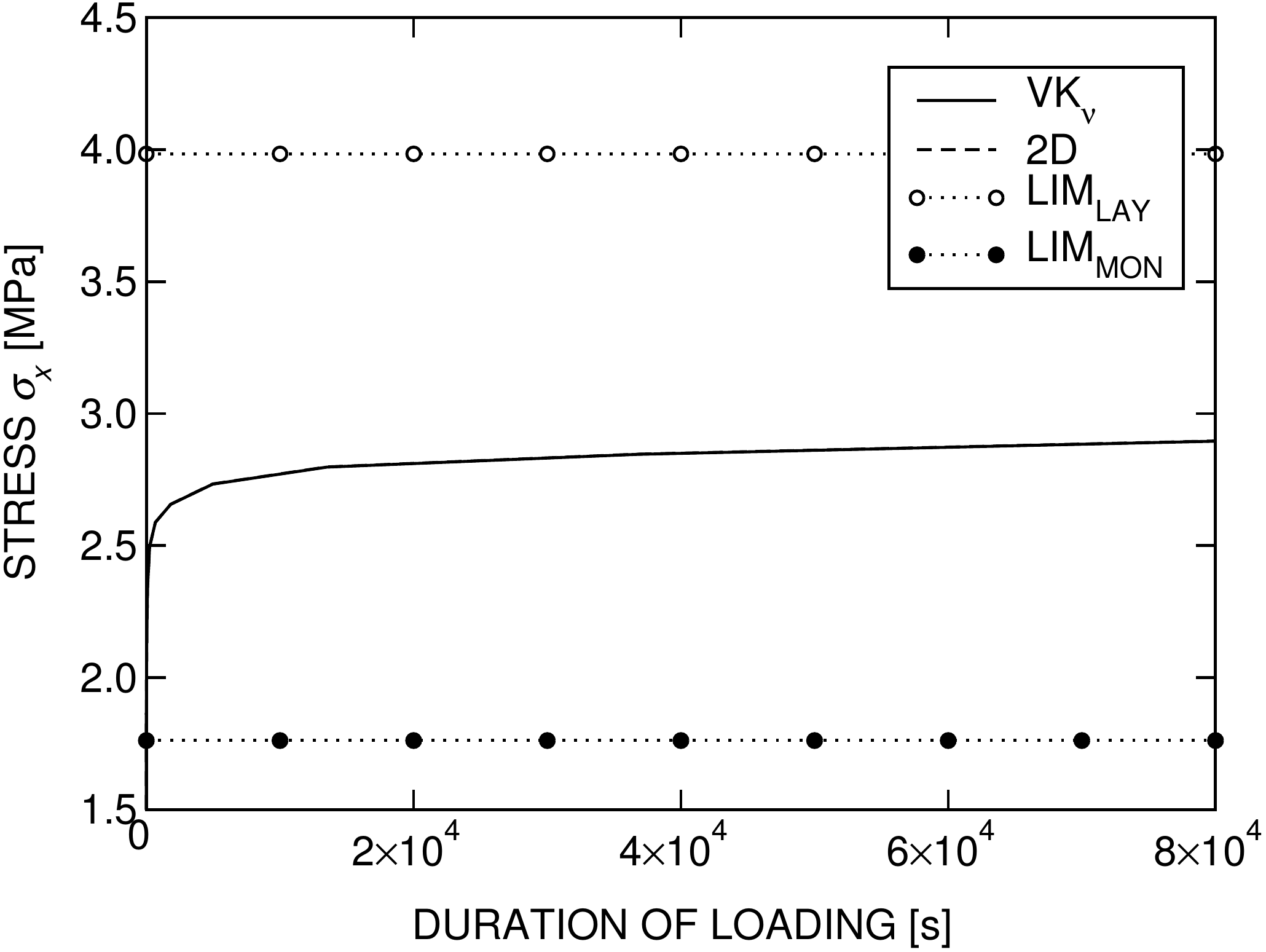}

\bigskip

\small (3a)\includegraphics[height=\myFigureHeight]{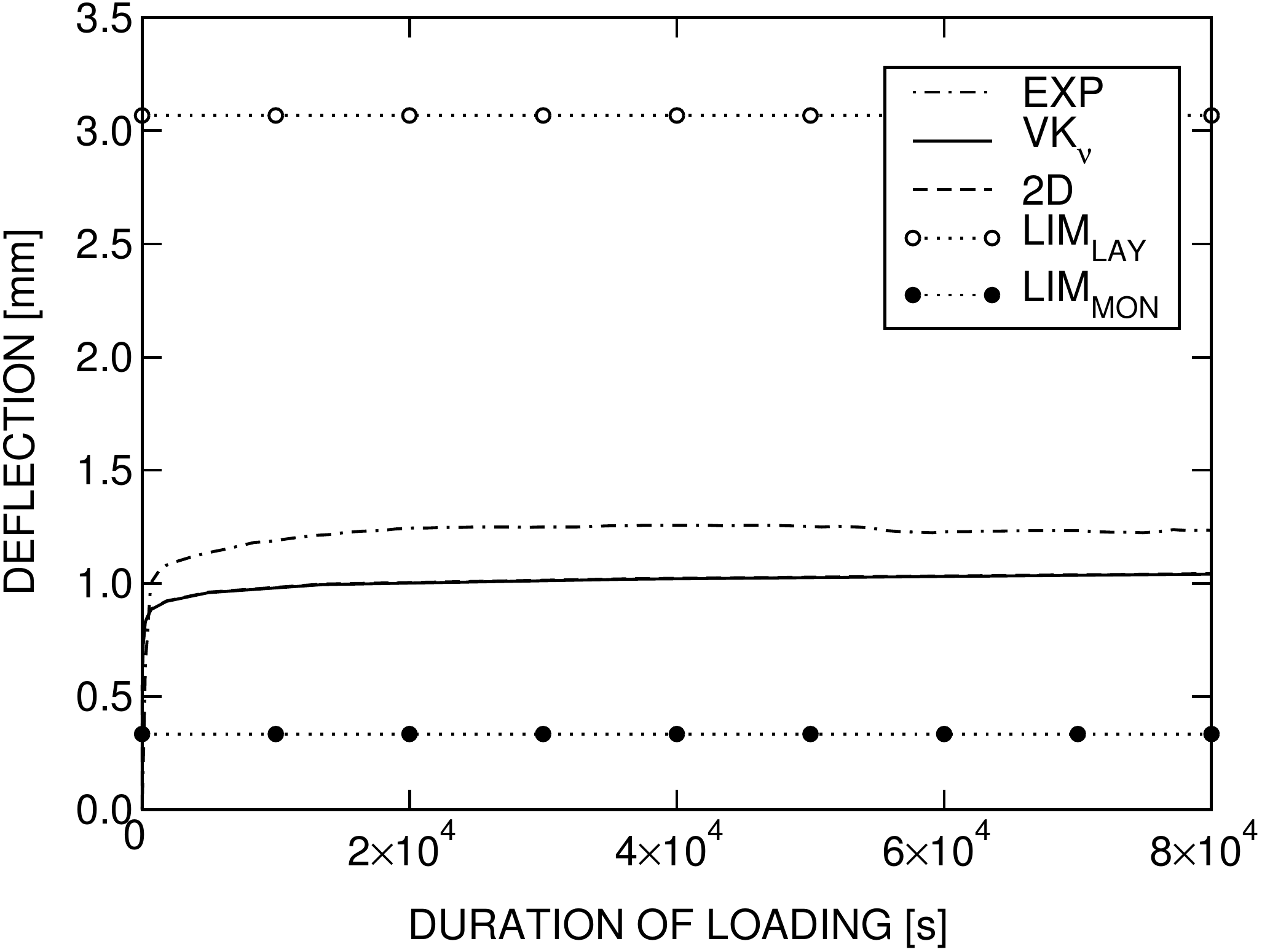}
\hfill
\small (3b)\includegraphics[height=\myFigureHeight]{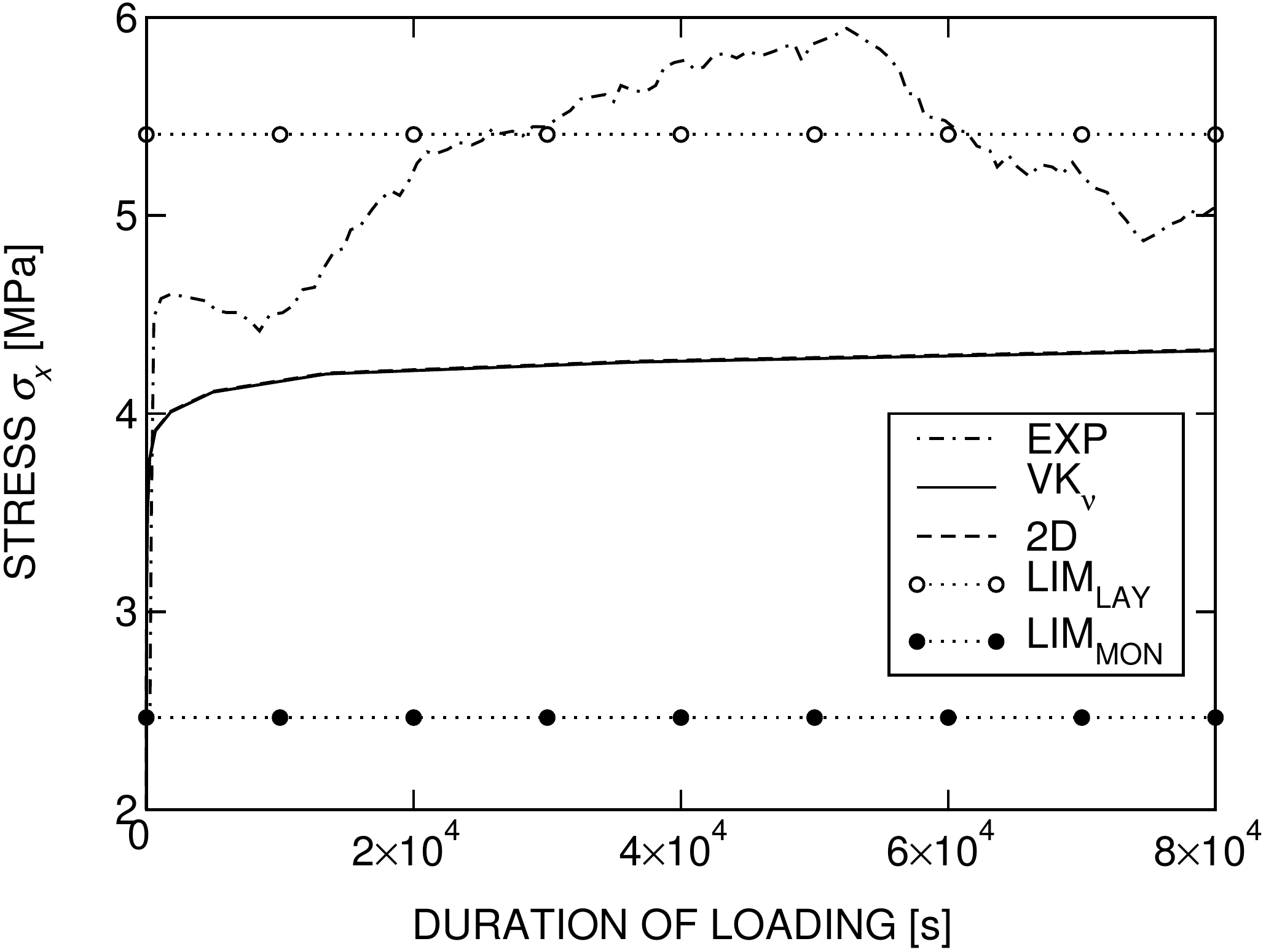}

\caption{%
		Comparison of maximum mid-span (a)~deflections and (b)~normal stresses for simply supported beams of span $1$~m and of width $0.1$~m composed of three glass/PVB/glass layers with thicknesses of (1) 4/0.38/8~mm or (2) 4/0.76/8~mm and (3)~a simply supported continuous beam of spans $2 \times 0.7$~m and of width $0.1$~m, composed of three 4/0.38/4~mm layers. The beam is subjected to a ramp load with uniform intensity of (1)~$38.25$~N/m, (2)~$38.25$~N/m, and (3)~$94.22$~N/m under average temperature of (1)~$17.4\,^\circ$C, (2)~$18.3\,^\circ$C, or (3)~$17.8\,^\circ$C. Response of VK$_\nu$ formulation is compared with experimental data (EXP), 2D FE model~(2D), and monolithic~(LIM$_\text{MON}$) and layered~(LIM$_\text{LAY}$) limits.
        }
	\label{fig:validation}
\end{figure}

Validation of the numerical model was performed using data from experimental campaign by~\citet{Aenlle:2013:ETC}, utilizing three types of simply-supported one- or two-span laminated glass beams (exact dimensions are specified in the caption of \Fref{fig:validation}). The beams were subjected to seven concentrated loads that approximated uniformly distributed loading and the evolution of mid-span deflections and of strains was measured with laser sensors and tensometers, respectively. In addition, the average temperature during the experiment was recorded. In order to generate verification data in ADINA system, we discretized each beam with a very fine mesh of quadratic finite strain elements of approximately square shape with edge size of $0.38$~mm, and determined its response under a ramp loading~(analogous to the one shown in \Fref{fig:load_hist}(a)) while assuming the constant value of the bulk modulus $K\lay{2} = 2$~GPa. 

Results of the simulations and experiments are summarized in \Fref{fig:validation} and involve the values of deflections, \Fref{fig:validation}(a), and stresses, \Fref{fig:validation}(b), as functions of loading time. Notice that the experimental results for stresses appear only in \Fref{fig:validation}(3b), in order to demonstrate that they are oscillatory and even do not fit in the layered-monolithic bounds. This is probably because the temperature was not kept constant during the experiment and the polymer interlayer shows very high sensitivity to temperature variations, as can be deduced from data in \Tref{tab:MaxwellSeries}.  Therefore, the experimental values of stresses are excluded from consideration. 

We conclude from the collected graphs that our computational model is successfully verified and only partially validated. Indeed, the results of the detailed two-dimensional finite element model~(2D) are indistinguishable from the results of VK$_\nu$ model, both for displacements and stresses. The best match between the measured displacements~(EXP) is achieved for non-symmetric 4/0.38/8 beam, \Fref{fig:validation}(1a); the other experiments show visible deviations, which we attribute again to only partial temperature control.  

\begin{table}[ht]
\caption{Comparison of mid-span deflections at 10~hours}
\label{tab:val_def}
\centerline{
\begin{tabular}{cccccccc}
\hline
thickness [mm] & \multicolumn{5}{c}{deflection [mm]}& \multicolumn{2}{c}{error [\%]} \\
glass/PVB/glass & EXP & VK$_\nu$ & 2D & LIM$_\text{MON}$ & LIM$_\text{LAY}$ & VK$_\nu$-EXP & VK$_\nu$-2D \\
\hline
4/0.38/8 & 0.8158 & 0.7839 & 0.7840 & 0.4370 & 1.441 & -4 & -0.01\\
4/0.76/8 & 0.9947 & 0.9234 & 0.9237 & 0.4000 & 1.441 & -7 & -0.03 \\
4/0.38/4 & 1.254 & 1.018 & 1.021 & 0.3340 & 3.068 & -19 & -0.29\\
\hline
\end{tabular}}
\caption{Comparison of mid-span maximum normal stresses at 10~hours}
\label{tab:val_stress}
\centerline{
\begin{tabular}{cccccc}
\hline
thickness [mm] & \multicolumn{4}{c}{stress $\stress_{x}$ [MPa]} & error [\%]\\
glass/PVB/glass & VK$_\nu$ & 2D & LIM$_\text{MON}$ & LIM$_\text{LAY}$ & VK$_\nu$-2D\\
\hline
4/0.38/8 & 2.567 & 2.567 & 1.872 & 3.984 & 0.00\\
4/0.76/8 & 2.846 & 2.847 & 1.762 & 3.984 & -0.04\\
4/0.38/4 & 4.261 & 4.266 & 2.465 & 5.410 & -0.12\\
\hline
\end{tabular}}
\end{table} 

In order to make these claims more quantitative, in Tables~\ref{tab:val_def} and~\ref{tab:val_stress} we provide detailed results of all tested beams at 10 hours. Note that, for example, the VK$_\nu$-EXP error is understood as $( w_{\mathrm{VK}_{\nu}} - w_{\mathrm{EXP}}) / w_{\mathrm{EXP}}$. The results confirm an excellent match between detailed FE model and VK$_\nu$ model, the errors do not exceed $0.3\%$ in deflections and $0.2\%$ in stresses. The difference between experimental data and VK$_\nu$ formulation is somewhat less satisfactory, and ranges from about $5$--$20$\%. However, even this lower accuracy is superior to the accuracy of monolithic~(LIM$_\mathrm{MON}$) and layered~(LIM$_\mathrm{LAY}$) limits: the monolithic bound predicts values that can be up to $50\%$ smaller than for the more refined models, whereas the overestimation by the layered limit can reach up to $200\%$. 

\subsection{Geometric nonlinearity and temperature}\label{sec:GNE_TE}

Herein, we complement the verification and validation results with a study on a structure exhibiting significant effects of geometric nonlinearity. Attention is also paid to the effect of temperature variation, accounted for by the Williams--Landel--Ferry formula~\eqref{eq6:shiftfactor}. In particular, we consider the fixed-end beam structure from \Tref{tab:geom} subjected to loading~A, recall \Fref{fig:load_hist}(a), under three constant temperatures $0\,^\circ$C, $25\,^\circ$C, and $50\,^\circ$C.

\begin{figure}[p]
\small (1a)\includegraphics[height=\myFigureHeight]{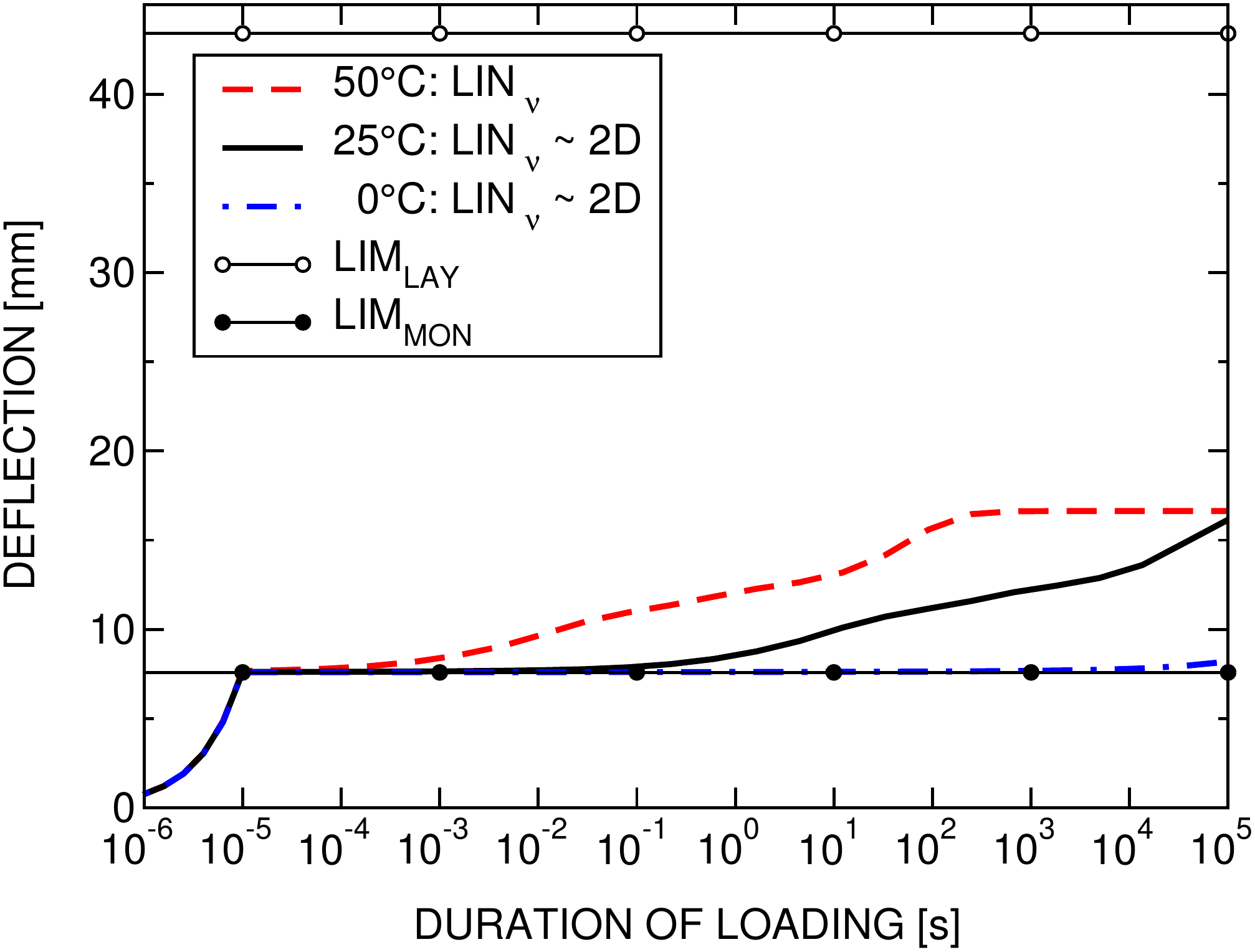}	
\hfill
\small (1b)\includegraphics[height=\myFigureHeight]{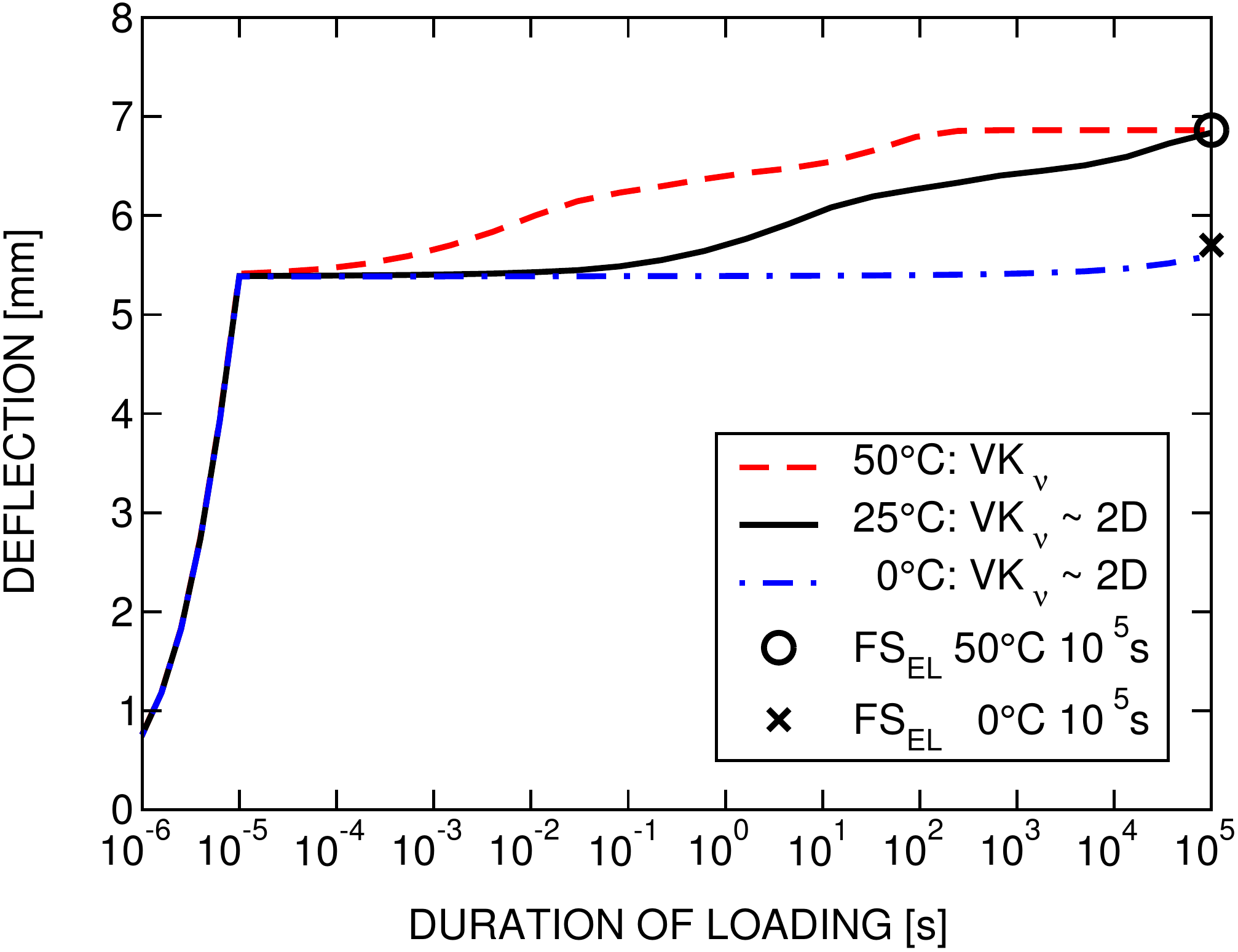}	

\bigskip

\small (2a)\includegraphics[height=\myFigureHeight]{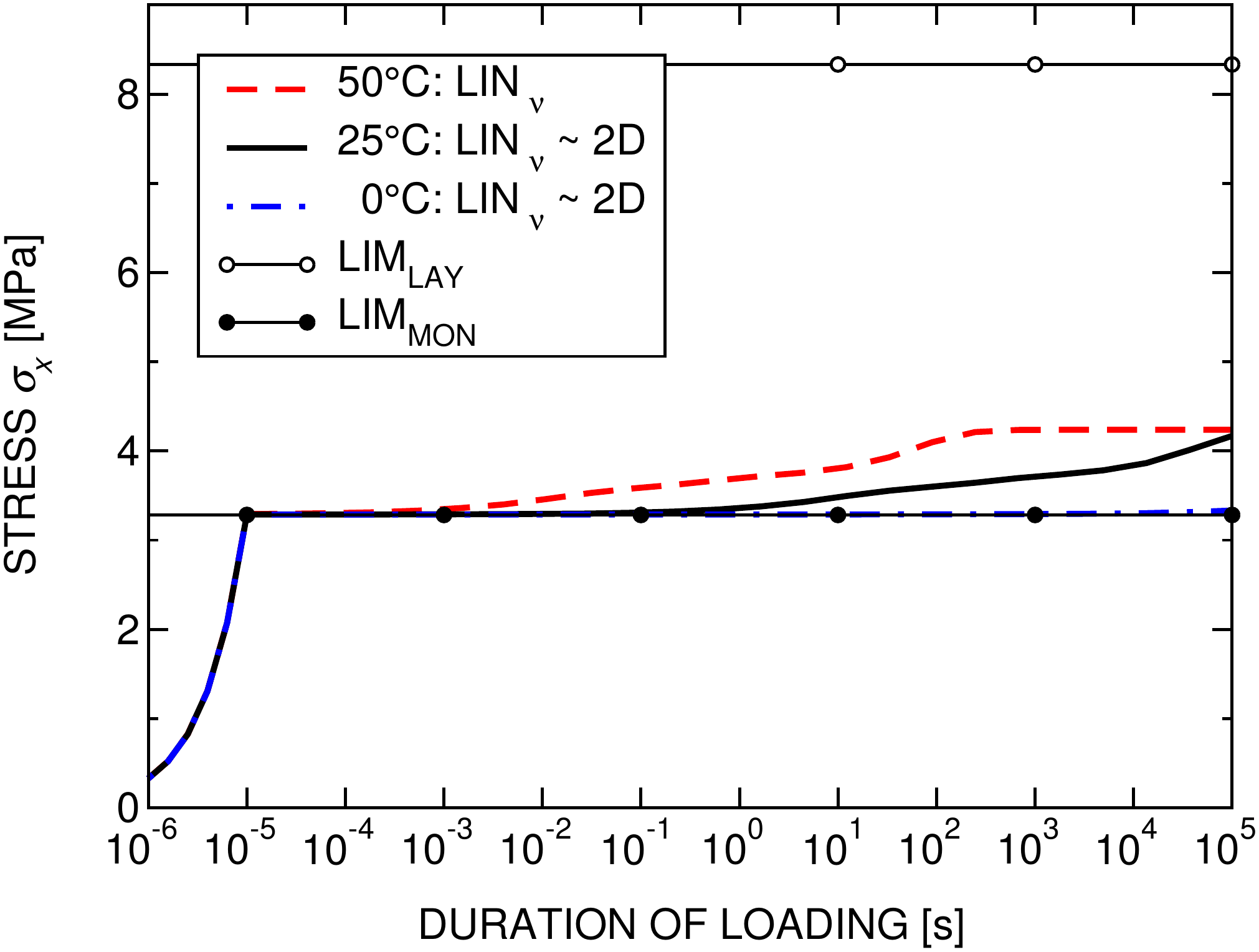}	
\hfill
\small (2b)\includegraphics[height=\myFigureHeight]{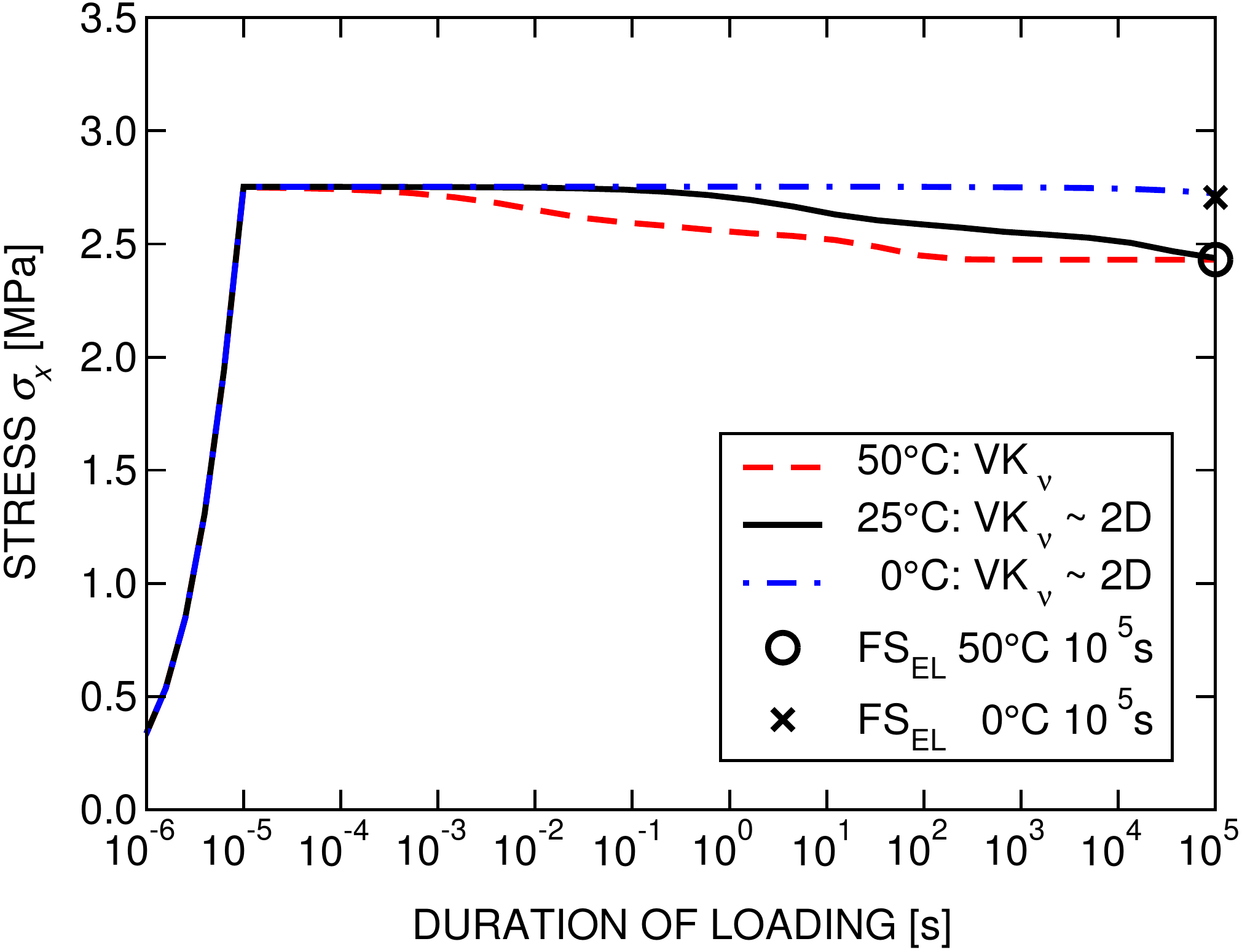}	

\bigskip

\small (3a)\includegraphics[height=\myFigureHeight]{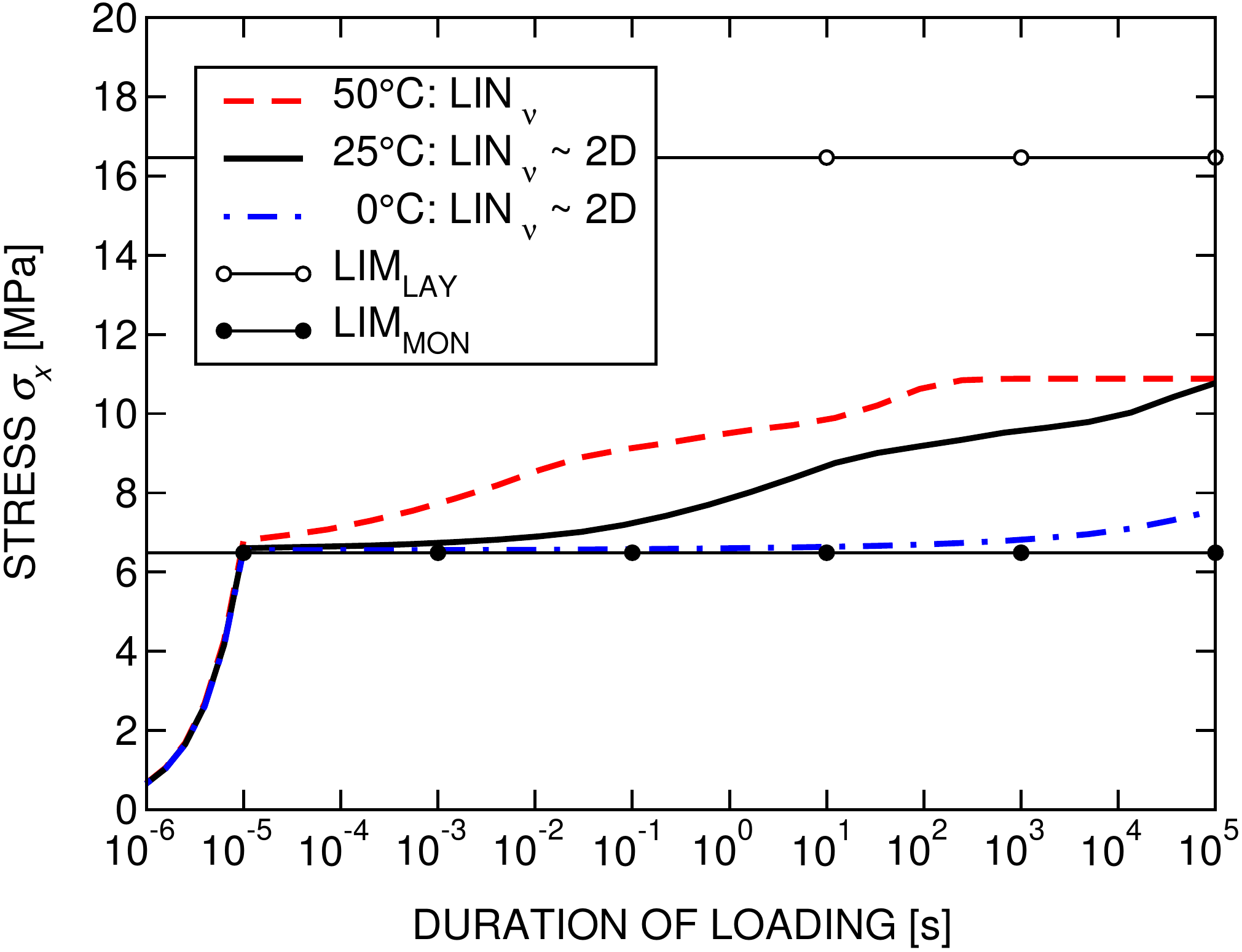}	
\hfill
\small (3b)\includegraphics[height=\myFigureHeight]{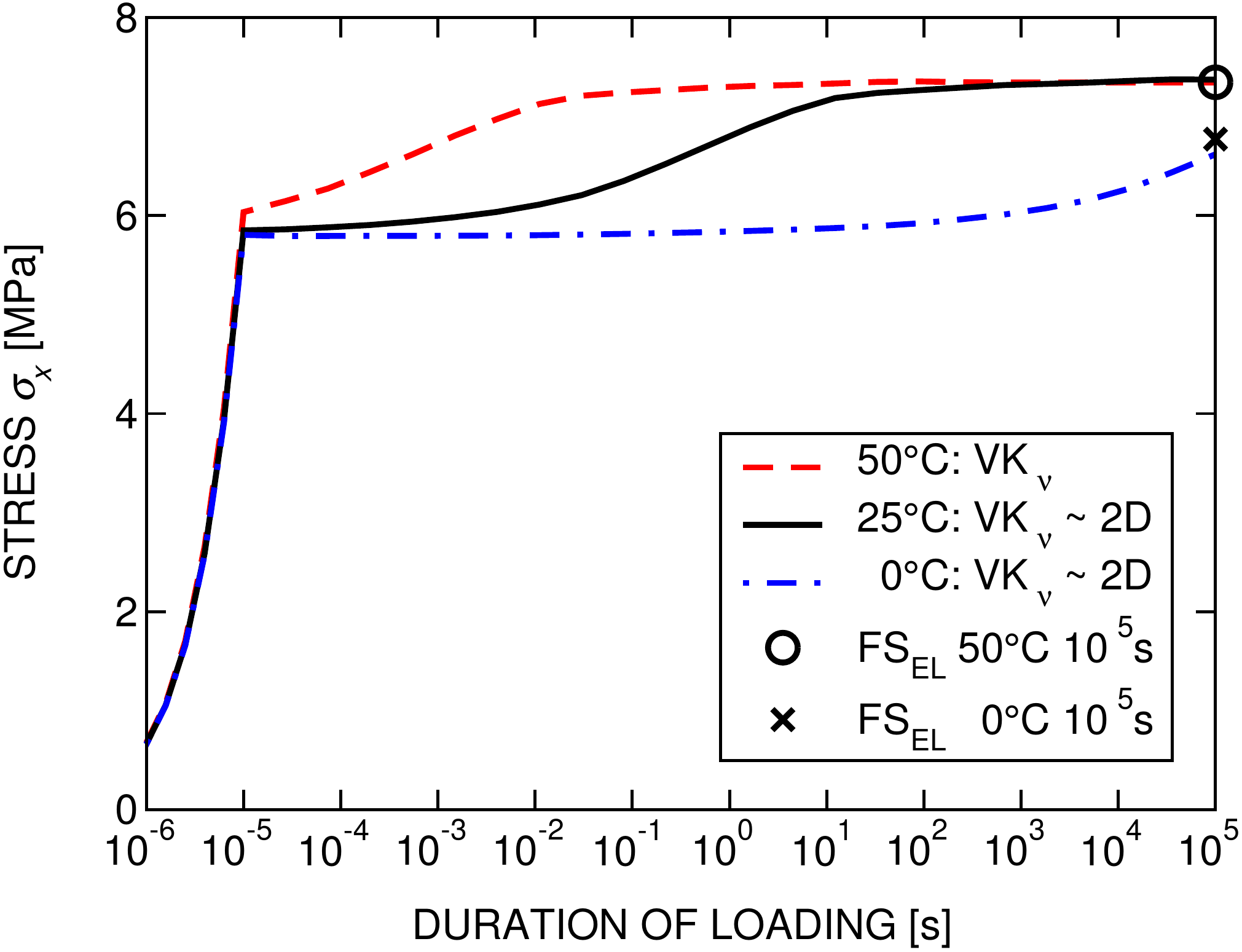}	
\caption{Fixed-end beam with geometry~I under loading~A: Mid-point (1)~deflections, (2)~maximum normal stresses, and (3) the largest values of maximum normal stresses at temperatures of $0$, $25$, and $50\,^\circ$C. Response of (a)~geometrically linear~(LIN$_\nu$) and (b)~nonlinear~(VK$_\nu$) beam models are compared with detailed model in ADINA (2D), monolithic (LIM$_\text{MON}$) and layered (LIM$_\text{LAY}$) limits, and elastic finite strain model (FS$_\text{EL}$), using the shear modulus of the interlayer corresponding to temperatures from $0$ to $50~^\circ$C and duration of loading $10^5$~s.}
\label{fig:fixed}
\end{figure}

The resulting time evolution of deflections and stresses is plotted in \Fref{fig:fixed} for geometrically linear and nonlinear models. The standard upper and lower limits, provided by the geometrically linear monolithic and layered approximations, are plotted in \Fref{fig:fixed}~(a) and they indeed bound the behavior of the laminated unit.
For geometrically nonlinear response, we also provide the elastic response (FS$_\text{EL}$), determined by elastic model from~\cite{Zemanova:2014:NMFS} with the interlayer shear modulus determined from~\eqref{eq6:PronyG} for the given temperature at the final simulation time of $10^5$~s. This is complemented with selected data of mid-point deflections, \Tref{tab:comp_def} and stresses, \Tref{tab:comp_def2}.

\begin{table}[ht]
\caption{Comparison of mid-span deflections at~$10^{5}$~s. The symbol -- indicates divergence of ADINA solver.}
\label{tab:comp_def}
\centerline{
\begin{tabular}{cccccccc}
\hline
& \multicolumn{2}{c}{linear response [mm]} & \multicolumn{3}{c}{nonlinear response [mm]} & \multicolumn{2}{c}{error [\%]}\\
temperature [$\,^\circ$C]& LIN$_\nu$ & 2D(LIN) & VK$_\nu$ & 2D & FS$_\text{EL}$ & VK$_\nu$-2D & FS$_\text{EL}$-VK$_\nu$ \\
\hline
 0 & 8.192 & 8.191 & 5.596 & 5.595 & 5.701 &0.02 & 1.88\\
25 & 16.15 & 16.15 & 6.838 & 6.838 & 6.857 &0.00 & 0.28\\
50 & 16.63 & -- & 6.863 & -- & 6.863 &-- & 0.00\\
\hline
\end{tabular}}

\caption{Comparison of maximum mid-span normal stresses $\stress_{x}$ at~$10^{5}$~s. The symbol -- indicates divergence of ADINA solver.}
\label{tab:comp_def2}
\centerline{
\begin{tabular}{cccccccc}
\hline
& \multicolumn{2}{c}{linear response [MPa]} & \multicolumn{3}{c}{nonlinear response [MPa]} & \multicolumn{2}{c}{error [\%]} \\
temperature [$\,^\circ$C]& LIN$_\nu$ & 2D(LIN) & VK$_\nu$ & 2D & FS$_\text{EL}$ & VK$_\nu$-2D & FS$_\text{EL}$-VK$_\nu$\\
\hline
 0 & 3.332 & 3.332 & 2.724 & 2.724 & 2.706 &0.00&-0.66 \\
25 & 4.170 & 4.170 & 2.437 & 2.438 & 2.433 &-0.04&-0.21\\
50 & 4.237 & -- & 2.431 & -- & 2.431 &--&0.00\\
\hline
\end{tabular}}
\end{table} 

The results in \Fref{fig:fixed}(a) are in full agreement with the outcomes presented in~\cite{Behr:1993:SBA}, namely that the laminated units exposed to
temperatures around $0\,^\circ$C effectively behave as the monolithic ones. In addition, even at $50^\circ$C, the shear stiffness is sufficient to ensure the inter-layer interaction, so that the overall response is much closer to the monolithic limit rather than to the layered limit. Behavior of geometrically non-linear models, \Fref{fig:fixed}(b), exhibits similar trends, but the displacement and stress magnitudes are reduced to $40\%$ and $60\%$, respectively; see also Tables~\ref{tab:comp_def} and~\ref{tab:comp_def2}. In addition, unlike in the geometrically linear model, the extreme stresses at the mid-span decrease with increasing time. We consider this to be a direct consequence of the development of membrane stresses in glass layers, which leads to the redistribution of extreme stresses from the beam mid-section towards the supports (the same mechanism was observed in the detailed 2D finite element model). 

Furthermore, the results in this section serve as an additional verification of VK$_\nu$ model, because the errors in displacements and stresses do not exceed $0.5\%$ for temperatures of $0$ and $25^\circ$C with respect to the results of detailed 2D finite element model, consult again Tables~\ref{tab:comp_def} and~\ref{tab:comp_def2}. Note that simulations in ADINA did not convergence at $50^\circ$C, because of poor conditioning of the stiffness matrix caused by a very high contrast in the interlayer and glass properties. In addition, the results reveal that the elastic model provides very good estimates at the final time of simulations: the errors stay below $2\%$ for all quantities of interest. This aspect will be studied in more detail in the next section.    

\subsection{Comparison with simplified elastic solution}\label{sec:Visco_simp}
%

Being inspired by earlier studies by~\citet{Galuppi:2013:DLG}, we close this section by comparing the full viscoelastic solution with a simplified "secant" solution, in which we determine the structural response at the relevant times by the geometrically non-linear model~\cite{Zemanova:2014:NMLG} with shear modulus determined from~\eqref{eq6:PronyG}. Similarly to~\cite{Galuppi:2013:DLG}, we consider the two-load history~B, see \Fref{fig:load_hist}(b), consisting of a sudden loading, sudden unloading, and holding parts. In addition, both simply supported and fixed-end beams are considered under constant temperatures of $25$ and $50^\circ$C. 

\begin{figure}[h]
\small (1a)\includegraphics[height=\myFigureHeight]{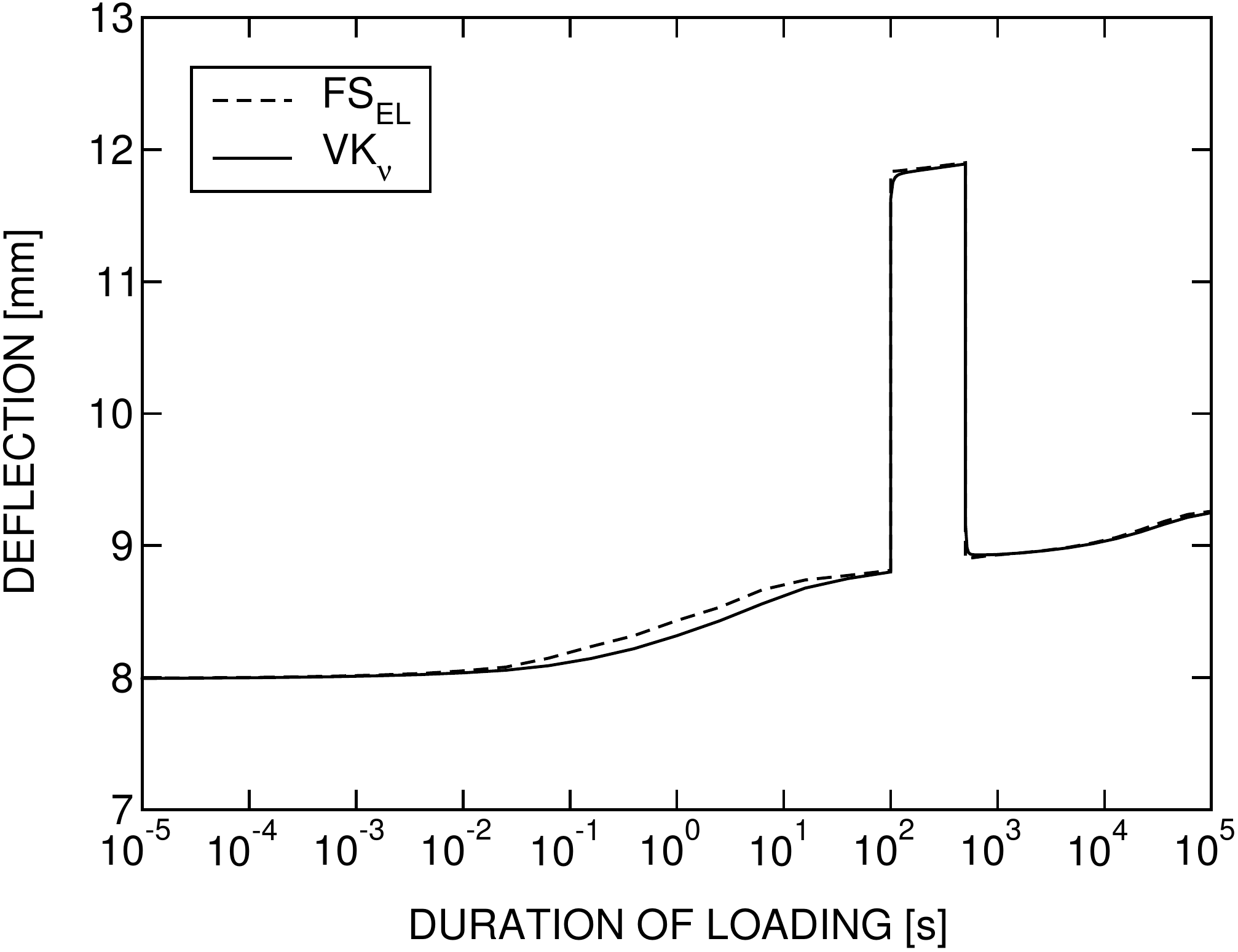}	
\hfill
\small (2a)\includegraphics[height=\myFigureHeight]{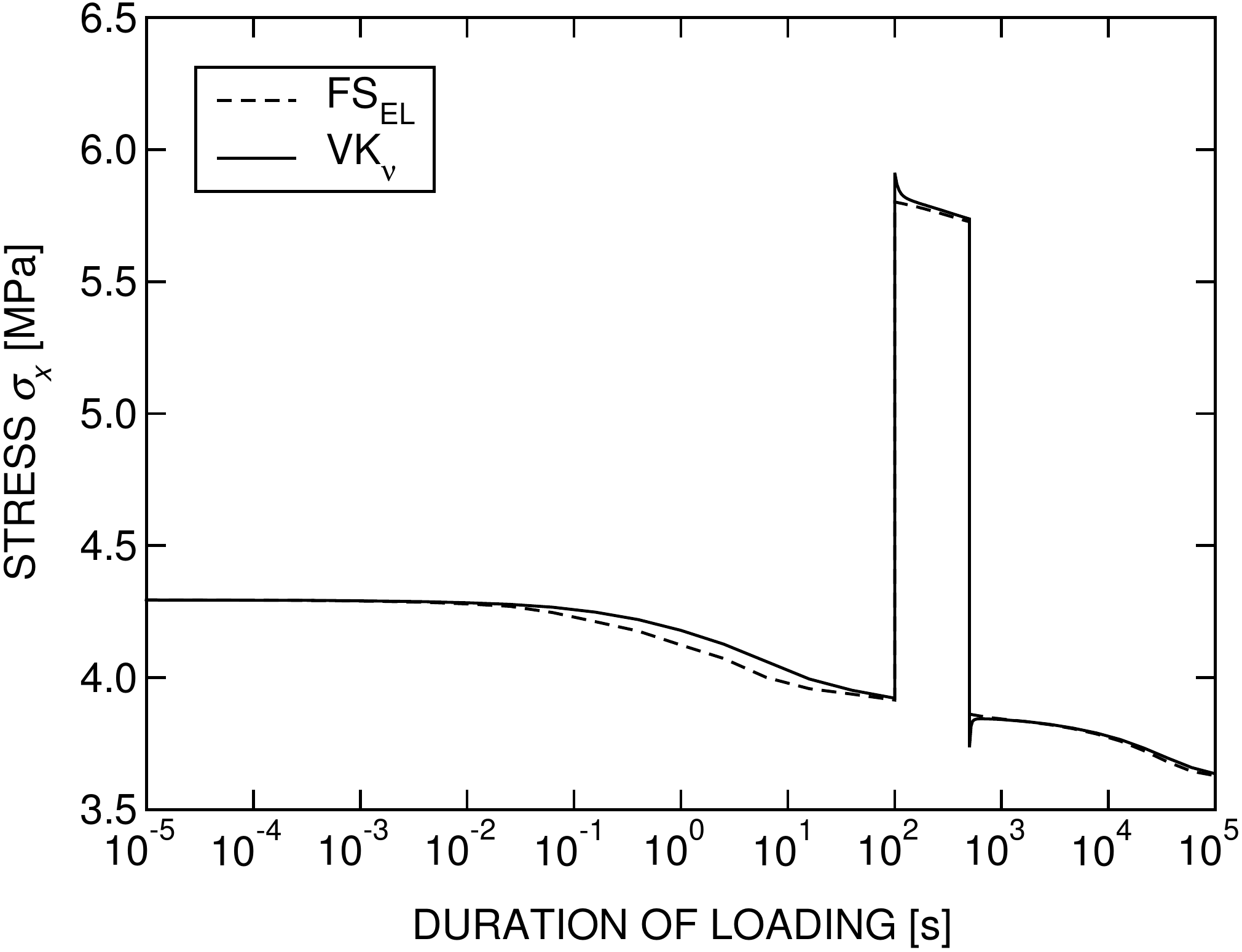}	

\bigskip

\small (1b)\includegraphics[height=\myFigureHeight]{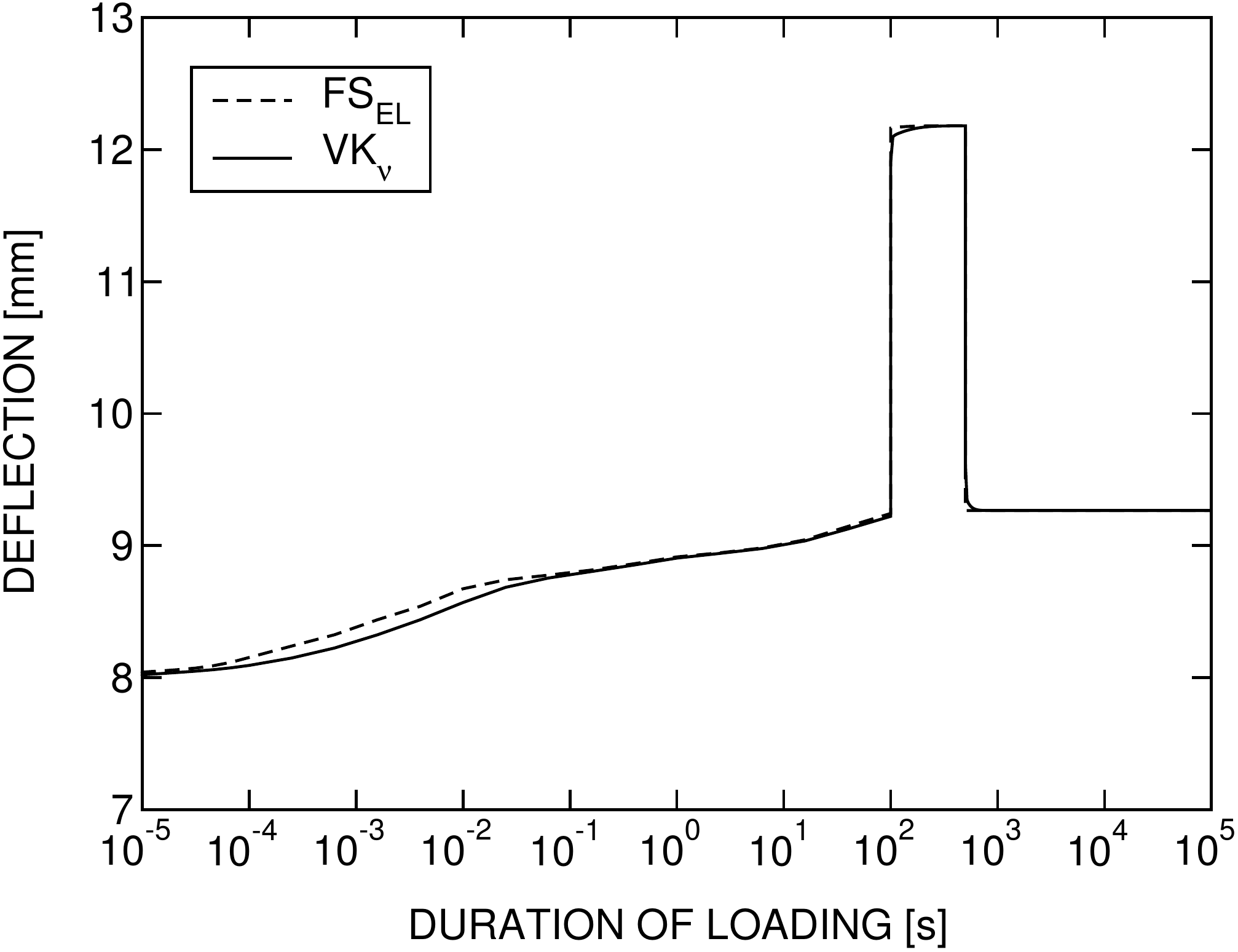}	
\hfill
\small (2b)\includegraphics[height=\myFigureHeight]{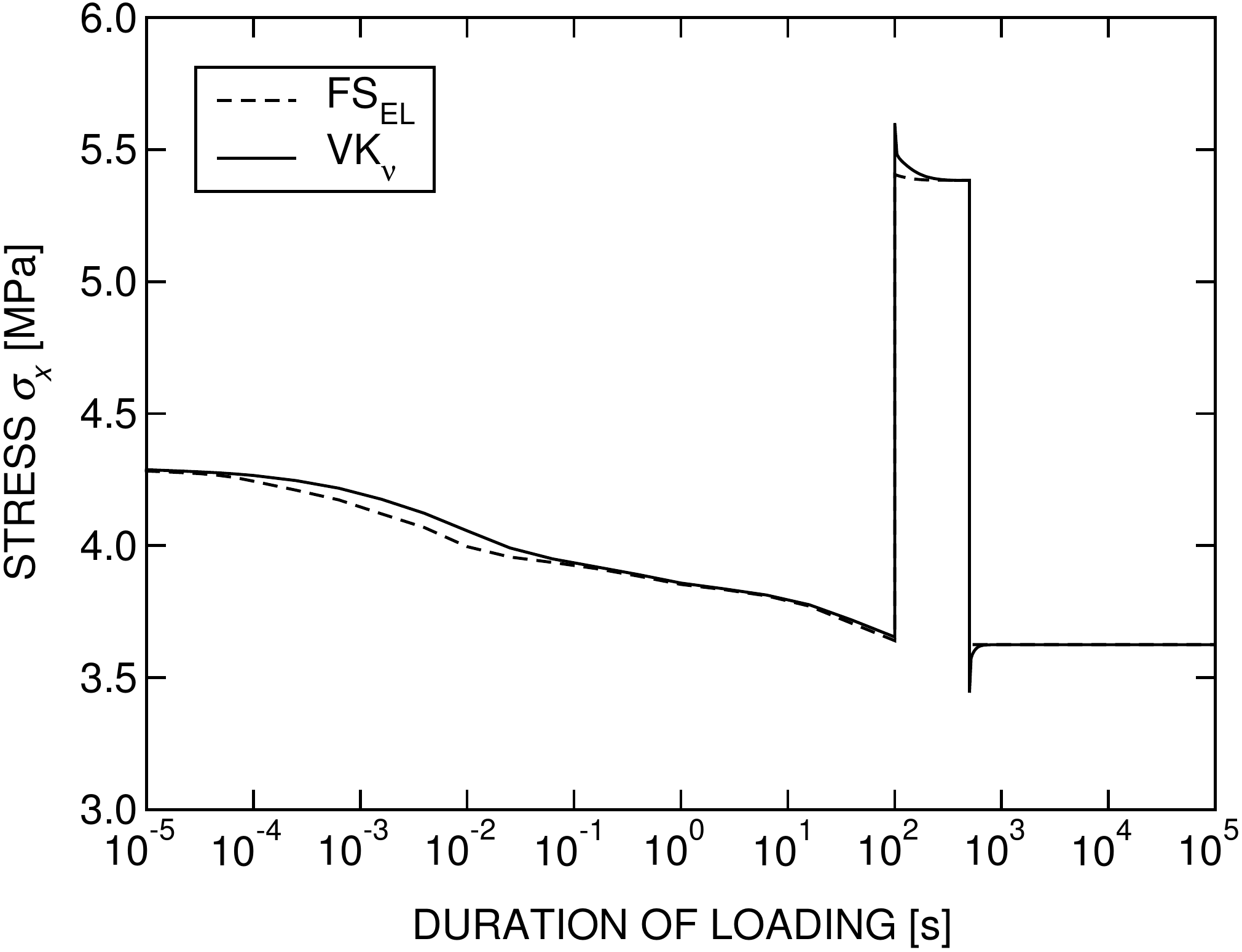}	
\caption{
	Fixed-end beam with geometry I under loading B: Mid-point (1)~deflections or (2)~maximum normal stresses at temperatures of (a)~$25\,^\circ$C and (b)~$50\,^\circ$C. Response of geometrically nonlinear VK$_\nu$ models is compared with simplified elastic model (FS$_\text{EL}$).} 
\label{fig:VxS_VV}
\end{figure}

\begin{figure}[h]
\small (1a)\includegraphics[height=\myFigureHeight]{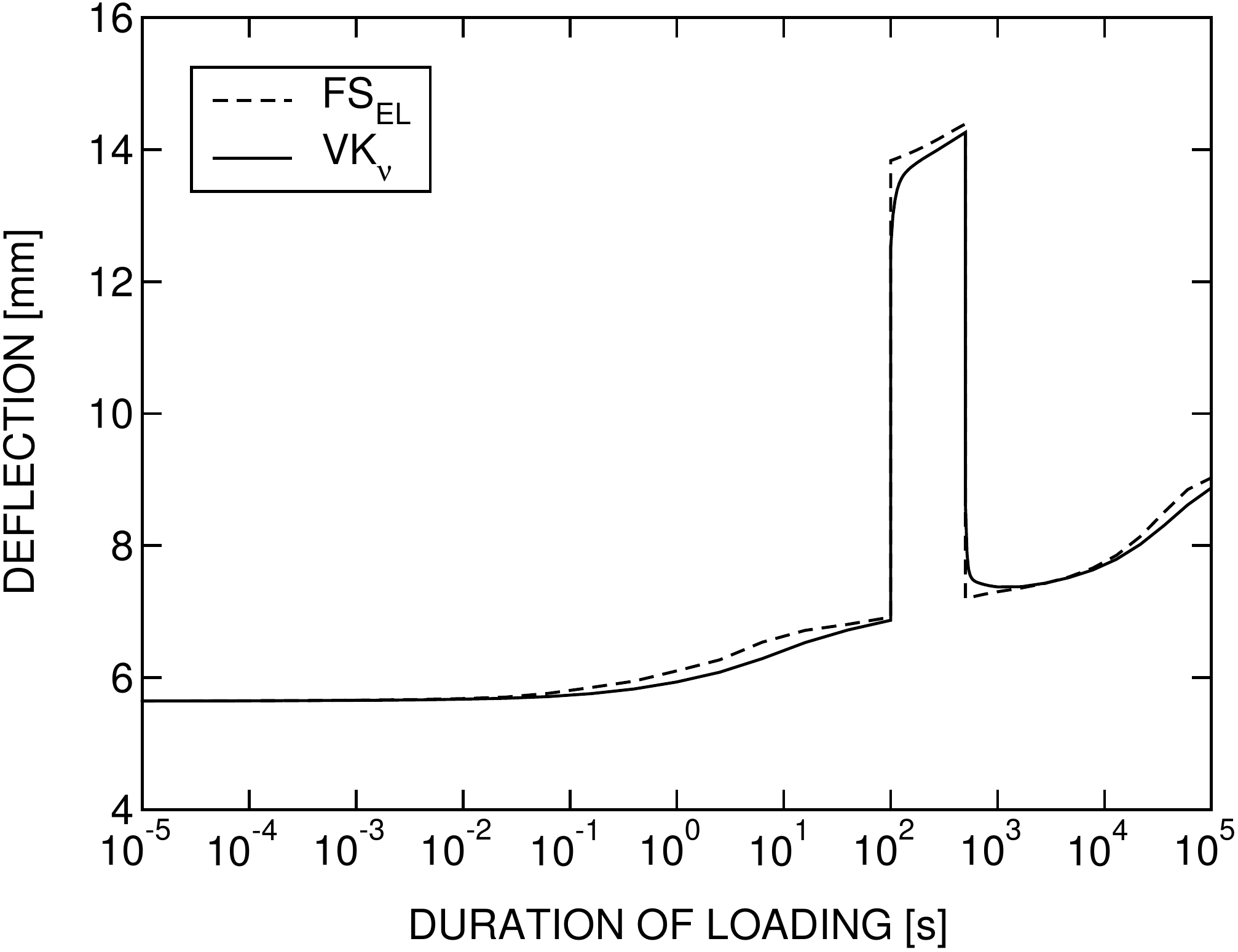}	
\hfill
\small (2a)\includegraphics[height=\myFigureHeight]{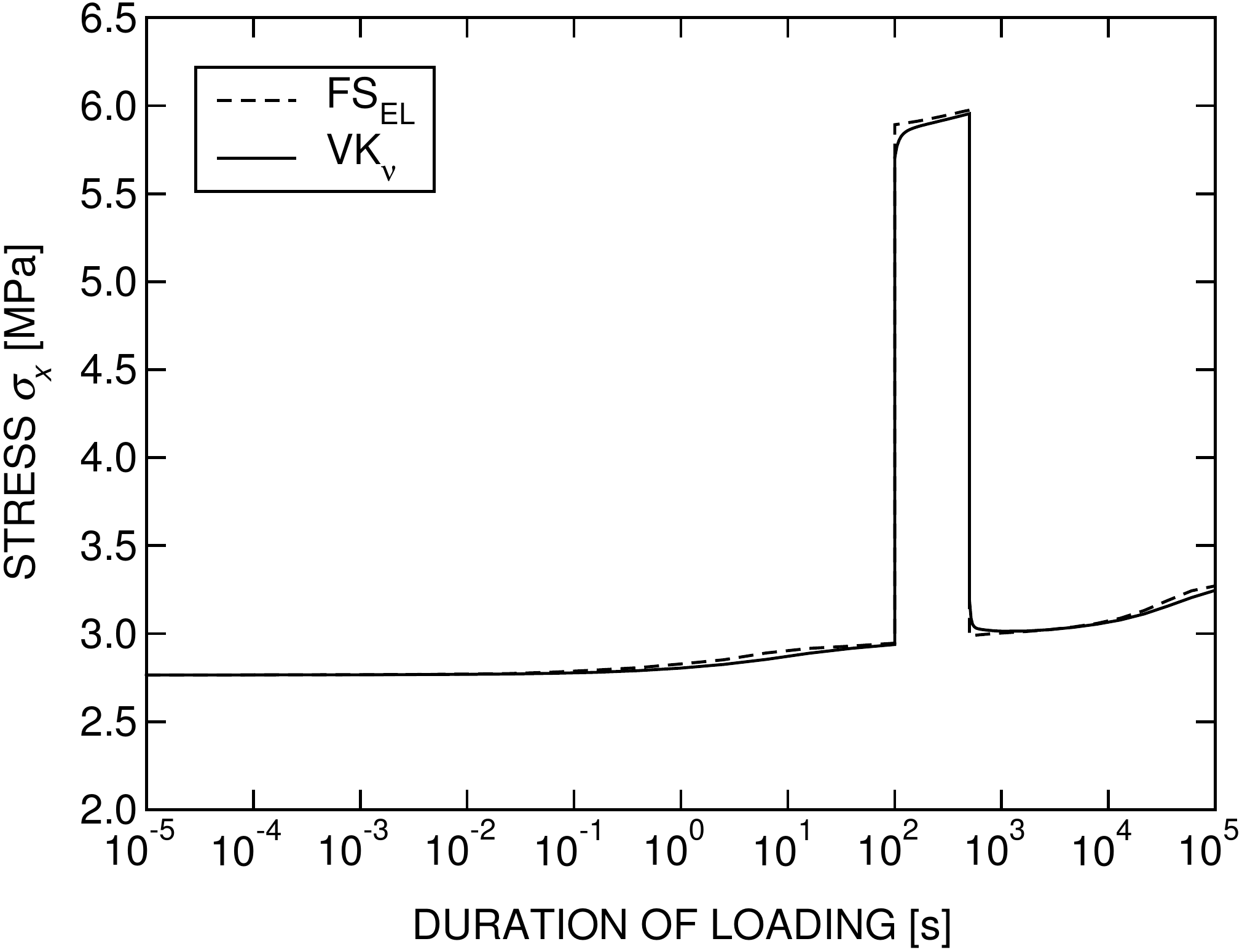}	

\bigskip

\small (1b)\includegraphics[height=\myFigureHeight]{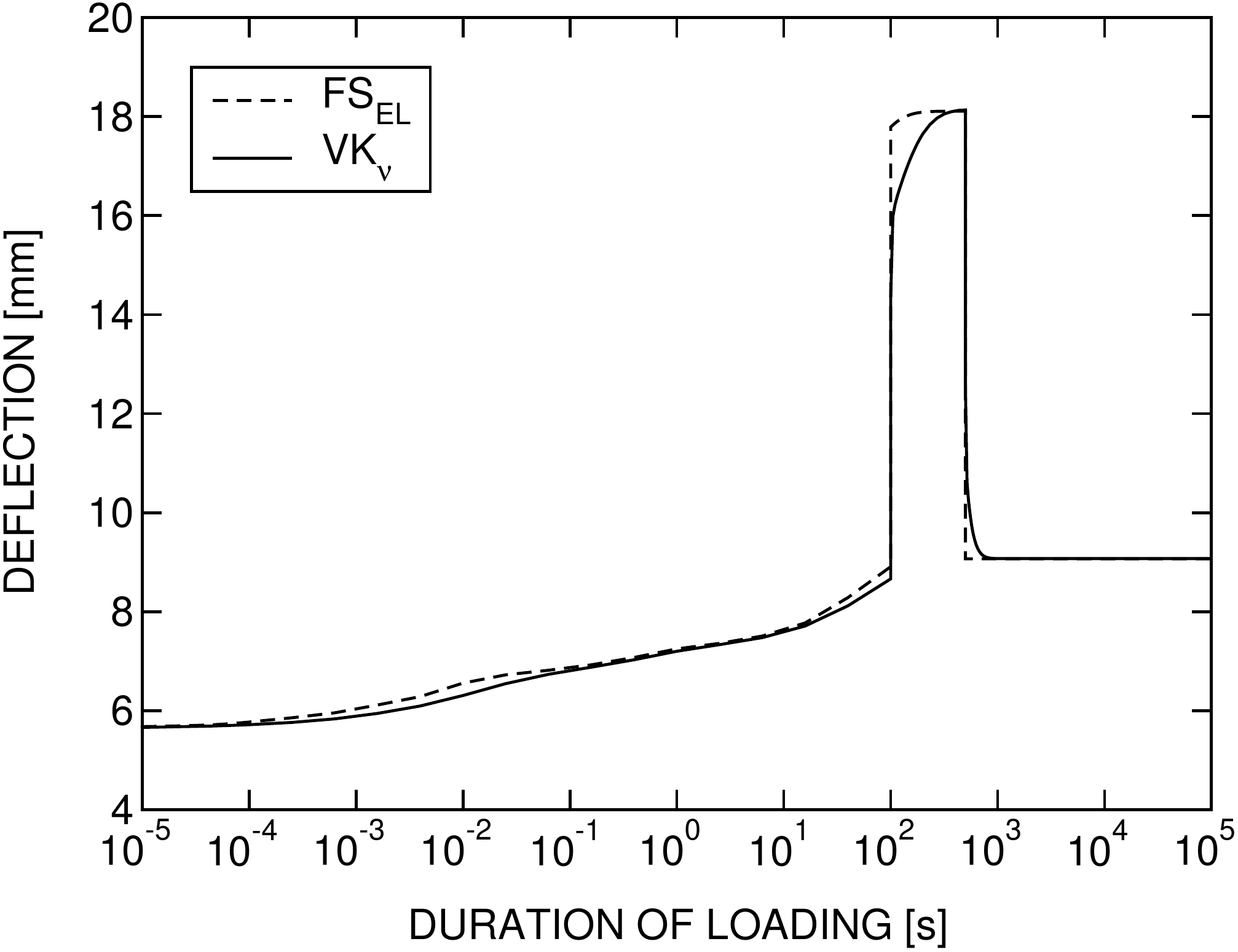}	
\hfill
\small (2b)\includegraphics[height=\myFigureHeight]{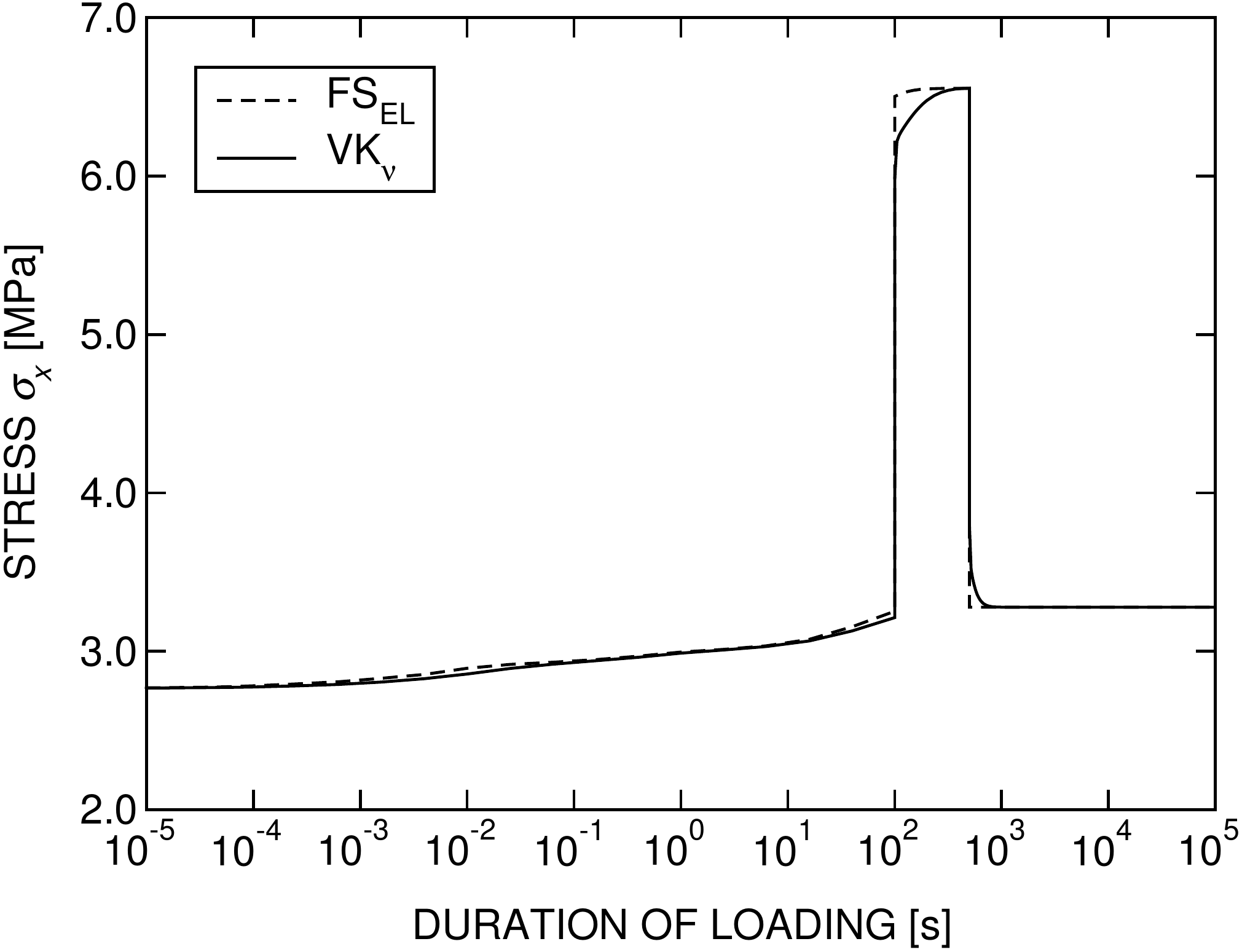}	
\caption{Simply-supported beam with geometry II under loading C: Mid-point (1)~deflections or (2)~maximum normal stresses at temperatures of (a)~$25\,^\circ$C and (b)~$50\,^\circ$C. Response of geometrically nonlinear VK$_\nu$ models is compared with simplified elastic model (FS$_\text{EL}$).} 
\label{fig:VxS_SS}
\end{figure}

The simulation results collected in Figures~\ref{fig:VxS_VV} and~\ref{fig:VxS_SS} reveal that the correspondence between the simplified model and full viscoelastic solution is rather satisfactory for holding phases of the loading, but differences appear in the vicinity of displacement jumps, where the simplified solution can lead to unsafe predictions. This effects is the most pronounced for stresses in the fixed-end beam and it is attributed to a different stress redistribution in the viscoelastic and elastic beams as discussed in the previous section. In addition, the magnitude seems to be highly temperature-dependent: the differences are negligible at $0^\circ$C~(not shown) and increase with increasing temperature. 

More quantitatively, for the fixed-end beam, \Fref{fig:VxS_VV}, the differences remain below $5\%$; at $25^\circ$C they do not exceed $2\%$ for both stresses and strains, whereas for $50^\circ$C they reach $4\%$ for displacements and $5\%$ for stresses. For the simply supported beam, the differences between elastic and viscoelastic solution increase and reach up to $27\%$ for displacements and $14\%$ for stresses. These results again highlight the complex interplay between structural heterogeneity, geometric nonlinearity, and viscoelastic effects.

Note in conclusion that all these findings agree well with the analytical study by~\citet{Galuppi:2013:DLG} for a simply supported beam with an interlayer modeled using simplified one- and three-unit Maxwell chains, disregarding the temperature effects.

\section{Conclusions}\label{sec:conclusions}

In this paper, we have introduced and thoroughly tested four layer-wise finite element formulations to determine displacements and stresses in laminated glass beams with interlayer(s) exhibiting time/temperature-dependent material response. The models differ in the adopted beam kinematics --- large-deflection von Karm\'{a}n formulation vs. finite strain Reissner formulation --- and the assumptions adopted in the constitutive model --- constant bulk modulus vs. constant Poisson ratio. By comparing their response with detailed two-dimensional finite element simulations, experimental data, and simplified approaches, we have reached the following conclusions:

\begin{enumerate}
	\item All four considered models provide almost identical response under monotone and non-monotone loading for deflections up to $1/50$ of span. Therefore, the most straightforward formulation, combining the assumptions of the von K\'{a}rm\'{a}n kinematics and of constant Poisson ratio, can be used to simulate more complex structures. 
	
	\item This formulation has been successfully verified against several detailed finite element simulations performed in ADINA system and partially validated against experimental data collected by~\citet{Aenlle:2013:ETC}. The model has also consistently reproduced typical features of mechanical response of laminated glass structures at different temperatures.
	
	\item A simplified model, based on an elastic formulation with the interlayer shear modulus adjusted to duration of loading and temperature, provides satisfactory agreement with full viscoelastic solutions at low temperatures or at holding periods of loading. At room and elevated temperatures, differences have been observed for rapid changes of loading and unloading that may reach the order of $10\%$ and/or may be on the unsafe side.   
\end{enumerate}

As the next step, we will enhance our elastic finite element solver for laminated glass plates~\cite{Zemanova:2015:FEM} with viscoelastic effects, closely following the formulations developed in this paper. Subsequently, we would like to extend the formulation to post-critical state, by incorporating glass fracture into the layer-wise formulation.  

\paragraph{Acknowledgments}
This work was initiated when Alena Zemanov\'{a} was a post-doctoral researcher, supported by the project No.~CZ.1.07/2.3.00/30.0034: Support of inter-sectoral mobility and quality enhancement of research teams at Czech Technical University in Prague (06/2014--06/2015). All authors acknowledge the support by the Czech Science Foundation, project No.~16-14770S. 

\section*{\refname}
\bibliography{liter}

\begin{thebibliography}{44}
\providecommand{\natexlab}[1]{#1}
\providecommand{\url}[1]{\texttt{#1}}
\providecommand{\urlprefix}{URL }
\expandafter\ifx\csname urlstyle\endcsname\relax
  \providecommand{\doi}[1]{doi:\discretionary{}{}{}#1}\else
  \providecommand{\doi}[1]{doi:\discretionary{}{}{}\begingroup
  \urlstyle{rm}\url{#1}\endgroup}\fi
\providecommand{\bibinfo}[2]{#2}

\bibitem[{Haldimann et~al.(2008)Haldimann, Luible, and
  Overend}]{Haldimann:2008:SUG}
\bibinfo{author}{M.~Haldimann}, \bibinfo{author}{A.~Luible},
  \bibinfo{author}{M.~Overend}, \bibinfo{title}{Structural Use of Glass},
  vol.~\bibinfo{volume}{10} of \emph{\bibinfo{series}{Structural Engineering
  Documents}}, \bibinfo{publisher}{{IABSE}}, \bibinfo{address}{Z\"{u}rich,
  Switzerland}, \bibinfo{year}{2008}.

\bibitem[{Serafinavi\v{c}ius et~al.(2013)Serafinavi\v{c}ius, Lebet, Louter,
  Lenkimas, and Kuranovas}]{Serafinaviciusa:2013:LTLG}
\bibinfo{author}{T.~Serafinavi\v{c}ius}, \bibinfo{author}{J.~P. Lebet},
  \bibinfo{author}{C.~Louter}, \bibinfo{author}{T.~Lenkimas},
  \bibinfo{author}{A.~Kuranovas}, \bibinfo{title}{Long-term laminated glass
  four point bending test with {PVB}, {EVA} and {SG} interlayers at different
  temperatures}, \bibinfo{journal}{Procedia Engineering} \bibinfo{volume}{57}
  (\bibinfo{year}{2013}) \bibinfo{pages}{996--1004},
  \doi{\bibinfo{doi}{10.1016/j.proeng.2013.04.126}}.

\bibitem[{Huang et~al.(2014)Huang, Liu, Liu, and Bennison}]{Huang:2014:ESLG}
\bibinfo{author}{X.~Huang}, \bibinfo{author}{G.~Liu}, \bibinfo{author}{Q.~Liu},
  \bibinfo{author}{S.~Bennison}, \bibinfo{title}{An experimental study on the
  flexural performance of laminated glass}, \bibinfo{journal}{Structural
  Engineering and Mechanics} \bibinfo{volume}{49}~(\bibinfo{number}{2})
  (\bibinfo{year}{2014}) \bibinfo{pages}{261--271},
  \doi{\bibinfo{doi}{10.12989/sem.2014.49.2.261}}.

\bibitem[{Shitanoki et~al.(2014)Shitanoki, Bennison, and
  Koike}]{Shitanoki:2014:PNM}
\bibinfo{author}{Y.~Shitanoki}, \bibinfo{author}{S.~Bennison},
  \bibinfo{author}{Y.~Koike}, \bibinfo{title}{A practical, nondestructive
  method to determine the shear relaxation modulus behavior of polymeric
  interlayers for laminated glass}, \bibinfo{journal}{Polymer Testing}
  \bibinfo{volume}{37} (\bibinfo{year}{2014}) \bibinfo{pages}{59--67},
  \doi{\bibinfo{doi}{10.1016/j.polymertesting.2014.04.011}}.

\bibitem[{Zemanov\'{a} et~al.(2014)Zemanov\'{a}, Zeman, and
  \v{S}ejnoha}]{Zemanova:2014:NMFS}
\bibinfo{author}{A.~Zemanov\'{a}}, \bibinfo{author}{J.~Zeman},
  \bibinfo{author}{M.~\v{S}ejnoha}, \bibinfo{title}{Numerical model of elastic
  laminated glass beams under finite strain}, \bibinfo{journal}{Archives of
  Civil and Mechanical Engineering} \bibinfo{volume}{14}~(\bibinfo{number}{4})
  (\bibinfo{year}{2014}) \bibinfo{pages}{734--744},
  \doi{\bibinfo{doi}{10.1016/j.acme.2014.03.005}}.

\bibitem[{Behr et~al.(1985)Behr, Minor, Linden, and Vallabhan}]{Behr:1985:LGU}
\bibinfo{author}{R.~A. Behr}, \bibinfo{author}{J.~E. Minor},
  \bibinfo{author}{M.~P. Linden}, \bibinfo{author}{C.~V.~G. Vallabhan},
  \bibinfo{title}{Laminated glass units under uniform lateral pressure},
  \bibinfo{journal}{Journal of Structural Engineering}
  \bibinfo{volume}{111}~(\bibinfo{number}{5}) (\bibinfo{year}{1985})
  \bibinfo{pages}{1037--1050},
  \doi{\bibinfo{doi}{10.1061/(ASCE)0733-9445(1985)111:5(1037)}}.

\bibitem[{Koutsawa and Daya(2007)}]{Koutsawa:2007:SFVA}
\bibinfo{author}{Y.~Koutsawa}, \bibinfo{author}{E.~Daya},
  \bibinfo{title}{Static and free vibration analysis of laminated glass beam on
  viscoelastic supports}, \bibinfo{journal}{International Journal of Solids and
  Structures} \bibinfo{volume}{44} (\bibinfo{year}{2007})
  \bibinfo{pages}{8735--8750},
  \doi{\bibinfo{doi}{10.1016/j.ijsolstr.2007.07.009}}.

\bibitem[{Benninson et~al.(2008)Benninson, Qin, and
  Davis}]{Benninson:2008:HPLG}
\bibinfo{author}{S.~J. Benninson}, \bibinfo{author}{M.~H.~X. Qin},
  \bibinfo{author}{P.~S. Davis}, \bibinfo{title}{High-performance laminated
  glass for structurally efficient glazing}, \bibinfo{journal}{Innovative
  light-weight structures and sustainable facades, Hong Kong} .

\bibitem[{Galuppi and Royer-Carfagni(2012{\natexlab{a}})}]{Galuppi:2012:ETL}
\bibinfo{author}{L.~Galuppi}, \bibinfo{author}{G.~F. Royer-Carfagni},
  \bibinfo{title}{Effective thickness of laminated glass beams: {New}
  expression via a variational approach}, \bibinfo{journal}{Engineering
  Structures} \bibinfo{volume}{38} (\bibinfo{year}{2012}{\natexlab{a}})
  \bibinfo{pages}{53--67},
  \doi{\bibinfo{doi}{10.1016/j.engstruct.2011.12.039}}.

\bibitem[{L\'{o}pez-Aenlle et~al.(2013)L\'{o}pez-Aenlle, Pelayo,
  Fern\'{a}ndez-Canteli, and Garc\'{i}a~Prieto}]{Aenlle:2013:ETC}
\bibinfo{author}{M.~L\'{o}pez-Aenlle}, \bibinfo{author}{F.~Pelayo},
  \bibinfo{author}{A.~Fern\'{a}ndez-Canteli}, \bibinfo{author}{M.~A.
  Garc\'{i}a~Prieto}, \bibinfo{title}{The effective-thickness concept in
  laminated-glass elements under static loading}, \bibinfo{journal}{Engineering
  Structures} \bibinfo{volume}{56} (\bibinfo{year}{2013})
  \bibinfo{pages}{1092--1102},
  \doi{\bibinfo{doi}{10.1016/j.engstruct.2013.06.018}}.

\bibitem[{Eisentraeger et~al.(2015)Eisentraeger, Naumenko, Altenbach, and
  Koeppe}]{Eisentraeger:2015:AFO}
\bibinfo{author}{J.~Eisentraeger}, \bibinfo{author}{K.~Naumenko},
  \bibinfo{author}{H.~Altenbach}, \bibinfo{author}{H.~Koeppe},
  \bibinfo{title}{Application of the first-order shear deformation theory to
  the analysis of laminated glasses and photovoltaic panels},
  \bibinfo{journal}{International Journal of Mechanical Sciences}
  \bibinfo{volume}{96--97} (\bibinfo{year}{2015}) \bibinfo{pages}{163--171},
  \doi{\bibinfo{doi}{10.1016/j.ijmecsci.2015.03.012}}.

\bibitem[{Hooper(1973)}]{Hooper:1973:BAL}
\bibinfo{author}{J.~Hooper}, \bibinfo{title}{On the bending of architectural
  laminated glass}, \bibinfo{journal}{International Journal of Mechanical
  Sciences} \bibinfo{volume}{15}~(\bibinfo{number}{4}) (\bibinfo{year}{1973})
  \bibinfo{pages}{309--323}, \doi{\bibinfo{doi}{10.1016/0020-7403(73)90012-X}}.

\bibitem[{Norville et~al.(1998)Norville, King, and
  Swofford}]{Norville:1998:BSL}
\bibinfo{author}{H.~S. Norville}, \bibinfo{author}{K.~W. King},
  \bibinfo{author}{J.~L. Swofford}, \bibinfo{title}{Behavior and strength of
  laminated glass}, \bibinfo{journal}{Journal of Engineering Mechanics}
  \bibinfo{volume}{124}~(\bibinfo{number}{1}) (\bibinfo{year}{1998})
  \bibinfo{pages}{46--53},
  \doi{\bibinfo{doi}{10.1061/(ASCE)0733-9399(1998)124:1(46)}}.

\bibitem[{Ivanov(2006)}]{Ivanov:2006:AMO}
\bibinfo{author}{I.~V. Ivanov}, \bibinfo{title}{Analysis, modelling, and
  optimization of laminated glasses as plane beam},
  \bibinfo{journal}{International Journal of Solids and Structures}
  \bibinfo{volume}{43}~(\bibinfo{number}{22--23}) (\bibinfo{year}{2006})
  \bibinfo{pages}{6887--6907},
  \doi{\bibinfo{doi}{10.1016/j.ijsolstr.2006.02.014}}.

\bibitem[{A\c{s}\i{}k and Tezcan(2005)}]{Asik:2005:MMB}
\bibinfo{author}{M.~Z. A\c{s}\i{}k}, \bibinfo{author}{S.~Tezcan},
  \bibinfo{title}{A mathematical model for the behavior of laminated glass
  beams}, \bibinfo{journal}{Computers \& Structures}
  \bibinfo{volume}{83}~(\bibinfo{number}{21--22}) (\bibinfo{year}{2005})
  \bibinfo{pages}{1742--1753},
  \doi{\bibinfo{doi}{10.1016/j.compstruc.2005.02.020}}.

\bibitem[{Schulze et~al.(2012)Schulze, Pander, Naumenko, and
  Altenbach}]{Schultze:2012:ALG}
\bibinfo{author}{S.-H. Schulze}, \bibinfo{author}{M.~Pander},
  \bibinfo{author}{K.~Naumenko}, \bibinfo{author}{H.~Altenbach},
  \bibinfo{title}{Analysis of laminated glass beams for photovoltaic
  applications}, \bibinfo{journal}{International Journal of Solids and
  Structures} \bibinfo{volume}{49}~(\bibinfo{number}{15--16})
  (\bibinfo{year}{2012}) \bibinfo{pages}{2027--2036},
  \doi{\bibinfo{doi}{10.1016/j.ijsolstr.2012.03.028}}.

\bibitem[{Ja\'{s}kowiec et~al.(2015)Ja\'{s}kowiec, Pluci\'{n}ski, and
  Pamin}]{Jaskowiec:2015:XFEM}
\bibinfo{author}{J.~Ja\'{s}kowiec}, \bibinfo{author}{P.~Pluci\'{n}ski},
  \bibinfo{author}{J.~Pamin}, \bibinfo{title}{Thermo-mechanical {XFEM}-type
  modeling of laminated structure with thin inner layer},
  \bibinfo{journal}{Engineering Structures} \bibinfo{volume}{100}
  (\bibinfo{year}{2015}) \bibinfo{pages}{511--521},
  \doi{\bibinfo{doi}{10.1016/j.engstruct.2015.06.035}}.

\bibitem[{Galuppi and Royer-Carfagni(2012{\natexlab{b}})}]{Galuppi:2012:LBV}
\bibinfo{author}{L.~Galuppi}, \bibinfo{author}{G.~F. Royer-Carfagni},
  \bibinfo{title}{Laminated beams with viscoelastic interlayer},
  \bibinfo{journal}{International Journal of Solids and Structures}
  \bibinfo{volume}{49}~(\bibinfo{number}{18})
  (\bibinfo{year}{2012}{\natexlab{b}}) \bibinfo{pages}{2637--2645},
  \doi{\bibinfo{doi}{10.1016/j.ijsolstr.2012.05.028}}.

\bibitem[{Galuppi and Royer-Carfagni(2013)}]{Galuppi:2013:DLG}
\bibinfo{author}{L.~Galuppi}, \bibinfo{author}{G.~Royer-Carfagni},
  \bibinfo{title}{The design of laminated glass under time-dependent loading},
  \bibinfo{journal}{International Journal of Mechanical Sciences}
  \bibinfo{volume}{68} (\bibinfo{year}{2013}) \bibinfo{pages}{67--75},
  \doi{\bibinfo{doi}{10.1016/j.ijmecsci.2012.12.019}}.

\bibitem[{Galuppi and Royer-Carfagni(2014)}]{Galuppi:2014:BTL}
\bibinfo{author}{L.~Galuppi}, \bibinfo{author}{G.~Royer-Carfagni},
  \bibinfo{title}{Buckling of three-layered composite beams with viscoelastic
  interaction}, \bibinfo{journal}{Composite Structures} \bibinfo{volume}{107}
  (\bibinfo{year}{2014}) \bibinfo{pages}{512--521},
  \doi{\bibinfo{doi}{10.1016/j.compstruct.2013.08.006}}.

\bibitem[{Galuppi and Royer-Carfagni(2015)}]{Galuppi:2015:LCS}
\bibinfo{author}{L.~Galuppi}, \bibinfo{author}{G.~Royer-Carfagni},
  \bibinfo{title}{Localized contacts, stress concentrations and transient
  states in bent-lamination with viscoelastic adhesion. An analytical study},
  \bibinfo{journal}{International Journal of Mechanical Sciences}
  \bibinfo{volume}{103} (\bibinfo{year}{2015}) \bibinfo{pages}{275--287},
  \doi{\bibinfo{doi}{10.1016/j.ijmecsci.2015.08.016}}.

\bibitem[{Wu et~al.(2016)Wu, Zhou, Liu, Wan, and Liu}]{Wu:2016:EST}
\bibinfo{author}{P.~Wu}, \bibinfo{author}{D.~Zhou}, \bibinfo{author}{W.~Liu},
  \bibinfo{author}{L.~Wan}, \bibinfo{author}{D.~Liu},
  \bibinfo{title}{Elasticity solution of two-layer beam with a viscoelastic
  interlayer considering memory effect}, \bibinfo{journal}{International
  Journal of Solids and Structures} \bibinfo{volume}{94--95}
  (\bibinfo{year}{2016}) \bibinfo{pages}{76--86},
  \doi{\bibinfo{doi}{10.1016/j.ijsolstr.2016.05.007}}.

\bibitem[{Duser et~al.(1999)Duser, Jagota, and Bennison}]{Duser:1999:AGBL}
\bibinfo{author}{A.~V. Duser}, \bibinfo{author}{A.~Jagota},
  \bibinfo{author}{S.~J. Bennison}, \bibinfo{title}{Analysis of glass/polyvinyl
  butyral laminates subjected to uniform pressure}, \bibinfo{journal}{Journal
  of Engineering Mechanics} \bibinfo{volume}{125}~(\bibinfo{number}{4})
  (\bibinfo{year}{1999}) \bibinfo{pages}{435--442},
  \doi{\bibinfo{doi}{10.1061/(ASCE)0733-9399(1999)125:4(435)}}.

\bibitem[{Bennison et~al.(1999)Bennison, Jagota, and Smith}]{Bennison:1999:FLB}
\bibinfo{author}{S.~J. Bennison}, \bibinfo{author}{A.~Jagota},
  \bibinfo{author}{C.~A. Smith}, \bibinfo{title}{Fracture of {Glass/Poly(vinyl
  butyral) (Butacite$^\text{\textregistered}$)} Laminates in Biaxial Flexure},
  \bibinfo{journal}{Journal of the American Ceramic Society}
  \bibinfo{volume}{82}~(\bibinfo{number}{7}) (\bibinfo{year}{1999})
  \bibinfo{pages}{1761--1770},
  \doi{\bibinfo{doi}{10.1111/j.1151-2916.1999.tb01997.x}}.

\bibitem[{Bedon et~al.(2014)Bedon, Belis, and Luible}]{Bedon:2014:AEA}
\bibinfo{author}{C.~Bedon}, \bibinfo{author}{J.~Belis},
  \bibinfo{author}{A.~Luible}, \bibinfo{title}{Assessment of existing
  analytical models for the lateral torsional buckling analysis of {PVB} and
  {SG} laminated glass beams via viscoelastic simulations and experiments},
  \bibinfo{journal}{Engineering Structures} \bibinfo{volume}{60}
  (\bibinfo{year}{2014}) \bibinfo{pages}{52--67},
  \doi{\bibinfo{doi}{10.1016/j.engstruct.2013.12.012}}.

\bibitem[{Mau(1973)}]{Mau:1973:RLP}
\bibinfo{author}{S.~T. Mau}, \bibinfo{title}{A refined laminated plate theory},
  \bibinfo{journal}{Journal of Applied Mechanics--Transactions of the ASME}
  \bibinfo{volume}{40}~(\bibinfo{number}{2}) (\bibinfo{year}{1973})
  \bibinfo{pages}{606--607}, \doi{\bibinfo{doi}{10.1115/1.3423032}}.

\bibitem[{Zienkiewicz et~al.(1968)Zienkiewicz, Watson, and
  King}]{Zienkiewicz:1968:NVS}
\bibinfo{author}{O.~C. Zienkiewicz}, \bibinfo{author}{M.~Watson},
  \bibinfo{author}{I.~P. King}, \bibinfo{title}{A numerical method of
  visco-elastic stress analysis}, \bibinfo{journal}{International Journal of
  Mechanical Sciences} \bibinfo{volume}{10}~(\bibinfo{number}{10})
  (\bibinfo{year}{1968}) \bibinfo{pages}{807--827},
  \doi{\bibinfo{doi}{10.1016/0020-7403(68)90022-2}}.

\bibitem[{Zemanov\'{a} et~al.(2015)Zemanov\'{a}, Zeman, and
  \v{S}ejnoha}]{Zemanova:2015:FEM}
\bibinfo{author}{A.~Zemanov\'{a}}, \bibinfo{author}{J.~Zeman},
  \bibinfo{author}{M.~\v{S}ejnoha}, \bibinfo{title}{Finite element model based
  on refined plate theories for laminated glass units}, \bibinfo{journal}{Latin
  American Journal of Solids and Structures}
  \bibinfo{volume}{15}~(\bibinfo{number}{6}) (\bibinfo{year}{2015})
  \bibinfo{pages}{1158--1180}, \doi{\bibinfo{doi}{10.1590/1679-78251676}}.

\bibitem[{Kruis et~al.(2013)Kruis, Zeman, and Gruber}]{Kruis:2013:MII}
\bibinfo{author}{J.~Kruis}, \bibinfo{author}{J.~Zeman},
  \bibinfo{author}{P.~Gruber}, \bibinfo{title}{Model of Imperfect Interfaces in
  Composite Materials and Its Numerical Solution by {FETI} Method}, in:
  \bibinfo{editor}{R.~Bank}, \bibinfo{editor}{M.~Holst},
  \bibinfo{editor}{O.~Widlund}, \bibinfo{editor}{J.~Xu} (Eds.),
  \bibinfo{booktitle}{Domain Decomposition Methods in Science and Engineering
  XX}, no.~\bibinfo{number}{91} in \bibinfo{series}{Lecture Notes in
  Computational Science and Engineering}, \bibinfo{publisher}{Springer Berlin
  Heidelberg}, \bibinfo{pages}{337--344},
  \doi{\bibinfo{doi}{10.1007/978-3-642-35275-1_39}}, \bibinfo{year}{2013}.

\bibitem[{Larcher et~al.(2012)Larcher, Solomos, Casadei, and
  Gebbeken}]{Larcher:2012:ENI}
\bibinfo{author}{M.~Larcher}, \bibinfo{author}{G.~Solomos},
  \bibinfo{author}{F.~Casadei}, \bibinfo{author}{N.~Gebbeken},
  \bibinfo{title}{Experimental and numerical investigations of laminated glass
  subjected to blast loading}, \bibinfo{journal}{International Journal of
  Impact Engineering} \bibinfo{volume}{39}~(\bibinfo{number}{1})
  (\bibinfo{year}{2012}) \bibinfo{pages}{42--50},
  \doi{\bibinfo{doi}{10.1016/j.ijimpeng.2011.09.006}}.

\bibitem[{Reissner(1972)}]{Reissner:1972:ODFSBT}
\bibinfo{author}{E.~Reissner}, \bibinfo{title}{On One-Dimensional Finite-Strain
  Beam Theory: the Plane Problem}, \bibinfo{journal}{Journal of Applied
  Mathematics and Physics} \bibinfo{volume}{23}~(\bibinfo{number}{5})
  (\bibinfo{year}{1972}) \bibinfo{pages}{795--804},
  \doi{\bibinfo{doi}{10.1007/BF01602645}}.

\bibitem[{Reddy(2004)}]{Reddy:2004:FEM}
\bibinfo{author}{J.~N. Reddy}, \bibinfo{title}{An Introduction to Nonlinear
  Finite Element Analysis}, \bibinfo{publisher}{Oxford University Press},
  \doi{\bibinfo{doi}{10.1093/acprof:oso/9780198525295.001.0001}},
  \bibinfo{year}{2004}.

\bibitem[{Pelayo et~al.(2013)Pelayo, L\'{o}pez-Aenlle, Hermans, and
  Fraile}]{Pelayo:2013:MSLGP}
\bibinfo{author}{F.~Pelayo}, \bibinfo{author}{M.~L\'{o}pez-Aenlle},
  \bibinfo{author}{L.~Hermans}, \bibinfo{author}{A.~Fraile},
  \bibinfo{title}{Modal Scaling of a Laminated Glass Plate},
  \bibinfo{journal}{5th International Operational Modal Analysis Conference}
  (\bibinfo{year}{2013})
  \bibinfo{pages}{1--10}\urlprefix\url{http://iomac.eu/iomac/2013/IOMAC_Guimaraes/files/papers/Model%20Updating/Paper-175-Fernandes.pdf}.

\bibitem[{Williams et~al.(1955)Williams, Landel, and Ferry}]{Williams:1955:TDR}
\bibinfo{author}{M.~Williams}, \bibinfo{author}{R.~Landel},
  \bibinfo{author}{J.~Ferry}, \bibinfo{title}{The Temperature Dependence of
  Relaxation Mechanisms in Amorphous Polymers and Other Glass-forming Liquids},
  \bibinfo{journal}{Journal of the American Chemical Society}
  \bibinfo{volume}{77}~(\bibinfo{number}{14}) (\bibinfo{year}{1955})
  \bibinfo{pages}{3701--3707}, \doi{\bibinfo{doi}{10.1021/ja01619a008}}.

\bibitem[{Christensen(1982)}]{Christensen:1982:TVI}
\bibinfo{author}{R.~Christensen}, \bibinfo{title}{Theory of Viscoelasticity:
  {An} Introduction}, \bibinfo{publisher}{Elsevier}, \bibinfo{edition}{second}
  edn., \bibinfo{year}{1982}.

\bibitem[{Jir\'{a}sek and Ba\v{z}ant(2002)}]{Jirasek:2002:IAS}
\bibinfo{author}{M.~Jir\'{a}sek}, \bibinfo{author}{Z.~P. Ba\v{z}ant},
  \bibinfo{title}{Inelastic analysis of structures}, \bibinfo{publisher}{John
  Wiley \& Sons, Ltd.}, \bibinfo{year}{2002}.

\bibitem[{Sokolnikoff(1956)}]{Sokolnikoff:1956:MTE}
\bibinfo{author}{I.~S. Sokolnikoff}, \bibinfo{title}{Mathematical theory of
  elasticity}, \bibinfo{publisher}{McGraw-Hill Book Company, Inc.},
  \bibinfo{address}{New York, Toronto, London}, \bibinfo{edition}{second} edn.,
  \bibinfo{year}{1956}.

\bibitem[{Birman and Bert(2002)}]{Birman:2002:CSC}
\bibinfo{author}{V.~Birman}, \bibinfo{author}{C.~Bert}, \bibinfo{title}{On the
  Choice of Shear Correction Factor in Sandwich Structures},
  \bibinfo{journal}{Journal of Sandwich Structures \& Materials}
  \bibinfo{volume}{4}~(\bibinfo{number}{1}) (\bibinfo{year}{2002})
  \bibinfo{pages}{83--95}, \doi{\bibinfo{doi}{10.1177/1099636202004001180}}.

\bibitem[{Bonnans et~al.(2003)Bonnans, Gilbert, Lemar\'{e}chal, and
  Sagastiz\'{a}bal}]{Bonnans:2003:NOTPA}
\bibinfo{author}{J.~F. Bonnans}, \bibinfo{author}{J.~C. Gilbert},
  \bibinfo{author}{C.~Lemar\'{e}chal}, \bibinfo{author}{C.~A.
  Sagastiz\'{a}bal}, \bibinfo{title}{Numerical Optimization: {Theoretical} and
  Practical Aspects}, \bibinfo{publisher}{Springer},
  \doi{\bibinfo{doi}{10.1007/978-3-540-35447-5}}, \bibinfo{year}{2003}.

\bibitem[{Zemanov\'{a} et~al.(2008)Zemanov\'{a}, Zeman, and
  \v{S}ejnoha}]{Zemanova:2008:SNM}
\bibinfo{author}{A.~Zemanov\'{a}}, \bibinfo{author}{J.~Zeman},
  \bibinfo{author}{M.~\v{S}ejnoha}, \bibinfo{title}{Simple numerical model of
  laminated glass beams}, \bibinfo{journal}{Acta Polytechnica}
  \bibinfo{volume}{48}~(\bibinfo{number}{6}) (\bibinfo{year}{2008})
  \bibinfo{pages}{22--26}.

\bibitem[{A\c{s}\i{}k(2003)}]{Asik:2003:LGP}
\bibinfo{author}{M.~Z. A\c{s}\i{}k}, \bibinfo{title}{Laminated glass plates:
  revealing of nonlinear behavior}, \bibinfo{journal}{Computers \& Structures}
  \bibinfo{volume}{81}~(\bibinfo{number}{28--29}) (\bibinfo{year}{2003})
  \bibinfo{pages}{2659--2671},
  \doi{\bibinfo{doi}{10.1016/S0045-7949(03)00325-0}}.

\bibitem[{{prEN 16612}(2013)}]{prEN:16612:2013}
\bibinfo{author}{{prEN 16612}}, \bibinfo{title}{Glass in building -
  {Determination} of the load resistance of glass panes by calculation and
  testing}, \bibinfo{type}{European Strandard}, \bibinfo{institution}{European
  Committee for Standardization}, \bibinfo{year}{2013}.

\bibitem[{Behr et~al.(1993)Behr, Minor, and Norville}]{Behr:1993:SBA}
\bibinfo{author}{R.~A. Behr}, \bibinfo{author}{J.~E. Minor},
  \bibinfo{author}{H.~S. Norville}, \bibinfo{title}{Structural Behavior of
  Architectural Laminated Glass}, \bibinfo{journal}{Journal of Structural
  Engineering} \bibinfo{volume}{119}~(\bibinfo{number}{1})
  (\bibinfo{year}{1993}) \bibinfo{pages}{202--222},
  \doi{\bibinfo{doi}{10.1061/(ASCE)0733-9445(1993)119:1(202)}}.

\bibitem[{Zemanov\'{a}(2014)}]{Zemanova:2014:NMLG}
\bibinfo{author}{A.~Zemanov\'{a}}, \bibinfo{title}{Numerical modeling of
  laminated glass structures}, Ph.D. thesis, \bibinfo{school}{Faculty of Civil
  Engineering, CTU}, \bibinfo{address}{Prague}, \bibinfo{year}{2014}.

\end{thebibliography}

\appendix
\renewcommand*{\thesection}{\Alph{section}}

\section{Numerical aspects}\label{app:sensitivity_analysis}

The finite element discretization is accomplished with two-node elements with linear basis functions for each layer, and all integrals are evaluated by the one-point quadrature to mitigate locking effects. Therefore, the kinematics for the $e$-th element of the $i$-th layer is specified by the column matrix of nodal displacements and rotations
\begin{align*}
\Md\el\lay{i} 
= 
\begin{bmatrix}
\uce{i}{e,1} & \wce{i}{e,1} & \rote{i}{e,1} & \uce{i}{e,2} & \wce{i}{e,2} & \rote{i}{e,2}
\end{bmatrix}
\trn,
\end{align*}
where, e.g., $\uce{i}{e,1}$ stands for the centerline displacements of the left node and $\rote{i}{e,2}$ denotes the rotation of the cross-section  at the right node. These choices enables us to express, in the explicit form, all matrices introduced in Eqs.~\eqref{eq6:system_terms} and~\eqref{eq:matrices_block_diagonal}, as shown in the remainder of this section.

\subsection{von K\'{a}rm\'{a}n beams}\label{app:sensitivity_analysisVK}

\paragraph{Nodal forces}
The matrix of internal nodal forces follows by differentiating the element internal energy function with respect to the nodal displacements, recall~\eqref{eq6:system_terms}. For the von K\'{a}rm\'{a}n formulation, we obtain 
\begin{subequations}\label{eq:element_internal_forces_VK}
\begin{align}
{f}_{\mathrm{int},e,1}\lay{i} 
& =
- \E\lay{i} \A{i} \strainxce{i}, 
&
{f}_{\mathrm{int},e,4}\lay{i} 
& =
- {f}_{\mathrm{int},e,1}\lay{i},
\\
{f}_{\mathrm{int},e,2}\lay{i} & =
{f}_{\mathrm{int},e,1}\lay{i}
\frac{\wce{i}{e,2} - \wce{i}{e,1}}{\Le}
- 
\G\lay{i} \As{i}\sstrainxze{i},  
& 
{f}_{\mathrm{int},e,5}\lay{i} 
& =
-
{f}_{\mathrm{int},e,2}\lay{i},
\\
{f}_{\mathrm{int},e,3}\lay{i} 
& =
\G\lay{i} \As{i} \sstrainxze{i}\frac{\Le}{2}  
- 
\E\lay{i} \I{i}_y \curve{i}{e},
&
{f}_{\mathrm{int},e,6}\lay{i} 
& =
{f}_{\mathrm{int},e,3}\lay{i}
+
2 \E\lay{i} \I{i}_y \curve{i}{e}, 
\end{align}
\end{subequations}
where we have introduced the generalized element strains associated with $\Md\el\lay{i}$:
%
\begin{align*}
\strainxce{i} 
& =
\frac{\uce{i}{e,2} - \uce{i}{e,1}}{\Le} 
+ 
\half 
\Bigl( \frac{\wce{i}{e,2} - \wce{i}{e,1}}{\Le} \Bigr)^2, 
& 
\sstrainxze{i}
& =
\half \left( \rote{i}{e,1} + \rote{i}{e,2} \right)
+
\frac{\wce{i}{e,2} - \wce{i}{e,1}}{\Le},
\\
\curve{i}{e}
& =
\frac{\rote{i}{e,2} - \rote{i}{e,1}}{\Le},
\end{align*}
%
according to~\eqref{eq:VKstrain} and~\eqref{eq:VK_normal_strain}. Note that Eq.~\eqref{eq:element_internal_forces_VK} in fact holds only for $i=1,3$, the matrix $\widehat{\M{f}}_{\mathrm{int},e}\lay{2}$ related to the interlayer must be obtained with effective values of the Young and shear moduli, $\widehat{\E}\lay{2}$ and $\widehat{\G}\lay{2}$, introduced in \Sref{sec:constitutive}.

In order to derive the expression for the additional contribution to the nodal force matrix $\Delta\widehat{\M{f}}_{\mathrm{int},e}\lay{2}$, it is convenient to introduce the following history variables
\begin{align*}
	\delta N_{x,e}\lay{2}(t_n) 
	& = 
	N_{x,e}\lay{2}(t_n) 
	+ 
	\Delta \widehat{N}_{x,e}\lay{2} 
	-
	\widehat{\E}\lay{2} \A{2} \strainxce{2}(t_n),
	\\
	\delta V_{z,e}\lay{2}(t_n) 
	& = 
	V_{z,e}\lay{2}(t_n) 
	+ 
	\Delta \widehat{V}_{z,e}\lay{2} 
	-	
	\widehat{\G}\lay{2} \As{2} \sstrainxze{2}(t_n),
	\\
	\delta M_{y,e}\lay{2}(t_n) 
	& =
    M_{y,e}\lay{2}(t_n) + \Delta \widehat{M}_{y,e}\lay{2} 
	-
	\widehat{\E}\lay{2} I_y\lay{2} \curve{2}{e}(t_n),
\end{align*}
that account for initial internal forces and generalized strains at the beginning of time step $t_n$ and for the relaxation effects. Using these quantities, $\Delta\widehat{\M{f}}_{\mathrm{int},e}\lay{2}$ receives a form similar to Eq.~\eqref{eq:element_internal_forces_VK}:
\begin{align*}
\Delta \widehat{f}_{\mathrm{int},e,1}\lay{2} 
& =
- \delta N_{x,e}\lay{2}(t_n),   
&
\Delta \widehat{f}_{\mathrm{int},e,4}\lay{2} 
& =
- \Delta \widehat{f}_{\mathrm{int},e,1}\lay{2},
\\
\Delta \widehat{f}_{\mathrm{int},e,2}\lay{2} 
& =
\Delta \widehat{f}_{\mathrm{int},e,1}\lay{2}
\frac{\wce{2}{e,2} - \wce{2}{e,1}}{\Le}
-
\delta V_{z,e}\lay{2}(t_n), 
&
\Delta \widehat{f}_{\mathrm{int},e,5}\lay{2} 
& =
- \Delta \widehat{f}_{\mathrm{int},e,2}\lay{2},
\\
\Delta \widehat{f}_{\mathrm{int},e,3}\lay{2} 
& =
\delta V_{z,e}\lay{2}(t_n)\frac{\Le}{2}
-
\delta M_{y,e}\lay{2}(t_n)
&
\Delta \widehat{f}_{\mathrm{int},e,6}\lay{2} & =
\Delta \widehat{f}_{\mathrm{int},e,3}\lay{2}
+ 2 \delta M_{y,e}\lay{2}(t_n).
\end{align*}

\paragraph{Stiffness matrices}

Upon differentiating nodal forces with respect to nodal displacements, we obtain the stiffness matrix in the form
\begin{equation*}
\M{K}_{\mathrm t, e}\lay{i} =
\left[
\begin{array}{cccccc}
 \Kt{e,11}\lay{i} &  \Kt{e,12}\lay{i} &         0         & - \Kt{e,11}\lay{i} & - \Kt{e,12}\lay{i} &         0         \\
 \Kt{e,12}\lay{i} &  \Kt{e,22}\lay{i} &  \Kt{e,23}\lay{i} & - \Kt{e,12}\lay{i} & - \Kt{e,22}\lay{i} &  \Kt{e,23}\lay{i} \\
         0        &  \Kt{e,23}\lay{i} &  \Kt{e,33}\lay{i} &         0          & - \Kt{e,23}\lay{i} &  \Kt{e,36}\lay{i} \\
-\Kt{e,11}\lay{i} & -\Kt{e,12}\lay{i} &         0         &   \Kt{e,11}\lay{i} &   \Kt{e,12}\lay{i} &         0         \\
-\Kt{e,12}\lay{i} & -\Kt{e,22}\lay{i} & -\Kt{e,23}\lay{i} &   \Kt{e,12}\lay{i} &   \Kt{e,22}\lay{i} & -\Kt{e,23}\lay{i} \\
         0        &  \Kt{e,23}\lay{i} &  \Kt{e,36}\lay{i} &         0          & - \Kt{e,23}\lay{i} &  \Kt{e,33}\lay{i} 
\end{array}
\right],
\label{eq:K_e}
\end{equation*}
where the individual entries read
\begin{align*}
\Kt{e,11}\lay{i} 
& =   
\frac{\E\lay{i} \A{i}}{\Le}, 
& 
\Kt{e,12}\lay{i} 
& = 
\E\lay{i} \A{i} 
\frac{\wce{i}{e,2} - \wce{i}{e,1}}{\Le^2} , 
\\
\Kt{e,22}\lay{i} 
& =  
\Kt{e,11}\lay{i} \strainxce{i}
+ 
\Kt{e,12}\lay{i} 
\frac{\wce{i}{e,2} - \wce{i}{e,1}}{\Le} 
+ 
\frac{\G\lay{i} \As{i}}{\Le}, 
&
\Kt{e,23}\lay{i} 
& = 
- \half \G\lay{i} \As{i},
\\
\Kt{e,33}\lay{i} 
& = 
\tfrac{1}{4} \G\lay{i} \As{i} \Le + \frac{\E\lay{i} \I{i}_y}{\Le},
&
\Kt{e,36}\lay{i} 
&= 
\Kt{e,33}\lay{i} - \frac{2 \E\lay{i} \I{i}_y}{\Le}.
\end{align*}
Likewise, the history-dependent increment of the nodal forces yields an additional contribution to the tangent stiffness matrix
\begin{equation*}
\Delta \widehat{\M{K}}_{\mathrm t, e}\lay{2} =
\left[
\begin{array}{cccccc}
 0 & 0 & 0 & 0 & 0 & 0 \\
 0 &   \Delta \widehat{{K}}_{\mathrm t, e,22}\lay{2} & 0 & 0 & - \Delta \widehat{{K}}_{\mathrm t, e,22}\lay{2} & 0 \\
 0 & 0 & 0 & 0 & 0 & 0 \\
 0 & 0 & 0 & 0 & 0 & 0 \\
 0 & - \Delta \widehat{{K}}_{\mathrm t, e,22}\lay{2} & 0 & 0 &   \Delta \widehat{{K}}_{\mathrm t, e,22}\lay{2} & 0 \\
 0 & 0 & 0 & 0 & 0 & 0
\end{array} 
\right],
\label{eq:K_e2}
\end{equation*}
with $\Delta \widehat{{K}}_{\mathrm t, e,22}\lay{2} = \delta N_{x,e}\lay{2}(t_n) / \Le$.

\paragraph{Compatibility}

The inter-layer compatibility~\eqref{eq:comp_conVK} conditions, written for the $j$-th node for the interface between the $i$-th and $(i+1)$-th layer, receives the form
\begin{align*}
\M{c}\lay{i,i+1}_j
=
\begin{bmatrix}
\uce{i}{j} - \uce{i+1}{j} 
+ 
\half\h{i}\rote{i}{j} + \half\h{i+1} \rote{i+1}{j} 
\\  
\wce{i}{j} - \wce{i+1}{j} 
\end{bmatrix}
=
\M{C}\lay{i,i+1}_j\Md,
\end{align*}
from which we obtain the corresponding block of the constraining matrix
\begin{align*}
\M{C}\lay{i,i+1}_{j} 
= 
\begin{bmatrix}
1 & 0 & \half \h{i} & \cdots & 
-1 & 0 & \half \h{i+1} \\ 
0 & 1 & 0 & \cdots &  
0 &-1 & 0
\end{bmatrix}.
\end{align*}
Note that this expression coincides with the geometrically linear formulation developed by~\citet{Zemanova:2008:SNM}.

\subsection{Reissner beams}\label{app:sensitivity_analysisFS}

\paragraph{Nodal forces}

Explicit expressions for nodal forces $\M{f}_{\mathrm{int},e}\lay{i}$ in the elastic formulation are available in~\cite[Eq.~(37)]{Zemanova:2014:NMFS}. Therefore, we provide here only the history-dependent contribution:
\begin{align*}
\Delta \widehat{f}_{\mathrm{int},e,1}\lay{2} 
& =
- \delta N_{x,e}\lay{2}(t_n) \cos \beta\lay{2}\el 
- \delta V_{z,e}\lay{2}(t_n) \sin\beta\lay{2}\el,
\quad 
\Delta \widehat{f}_{\mathrm{int},e,4}\lay{2} =
- \Delta \widehat{f}_{\mathrm{int},e,1}\lay{2},
\\
\Delta \widehat{f}_{\mathrm{int},e,2}\lay{2} 
& =
\delta N_{x,e}\lay{2}(t_n) \sin\beta\lay{2}\el 
- \delta V_{z,e}\lay{2}(t_n) \cos\beta\lay{2}\el,
\quad
\Delta \widehat{f}_{\mathrm{int},e,5}\lay{2} 
=
- \Delta \widehat{f}_{\mathrm{int},e,2}\lay{2}, 
\\
\Delta \widehat{f}_{\mathrm{int},e,3}\lay{2} 
& =
- \half \left( \eL{e}{2} + \uce{2}{e,2} - \uce{2}{e,1}\right) 
\Delta \widehat{f}_{\mathrm{int},e,2}\lay{2} 
+ 
\half 
\left( \wce{2}{e,2} - \wce{2}{e,1} \right)
\Delta \widehat{f}_{\mathrm{int},e,1}\lay{2}
- \delta M_{y,e}\lay{2}(t_n),
\\
\Delta \widehat{f}_{\mathrm{int},e,6}\lay{2} & =
\Delta \widehat{f}_{\mathrm{int},e,3}\lay{2}
+ 2 \delta M_{y,e}\lay{2}(t_n),
\end{align*}
with $\beta\lay{i}\el = \half (\rote{i}{e,1} + \rote{i}{e,2})$.

\paragraph{Stiffness matrices}

Again, the expressions for the tangent stiffness matrix $\M{K}_{\mathrm t}$ and the contribution of the Lagrange multipliers $\M{K}_\lambda$ are available in~\cite[Eqs.~(38) and~(40)]{Zemanova:2014:NMFS}, therefore they are omitted here for the sake of brevity. The history-dependent contribution attains the form
\begin{equation*}
\Delta \widehat{\M{K}}_{\mathrm t, e}\lay{2} =
\left[
\begin{array}{cccccc}
 0 & 0 & \Delta \widehat{{K}}_{\mathrm t, e,13}\lay{2} & 0 & 0 & \Delta \widehat{{K}}_{\mathrm t, e,13}\lay{2} \\
 0 & 0 & \Delta \widehat{{K}}_{\mathrm t, e,23}\lay{2} & 0 & 0 & \Delta \widehat{{K}}_{\mathrm t, e,23}\lay{2} \\
 \Delta \widehat{{K}}_{\mathrm t, e,13}\lay{2} & \Delta \widehat{{K}}_{\mathrm t, e,23}\lay{2} & \Delta \widehat{{K}}_{\mathrm t, e,33}\lay{2} & -\Delta \widehat{{K}}_{\mathrm t, e,13}\lay{2} & -\Delta \widehat{{K}}_{\mathrm t, e,23}\lay{2} & \Delta \widehat{{K}}_{\mathrm t, e,33}\lay{2} \\
 0 & 0 & -\Delta \widehat{{K}}_{\mathrm t, e,13}\lay{2} & 0 & 0 & -\Delta \widehat{{K}}_{\mathrm t, e,13}\lay{i} \\
 0 & 0 & -\Delta \widehat{{K}}_{\mathrm t, e,23}\lay{2} & 0 & 0 & -\Delta \widehat{{K}}_{\mathrm t, e,23}\lay{i} \\
 \Delta \widehat{{K}}_{\mathrm t, e,13}\lay{2} & \Delta \widehat{{K}}_{\mathrm t, e,23}\lay{2} & \Delta \widehat{{K}}_{\mathrm t, e,33}\lay{2} & -\Delta \widehat{{K}}_{\mathrm t, e,13}\lay{2} & -\Delta \widehat{{K}}_{\mathrm t, e,23}\lay{2} & \Delta \widehat{{K}}_{\mathrm t, e,33}\lay{2}
\end{array} 
\right],
\label{eq:K_e3}
\end{equation*}
where
\begin{align*}
\Delta \widehat{{K}}_{\mathrm t, e,13}\lay{2} 
&= 
\frac{1}{2}
\left( 
\delta N_{x,e}\lay{2}(t_n) \sin\beta\lay{2}\el 
-
\delta V_{z,e}\lay{2}(t_n) \cos\beta\lay{2}\el 
\right),
\\
\Delta \widehat{{K}}_{\mathrm t, e,23}\lay{2} 
&= 
\frac{1}{2}
\left(
\delta N_{x,e}\lay{2}(t_n) \cos\beta\lay{2}\el 
+
\delta V_{z,e}\lay{2}(t_n) \sin\beta\lay{2}\el
\right),
\\
\Delta \widehat{{K}}_{\mathrm t, e,33}\lay{2} &= 
\frac{1}{2}
\left( -\left( 
  \eL{e}{2} 
  + 
  \uce{2}{e,2} - \uce{2}{e,1}
\right) 
\Delta \widehat{{K}}_{\mathrm t, e,23}\lay{2} 
+ 
\left( \wce{2}{e,2} - \wce{2}{e,1} \right)
\Delta \widehat{{K}}_{\mathrm t, e,13}\lay{2} \right).
\end{align*}

\paragraph{Compatibility}\label{app:sensitivity_analysisVK_c}

Expressions for the nodal block of the compatibility condition $\M{c}\lay{i,i+1}_j$ and its gradients $\M{C}\lay{i,i+1}_j$ can be found in~\cite[Eqs.~(23) and~(39)]{Zemanova:2014:NMFS}.

\end{document}